\pdfoutput=1
\RequirePackage{fix-cm}
\documentclass[reqno]{amsproc}
\usepackage{amssymb}
\usepackage{amsthm}
\usepackage{mathtools}
\usepackage{lmodern}

\usepackage[unicode,psdextra]{hyperref}

\usepackage{tikz}
\usetikzlibrary{arrows.meta}
\tikzset{> = {Straight Barb[scale=.75]}}

\usepackage[citation-order,nobysame]{amsrefs}
\usepackage{xyzbib}

\usepackage{graphicx,subcaption}

\mathtoolsset{showonlyrefs}

\numberwithin{equation}{section}
\let\cite=\cites

\newtheorem{theorem}{Theorem}[section]
\newtheorem{proposition}{Proposition}[section]
\newtheorem{corollary}{Corollary}[proposition]

\newtheorem{remark}{Remark}[section]

\DeclareMathOperator\sgn{\mathrm{sgn}}
\DeclareMathOperator\tr{\mathrm{tr}}

\newcommand\PL{\mathrm{PL}}
\newcommand\xc{x_\mathrm{c}}
\newcommand\yc{y_\mathrm{c}}
\newcommand\eps{\varepsilon}

\newcommand\wt{\widetilde}
\newcommand\pd{\partial}

\newcommand{\Ftwoone}[4]{%
\,{}_{2}F_{1}\bigg(\genfrac{}{}{0pt}{}{#1,\, #2}{#3} \bigg\vert #4\bigg)}

\newcommand{\Fthreetwo}[5]{%
\,{}_{3}F_{2}\bigg(\genfrac{}{}{0pt}{}{#1,\, #2,\, #3}{#4,\, #5} \bigg\vert 1\bigg)}

\newcommand\rmd{\mathrm{d}}
\newcommand\rme{\mathrm{e}}
\renewcommand{\leq}{\leqslant}
\renewcommand{\geq}{\geqslant}

\begin{document}
\vspace*{.5in}

\title{Thermodynamics of the five-vertex model with scalar-product 
boundary conditions}

\author{Ivan N. Burenev}
\address{Steklov Mathematical Institute, 
Fontanka 27, St.~Petersburg, 191023, Russia}
\email{inburenev@gmail.com}

\author{Andrei G. Pronko}
\address{Steklov Mathematical Institute, 
Fontanka 27, St.~Petersburg, 191023, Russia}
\email{agp@pdmi.ras.ru}

\begin{abstract} 
We consider the homogeneous five-vertex model on a rectangle domain of the
square lattice with so-called scalar-product boundary conditions. Peculiarity 
of these boundary conditions is that the
configurations of the model are in an one-to-one correspondence with the
3D Young diagrams limited by a box of a given size.  
We address the thermodynamics of the model using 
a connection of the partition function with the $\tau$-function of 
the sixth Painlev\'e equation. We compute an expansion of the logarithm 
of the partition function to the order of a constant in the size of the system. 
We find that the geometry of the domain 
is crucial for phase transition phenomena.  
Two cases need to be considered separately: 
one is where the region has an asymptotically 
square shape and the second one is where it is of an arbitrary rectangle, but not
square, shape. In the first case there are three regimes, which can be attributed to  
dominance in the configurations of a ferroelectric order, disorder, and 
anti-ferroelectric order. In the second case the third regime is absent. 
\end{abstract}
\maketitle
\tableofcontents

\section{Introduction}

The five-vertex model had originally emerged for modeling of crystal growth
and evaporation in two dimensions \cite{G-90,GLT-90,GS-92}. For periodic
boundary conditions its thermodynamic properties, including the phase
diagram, have been completely understood by Bethe ansatz methods 
\cite{GvBL-93,HWKK-96}. As the same time, it is known that the six-vertex model
(and hence the five-vertex model as its descendant) is sensitive to boundary
conditions. A paradigmatic example here is the six-vertex model with domain
wall boundary conditions \cite{KZj-00,Zj-00,BL-13}.  

As for the five-vertex model, interesting boundary conditions   
are such that the configurations of the model appear
to be in a one-to-one correspondence with the 3D Young diagrams limited by a
box of a given size (or, ``boxed'' plane partitions). These
boundary conditions are special fixed boundary conditions imposed to a
finite-size domain of the square lattice of a rectangular shape. 
They can be seen as a 
generalization of domain wall boundary conditions and called   
``scalar-product'' boundary conditions, as  
they arise when scalar products off-shell Bethe states are interpreted as
partition functions of related vertex models 
\cite{KBI-93,B-10,MSa-13}.        

Recently, a notable progress had been achieved in understanding scaling
properties of the five-vertex model in a rather general setup 
by variational methods, with the focus on phase separation and limit shape phenomena 
\cite{GKW-21,KP-21,KP-22,KP-24}. On the other hand, for the case of scalar-product
boundary conditions an important problem consists in
constructing expansions of the partition function in the limit of large
system size. In the free-fermion case, equivalent to the dimer model on 
a hexagonal domain (boxed 3D Young diagrams), a solution of this problem has been
provided in \cite{NP-20}. 

In the present paper, we consider the five-vertex model
with scalar-product boundary conditions 
and derive an expansion for the logarithm of 
the partition function for large lattice sizes. We obtain explicitly terms 
to the order of a constant, including the logarithmic terms.
In \cite{BP-21}, we have derived various determinant formulas for the partition function 
of the five-vertex model with scalar-product boundary conditions and 
showed that one of these representations   
coincides with the $\tau$-function of the sixth Painlev\'e 
equation. 
To derive the asymptotic expansion, we apply here 
the method originally proposed in \cite{KP-16}
which is based on use of the sixth Painlev\'e equation 
in its $\sigma$-form \cite{JM-81,O-87}. Similarly to \cite{KP-16}, we deal with 
an asymptotic expansion of the $\sigma$-function where the coefficients 
are large while the argument is a finite parameter. 

It has to be mentioned that to address the problem of finding 
asymptotic expansions for solutions of Painlev\'e equations  
one can use methods such as the isomonodromy 
deformation techniques \cite{J-82,IN-86} or the asymptotic analysis of the corresponding 
Riemann-Hilbert problem \cite{DZh-93}. Somewhat equivalently, 
one can construct asymptotic expansions by relating the $\tau$-function with 
a random matrix model \cite{FW-04} and formulating the Riemann-Hilbert 
problem for the orthogonal polynomials associated 
to the weight measure \cite{FIK-92}. Specifically, for the present problem 
the matrix model appearing 
on this route has a discrete measure, the corresponding polynomials 
have been studied in \cite{BKMM-07}. 
The method of \cite{KP-16} which we exploit here can be seen as 
an alternative to these approaches, 
and it relies on the theory of asymptotic expansions for solutions 
of ordinary differential equations \cite{W-87}. 

Our main result is collected in two theorems about the 
thermodynamic limit expansion for the logarithm of certain polynomial completely 
determining the partition function. We find that this expansion 
significantly depends on an 
asymptotic form of the domain, namely, whether the region has an asymptotically 
square shape, or the region has an arbitrary rectangle, but not
square, shape. In the former case there are three regimes, which can be attributed to  
a ferroelectric order, disorder, and anti-ferroelectric order. In the latter case 
the third regime is absent. We also illustrate that this extra phase 
transition between the disorder and anti-ferroelectric order for the 
square-shaped domain can be seen as a ``merger transition'' 
discussed recently in \cite{PGA-22}.

\subsection{The model}\label{sec:MD}

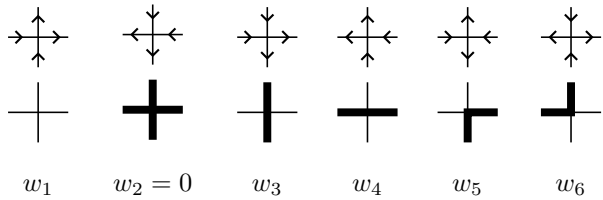
\begin{figure}
\centering
\begin{tikzpicture}[scale=.5]
\draw [semithick] (0.2,3)--(1.8,3);
\draw [semithick] (1,2.2)--(1,3.8);
\draw [thick] [->] (0.5,3)--(.6,3);
\draw [thick] [->] (1.5,3)--(1.6,3);
\draw [thick] [->] (1,2.5)--(1,2.6);
\draw [thick] [->] (1,3.5)--(1,3.6);
\draw [semithick] (0.2,1)--(1.8,1);
\draw [semithick] (1,0.2)--(1,1.8);
\node at (1,-1) {$w_1$};
\end{tikzpicture}
\quad
\begin{tikzpicture}[scale=.5]
\draw [semithick] (0.2,3)--(1.8,3);
\draw [semithick] (1,2.2)--(1,3.8);
\draw [thick] [->] (0.5,3)--(0.4,3);
\draw [thick] [->] (1.5,3)--(1.4,3);
\draw [thick] [->] (1,2.5)--(1,2.4);
\draw [thick] [->] (1,3.5)--(1,3.4);
\draw [line width=3] (0.2,1)--(1.8,1);
\draw [line width=3] (1,0.2)--(1,1.8);
\node at (1,-1) {$w_2=0$};
\end{tikzpicture}
\quad
\begin{tikzpicture}[scale=.5]
\draw [semithick] (0.2,3)--(1.8,3);
\draw [semithick] (1,2.2)--(1,3.8);
\draw [thick] [->] (0.5,3)--(0.6,3);
\draw [thick] [->] (1.5,3)--(1.6,3);
\draw [thick] [->] (1,2.5)--(1,2.4);
\draw [thick] [->] (1,3.5)--(1,3.4);
\draw [semithick] (0.2,1)--(1.8,1);
\draw [line width=3] (1,0.2)--(1,1.8);
\node at (1,-1) {$w_3$};
\end{tikzpicture}
\quad
\begin{tikzpicture}[scale=.5]
\draw [semithick] (0.2,3)--(1.8,3);
\draw [semithick] (1,2.2)--(1,3.8);
\draw [thick] [->] (0.5,3)--(0.4,3);
\draw [thick] [->] (1.5,3)--(1.4,3);
\draw [thick] [->] (1,2.5)--(1,2.6);
\draw [thick] [->] (1,3.5)--(1,3.6);
\draw [line width=3] (0.2,1)--(1.8,1);
\draw [semithick] (1,0.2)--(1,1.8);
\node at (1,-1) {$w_4$};
\end{tikzpicture}
\quad
\begin{tikzpicture}[scale=.5]
\draw [semithick] (0.2,3)--(1.8,3);
\draw [semithick] (1,2.2)--(1,3.8);
\draw [thick] [->] (0.5,3)--(.6,3);
\draw [thick] [->] (1.5,3)--(1.4,3);
\draw [thick] [->] (1,2.5)--(1,2.4);
\draw [thick] [->] (1,3.5)--(1,3.6);
\draw [semithick] (0.2,1)--(1,1)--(1,1.8);
\draw [line width=3] (1,0.2)--(1,1)--(1.8,1);
\node at (1,-1) {$w_5$};
\end{tikzpicture}
\quad
\begin{tikzpicture}[scale=.5]
\draw [semithick] (0.2,3)--(1.8,3);
\draw [semithick] (1,2.2)--(1,3.8);
\draw [thick] [->] (0.5,3)--(.4,3);
\draw [thick] [->] (1.5,3)--(1.6,3);
\draw [thick] [->] (1,2.5)--(1,2.6);
\draw [thick] [->] (1,3.5)--(1,3.4);
\draw [line width=3] (0.2,1)--(1,1)--(1,1.8);
\draw [semithick] (1,0.2)--(1,1)--(1.8,1);
\node at (1,-1) {$w_6$};
\end{tikzpicture}
\caption{The six vertices of the six-vertex model in terms of 
arrows (first row) or lines (second row), and their Boltzmann weights 
in the five-vertex model (third row)}
\label{fig-SixVertices}
\end{figure}

The five-vertex model is defined on a square lattice
in terms of arrows placed on edges or, equivalently, in terms of
lines  ``flowing'' through the lattice. The standard convention 
\cite{LW-72,B-82} between the arrow and line pictures is that 
if an arrow points down or left, then
this edge contains a line, otherwise the edge is empty. 
In the six-vertex model the admissible vertices are only those which have 
equal number of incoming and outgoing arrows, see Fig.~\ref{fig-SixVertices}.  
The five-vertex model can be obtained by  
requiring that only those vertices are admitted which contain non-intersecting 
lines, that is, the vertex of the second type is excluded. 

In this paper we consider the model on a lattice obtained by intersection of 
$L$ vertical and $M$ horizontal lines (the $M\times L$ lattice). The boundary 
conditions are the following: the $N$ first (last) arrows 
at the bottom (top) boundary point down, and the remaining arrows
point up or right, see Fig.~\ref{fig-LxMlattice}. 

\begin{figure}

\usetikzlibrary{decorations.pathreplacing}

\begin{tikzpicture}[scale=.45]
\draw [semithick] (0.2,1)--(8.8,1);
\draw [semithick] (0.2,2)--(8.8,2);
\draw [semithick] (0.2,3)--(8.8,3);
\draw [semithick] (0.2,4)--(8.8,4);
\draw [semithick] (0.2,5)--(8.8,5);
\draw [semithick] (0.2,6)--(8.8,6);
\draw [semithick] (0.2,7)--(8.8,7);
\draw [semithick] (0.2,8)--(8.8,8);
\draw [semithick] (0.2,9)--(8.8,9);
\draw [semithick] (1,0.2)--(1,9.8);
\draw [semithick] (2,0.2)--(2,9.8);
\draw [semithick] (3,0.2)--(3,9.8);
\draw [semithick] (4,0.2)--(4,9.8);
\draw [semithick] (5,0.2)--(5,9.8);
\draw [semithick] (6,0.2)--(6,9.8);
\draw [semithick] (7,0.2)--(7,9.8);
\draw [semithick] (8,0.2)--(8,9.8);
\draw [thick] [->] (.5,1)--(.6,1);
\draw [thick] [->] (.5,2)--(.6,2);
\draw [thick] [->] (.5,3)--(.6,3);
\draw [thick] [->] (.5,4)--(.6,4);
\draw [thick] [->] (.5,5)--(.6,5);
\draw [thick] [->] (.5,6)--(.6,6);
\draw [thick] [->] (.5,7)--(.6,7);
\draw [thick] [->] (.5,8)--(.6,8);
\draw [thick] [->] (.5,9)--(.6,9);
\draw [thick] [->] (1,9.5)--(1,9.6);
\draw [thick] [->] (2,9.5)--(2,9.6);
\draw [thick] [->] (3,9.5)--(3,9.6);
\draw [thick] [->] (4,9.5)--(4,9.6);
\draw [thick] [->] (5,9.5)--(5,9.6);
\draw [thick] [->] (6,9.5)--(6,9.4);
\draw [thick] [->] (7,9.5)--(7,9.4);
\draw [thick] [->] (8,9.5)--(8,9.4);
\draw [thick] [->] (8.5,1)--(8.6,1);
\draw [thick] [->] (8.5,2)--(8.6,2);
\draw [thick] [->] (8.5,3)--(8.6,3);
\draw [thick] [->] (8.5,4)--(8.6,4);
\draw [thick] [->] (8.5,5)--(8.6,5);
\draw [thick] [->] (8.5,6)--(8.6,6);
\draw [thick] [->] (8.5,7)--(8.6,7);
\draw [thick] [->] (8.5,8)--(8.6,8);
\draw [thick] [->] (8.5,9)--(8.6,9);
\draw [thick] [->] (1,.5)--(1,.4);
\draw [thick] [->] (2,.5)--(2,.4);
\draw [thick] [->] (3,.5)--(3,.4);
\draw [thick] [->] (4,.5)--(4,.6);
\draw [thick] [->] (5,.5)--(5,.6);
\draw [thick] [->] (6,.5)--(6,.6);
\draw [thick] [->] (7,.5)--(7,.6);
\draw [thick] [->] (8,.5)--(8,.6);
\draw [decorate,decoration={brace}]
(5.8,10.1) -- (8.2,10.1) node [midway,yshift=9pt] {$N$};
\draw [decorate,decoration={brace}]
(-.1,0.8) -- (-.1,9.2) node [midway,xshift=-9pt] {$M$};
\draw [decorate,decoration={brace}]
(0.8,11.4) -- (8.2,11.4) node [midway,yshift=9pt] {$L$};
\draw [decorate,decoration={brace,mirror}]
(.8,-.1) -- (3.2,-.1) node [midway,yshift=-10pt] {$N$};
\node at (4.5,-2) {(a)};
\end{tikzpicture}
\qquad
\begin{tikzpicture}[scale=.45]
\draw [semithick] (0.2,1)--(8.8,1);
\draw [semithick] (0.2,2)--(8.8,2);
\draw [semithick] (0.2,3)--(8.8,3);
\draw [semithick] (0.2,4)--(8.8,4);
\draw [semithick] (0.2,5)--(8.8,5);
\draw [semithick] (0.2,6)--(8.8,6);
\draw [semithick] (0.2,7)--(8.8,7);
\draw [semithick] (0.2,8)--(8.8,8);
\draw [semithick] (0.2,9)--(8.8,9);
\draw [semithick] (1,0.2)--(1,9.8);
\draw [semithick] (2,0.2)--(2,9.8);
\draw [semithick] (3,0.2)--(3,9.8);
\draw [semithick] (4,0.2)--(4,9.8);
\draw [semithick] (5,0.2)--(5,9.8);
\draw [semithick] (6,0.2)--(6,9.8);
\draw [semithick] (7,0.2)--(7,9.8);
\draw [semithick] (8,0.2)--(8,9.8);
\draw [line width=3] (1,0.2)--(1,4)--(3,4)--(3,5)--(4,5)--(4,6)--(5,6)--(5,9)--(6,9)--(6,9.8);
\draw [line width=3] (2,0.2)--(2,3)--(5,3)--(5,5)--(6,5)--(6,7)--(7,7)--(7,9.8);
\draw [line width=3] (3,0.2)--(3,1)--(6,1)--(6,3)--(8,3)--(8,9.8);
\node at (4.5,-2) {(b)};
\end{tikzpicture}
\caption{The boundary conditions (a) and one of the possible configurations (b).}
\label{fig-LxMlattice}
\end{figure}	

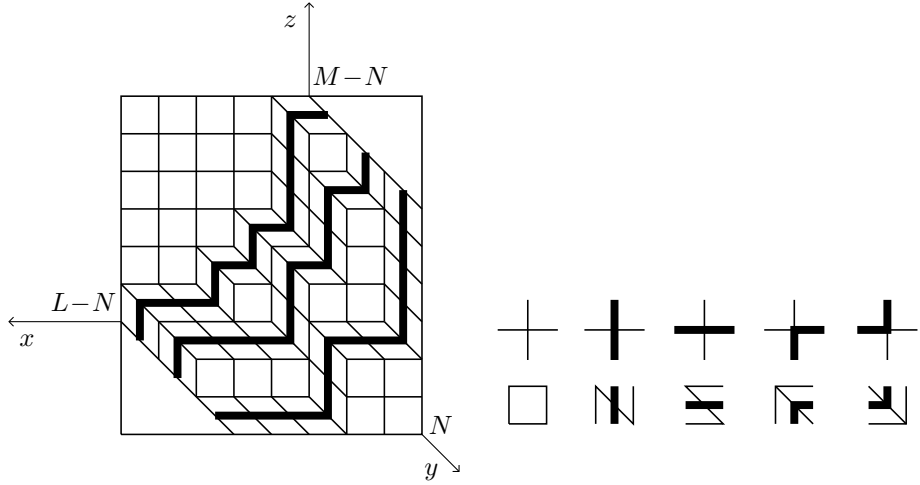
\begin{figure}


\begin{tikzpicture}[scale=.5]
\draw [line width=3] (1,3)--(1,4)--(3,4)--(3,5)--(4,5)--(4,6)--(5,6)--(5,9)--(6,9)--(6,9);
\draw [line width=3] (2,2)--(2,3)--(5,3)--(5,5)--(6,5)--(6,7)--(7,7)--(7,8);
\draw [line width=3] (3,1)--(3,1)--(6,1)--(6,3)--(8,3)--(8,7);
\draw [semithick] (.5,.5)--(.5,9.5)--(8.5,9.5)--(8.5,.5)--(.5,.5);
\draw [semithick] (.5,3.5)--(3.5,.5);
\draw [semithick] (.5,4.5)--(2.5,2.5);
\draw [semithick] (1.5,4.5)--(3.5,2.5);
\draw [semithick] (2.5,4.5)--(4.5,2.5);
\draw [semithick] (2.5,5.5)--(3.5,4.5);
\draw [semithick] (3.5,5.5)--(6.5,2.5);
\draw [semithick] (3.5,6.5)--(5.5,4.5);
\draw [semithick] (4.5,6.5)--(6.5,4.5);
\draw [semithick] (4.5,7.5)--(6.5,5.5);
\draw [semithick] (4.5,8.5)--(6.5,6.5);
\draw [semithick] (4.5,9.5)--(5.5,8.5);
\draw [semithick] (5.5,9.5)--(8.5,6.5);
\draw [semithick] (6.5,7.5)--(8.5,5.5);
\draw [semithick] (7.5,5.5)--(8.5,4.5);
\draw [semithick] (7.5,4.5)--(8.5,3.5);
\draw [semithick] (7.5,3.5)--(8.5,2.5);
\draw [semithick] (6.5,3.5)--(7.5,2.5);
\draw [semithick] (4.5,3.5)--(6.5,1.5);
\draw [semithick] (5.5,1.5)--(6.5,0.5);
\draw [semithick] (4.5,1.5)--(5.5,0.5);
\draw [semithick] (3.5,1.5)--(4.5,0.5);
\draw [semithick] (1.5,9.5)--(1.5,4.5);
\draw [semithick] (2.5,9.5)--(2.5,4.5);
\draw [semithick] (3.5,9.5)--(3.5,5.5);
\draw [semithick] (4.5,9.5)--(4.5,6.5);
\draw [semithick] (5.5,8.5)--(5.5,5.5);
\draw [semithick] (6.5,8.5)--(6.5,7.5);
\draw [semithick] (3.5,4.5)--(3.5,3.5);
\draw [semithick] (4.5,5.5)--(4.5,3.5);
\draw [semithick] (5.5,4.5)--(5.5,1.5);
\draw [semithick] (6.5,6.5)--(6.5,3.5);
\draw [semithick] (7.5,7.5)--(7.5,3.5);
\draw [semithick] (7.5,2.5)--(7.5,0.5);
\draw [semithick] (6.5,2.5)--(6.5,0.5);
\draw [semithick] (4.5,2.5)--(4.5,1.5);
\draw [semithick] (3.5,2.5)--(3.5,1.5);
\draw [semithick] (2.5,2.5)--(2.5,1.5);
\draw [semithick] (1.5,3.5)--(1.5,2.5);
\draw [semithick] (0.5,8.5)--(4.5,8.5);
\draw [semithick] (5.5,8.5)--(6.5,8.5);
\draw [semithick] (0.5,7.5)--(4.5,7.5);
\draw [semithick] (5.5,7.5)--(6.5,7.5);
\draw [semithick] (0.5,6.5)--(4.5,6.5);
\draw [semithick] (6.5,6.5)--(7.5,6.5);
\draw [semithick] (0.5,5.5)--(3.5,5.5);
\draw [semithick] (4.5,5.5)--(5.5,5.5);
\draw [semithick] (6.5,5.5)--(7.5,5.5);
\draw [semithick] (0.5,4.5)--(2.5,4.5);
\draw [semithick] (3.5,4.5)--(4.5,4.5);
\draw [semithick] (5.5,4.5)--(7.5,4.5);
\draw [semithick] (1.5,3.5)--(4.5,3.5);
\draw [semithick] (5.5,3.5)--(7.5,3.5);
\draw [semithick] (2.5,2.5)--(5.5,2.5);
\draw [semithick] (6.5,2.5)--(8.5,2.5);
\draw [semithick] (2.5,1.5)--(5.5,1.5);
\draw [semithick] (6.5,1.5)--(8.5,1.5);
\draw [->] (0.5,3.5)--(-2.5,3.5);
\draw [->] (8.5,0.5)--(9.5,-.5);
\draw [->] (5.5,9.5)--(5.5,12);
\node at (-2,3) {$x$};
\node at (8.75,-.5) {$y$};
\node at (5,11.5) {$z$};
\node at (-.5,4) {$L\!-\!N$};
\node at (9,.75) {$N$};
\node at (6.6,10) {$M\!-\!N$};
\end{tikzpicture}
\quad 
\begin{tikzpicture}[scale=.5]
\draw [semithick] (0.5,0.5)--(0.5,1.5)--(1.5,1.5)--(1.5,.5)--(.5,.5);
\draw [semithick] (0.2,3)--(1.8,3);
\draw [semithick] (1,2.2)--(1,3.8);
\node at (1,-1) {};
\end{tikzpicture}
\;
\begin{tikzpicture}[scale=.5]
\draw [semithick] (0.5,0.5)--(0.5,1.5)--(1.5,.5)--(1.5,1.5);
\draw [line width=3] (1,.5)--(1,1.5);
\draw [semithick] (.2,3)--(1.8,3);
\draw [line width=3] (1,2.2)--(1,3.8);
\node at (1,-1) {};
\end{tikzpicture}
\;
\begin{tikzpicture}[scale=.5]
\draw [semithick] (.5,.5)--(1.5,.5)--(.5,1.5)--(1.5,1.5);
\draw [line width=3] (.5,1)--(1.5,1);
\draw [line width=3] (.2,3)--(1.8,3);
\draw [semithick] (1,2.2)--(1,3.8);
\node at (1,-1) {};
\end{tikzpicture}
\;
\begin{tikzpicture}[scale=.5]
\draw [semithick] (.5,.5)--(.5,1.5)--(1.5,1.5);
\draw [semithick] (1.5,.5)--(.5,1.5);
\draw [line width=3] (1,.5)--(1,1)--(1.5,1);
\draw [semithick] (.2,3)--(1,3)--(1,3.8);
\draw [line width=3] (1,2.2)--(1,3)--(1.8,3);
\node at (1,-1) {};
\end{tikzpicture}
\;
\begin{tikzpicture}[scale=.5]
\draw [semithick] (.5,.5)--(1.5,.5)--(1.5,1.5);
\draw [semithick] (1.5,.5)--(.5,1.5);
\draw [line width=3] (.5,1)--(1,1)--(1,1.5);
\draw [line width=3] (.2,3)--(1,3)--(1,3.8);
\draw [semithick] (1,2.2)--(1,3)--(1.8,3);
\node at (1,-1) {};
\end{tikzpicture}

\caption{The five-vertex model configuration of Fig.~\ref{fig-LxMlattice}b as  
a 3D Young diagram (left) and mapping of the five vertices to flat fragments 
of images of 3D Young diagrams (right).}
\label{fig-3DYoung}
\end{figure}

An interesting property of these boundary conditions is that 
there exists an one-to-one 
correspondence between the configurations of the five-vertex model
with the 3D Young diagrams, which fit into $(L-N)\times N\times (M-N)$  
box, see Fig.~\ref{fig-3DYoung}. 
In this correspondence, the lines of the vertex model are gradient lines;
there also exists the one-to-one correspondence between vertices 
and flat fragments of images of 3D Young diagrams (see Fig.~\ref{fig-3DYoung}, right).
In a rather general setup the
boundary conditions defined above are related to the scalar products of off-shell
Bethe states and their generalizations \cite{KBI-93,B-10,MSa-13,BP-21}. For this reason 
we refer to them as scalar-product boundary conditions.

The partition function is defined as 
\begin{equation}\label{Z-def}
Z = \sum_{\mathcal{C}} \prod_{i=1,3,\ldots,6} w_i^{l_i(\mathcal{C})} 
\end{equation}
where the sum is taken over all admissible configurations $\mathcal{C}$ and 
$l_i(\mathcal{C})$ denotes the number of vertices of the $i$th type in the
configuration $\mathcal{C}$. Note that in all configurations the number of
vertices of the first type is fixed, $l_1(\mathcal{C})=(L-N)(M-N)$, 
the vertices of the third and fourth types
appear in pairs, $l_3(\mathcal{C})-l_4(\mathcal{C})=N(M+N-L)$, and 
the number of vertices of the fifth type is equal to the number of
vertices of the sixth type, $l_5(\mathcal{C})=l_6(\mathcal{C})$. 

A standard way to parametrize the Boltzmann weights (see, e.g., \cite{BP-21})
is the following:
\begin{equation}\label{weights}
w_1 = 
\frac{\alpha}{\sqrt{x}}\frac{x-1}{\Delta},
\qquad
w_3 = \frac{\sqrt{x}}{\alpha},
\qquad
w_4 = \alpha \, \sqrt{x},
{}\qquad
w_5 = w_6 = 1.
\end{equation}
Here, $x\in (1,\infty)$ for $\Delta>0$, and $x\in(0,1)$ for $\Delta<0$. 
The parameter $\Delta$ can be defined independently of the 
parameterization as follows:
\begin{equation}\label{Delta}
\Delta=\frac{w_3w_4-w_5w_6}{w_1w_3}. 
\end{equation}
The case $\Delta=0$ can be approached in the limit $x\to 1$; this is the 
free-fermion point of the model (for further details, see Sect.~\ref{sec:FV}). 
The parameter $\alpha$ is real and positive, it has the meaning of 
an external field.

The partition function $Z=Z(x;\Delta,\alpha)$ has the structure
\begin{equation}\label{Z=EP}
Z=E \,\wt Z.
\end{equation}
Here, $E=E(x;\Delta,\alpha)$ is a factor giving the weight of the configuration 
corresponding to the ``empty'' 3D Young diagram,  
\begin{equation}
E = \bigg(\frac{x-1}{\Delta} \bigg)^{(L-N)(M-N)}
\bigg(\frac{\alpha}{\sqrt{x}}\bigg)^{M(L-2N)} x^{N(L-N-1)}.
\end{equation}
The quantity $\wt Z=\wt Z(x)$ is independent of $\Delta$ and $\alpha$, 
and it has the form 
\begin{equation}\label{wtZ}
\wt Z = \binom{M}{N} \, P_{N,M,L} \left(x^{-1}\right).
\end{equation}
Here, $\binom{M}{N}$ is the binomial coefficient and $P_{N,M,L}\left(x^{-1}\right)$  
is a polynomial of its variable satisfying the normalization condition
\begin{equation}
P_{N,M,L}\left(0\right) = 1.
\end{equation}
The degree of $P_{N,M,L}\left(x^{-1}\right)$ is equal to the difference 
between the maximum and minimum number of pairs of vertices of the fifth and sixth types,
\begin{equation}\label{degP}
\deg P_{N,M,L} = N \min (M-N, L-N-1). 
\end{equation} 
A highly nontrivial and remarkable property of this polynomial is that 
all its coefficients are symmetric under exchange $L\leftrightarrow M+1$, i.e., 
\begin{equation}\label{Psym}
P_{N,M,L}\left(x^{-1}\right) = P_{N,L-1,M+1} \left(x^{-1}\right). 
\end{equation}
Though there seems no simple explanation of this property from the definition of 
the model, it is transparent in explicit expressions (see, e.g., 
representation \eqref{Pnew} below) discussed in the text.   

The polynomial $P_{N,M,L}\left(x^{-1}\right)$ can be seen as a generating
function which counts configurations with a fixed number of turns of `solid'
lines (vertices of the fifth and sixth types). Indeed, due to the combinatorial
restrictions (i.e., the fixed numbers $\ell_1(\mathcal{C})$ and 
$\ell_3(\mathcal{C})-\ell_4(\mathcal{C})$), one can take
weights, instead of \eqref{weights}, equal to $w_1=w_3=w_4=1$ and
$w_5=w_6=1/\sqrt{x}$. These are the weights which have been considered
in \cite{GKW-21} (where $1/\sqrt{x}$ has been denoted by $r$).

\subsection{Main result}

The aim of this paper is to study the thermodynamic limit of the model, 
i.e., the limit where the size of the 
domain tends to infinity with its geometry being fixed. 
To treat the general case one can introduce two 
``macroscopic'' parameters $p,q \in[0,\infty)$, 
which will describe the side lengths of the rectangle-shaped 
domain in the scale of $N$. Specifically, for the reasons 
explained below, we define them as follows:
\begin{equation}\label{def-p-q}
p N= M-N+\frac{1}{2}, \qquad q N= L-N-\frac{1}{2}.
\end{equation}
We are interested in the limit $N,M,L\to\infty$ with $p$ and $q$ being fixed. 
The main thermodynamic quantity of interest is the 
free energy per site $F=F(x;\Delta,\alpha)$, defined as
\begin{equation}
F=-\lim_{N,M,L\to\infty} \frac{\log Z}{ML}. 
\end{equation}
It can be given in the form 
\begin{multline}\label{FreeEn}
F=- \frac{f_2(x)}{(p+1)(q+1)}
-\frac{pq}{(p+1)(q+1)}\log \frac{x-1}{\Delta} 
+\left(\frac{1}{2}-\frac{1}{(p+1)(q+1)}\right)\log x 
\\
-\frac{q-1}{q+1} \log \alpha.
\end{multline}
The function $f_2(x)$ describes the leading large $N$ behavior 
of the nontrivial factor in \eqref{wtZ}, 
\begin{equation}
f_2(x) = \lim_{N,M,L\to\infty} \frac{\log P_{N,M,L}(x^{-1})}{N^2}.
\end{equation}  
Our main results concern the function $f_2(x)$ and all the sub-leading 
corrections for $\log P_{N,M,L}(x^{-1})$ up to $O(1)$ in the limit $N,M,L\to\infty$.   
We often call it below simply ``large $N$ limit'', assuming that 
$L$ and $M$ are connected to $N$ via \eqref{def-p-q}.
 
To treat the special case where the domain has an asymptotic square shape,
we find it convenient to use a ``macroscopic'' parameter $r\in[0,\infty)$ and
a ``microscopic'' parameter $\epsilon=0,\pm 1, \pm 2,\ldots$, defined as follows:
\begin{equation}\label{def-r-e}
rN=\frac{M+L}{2}-N,\qquad
\epsilon =M-L+1.
\end{equation}
In view of the symmetry \eqref{Psym}, the parameter $\epsilon$
appears below only via its absolute value, $|\epsilon|$. It describes a microscopic 
deformation from the ``perfect'' square shape which is in our problem corresponds to
the relation $M-L+1=0$. Our results for the square-shaped domain are 
obtained under the assumption that the parameter $\epsilon$ is fixed in the large $N$
limit, i.e., that $\epsilon$ is of $O(1)$. 

More broadly, the square-shaped domain asymptotically can be obtained under
the assumption that $M/L\to 1$ with $\epsilon$ slowly increasing\footnote{We thank 
the anonymous referee for pointing this possibility to our attention.}.
As it follows from our results, $\epsilon$ can be taken  
to be of $o(N)$ (see Remark~\ref{rem:egrow} below).    
This agrees with the fact that in \eqref{def-r-e} both $M$ and $L$ 
are of the order $N$ and the rectangle-shaped domain would correspond 
to $\epsilon$ to be of the order $N$ as well. 

In what follows we shall often call the square domain case simply as ``symmetric case'' 
and sometimes refer to it as ``the case $p=q$'', 
in view of \eqref{def-p-q}. Clearly, in this 
case the free energy is just given by \eqref{FreeEn} with $p=q=: r$. 

Our main finding about the thermodynamics of the model is that 
it depends strongly on whether the domain takes asymptotically a square or 
rectangular (but not square) shape. If $p=q$, then the model exhibits three different 
phases depending on the value of $x$. If $p\ne q$, then only two phases exist.
All the transitions between the phases are of the third order, that are 
characterized by discontinuities of $f_2'''(x)$ with continuous first- and 
second-order derivatives at these points. 
We summarize the main result in two theorems.

The first theorem concerns the case of a square-shaped domain. 
\begin{theorem}\label{th:square}
If $M,L,N\to \infty$, such that $r=(M+L-2N)/2N$ and $\epsilon=M-L+1$ are kept fixed,
there exist three asymptotic regimes which are 
separated by the critical values $x=\xc^{-1}$ and $x=\xc$, where
\begin{equation}
\xc=(2r+1)^2,\qquad r\in(0,\infty).
\end{equation}
If $x\in[\xc,\infty)$, then
\begin{equation}
\log P_{N,M,L}\left(x^{-1}\right) 
= N^2 f_2^{\mathrm{I}}(x) + N f_1^{\mathrm{I}}(x) + f_0^{\mathrm{I}}(x)
+ O\left(N^{-1}\right),
\end{equation}
where 
\begin{align}
f_2^{\mathrm{I}}(x) 
& 
= r^2 \log \frac{x}{x-1},
\\
f_1^{\mathrm{I}}(x) 
& 
= (2r+1) \log \frac{\sqrt{\xc(x-1)} + \sqrt{x-\xc}}{\big(1+\sqrt{\xc}\big)\sqrt{x}}
- \log \frac{\sqrt{x-1} + \sqrt{x-\xc}}{2\sqrt{x}},
\\
f_0^{\mathrm{I}}(x) 
& 
= \frac{1}{4} \log \frac{x}{x-\xc}-\frac{\epsilon^2}{4}\log\frac{x}{x-1}.
\end{align}
If  $x\in [\xc^{-1},\xc]$, then
\begin{equation}
\log P_{N,M,L}\left(x^{-1}\right) 
=  N^2 f_2^{\mathrm{II}}(x) + N f_1^{\mathrm{II}}(x)
+\frac{5}{12}\log N + f_0^{\mathrm{II}}(x) + O\left(N^{-1}\right),
\end{equation}
where
\begin{align}
f_2^\mathrm{II}(x) 
& 
= (2r+1) \log\frac{1+\sqrt{x}}{1+\sqrt{\xc}}
- \left(r+\frac{1}{4}\right)\log \frac{x}{\xc}
+r^2\log \frac{\xc}{\xc-1},
\\
f_1^{\mathrm{II}}(x) 
&
= \log\frac{2\sqrt{x}}{\sqrt{\xc}+1}+r\log\frac{\sqrt{\xc}-1}{\sqrt{\xc}+1}, 
\\
f_0^{\mathrm{II}}(x) 
&	
= \frac{1}{8}\log \frac{(\sqrt{\xc}-\sqrt{x})^3\sqrt{x}}{(\sqrt{\xc x}-1)}-
\frac{1}{12} \log \left(\sqrt{\xc}\big(\xc-1\big)\right)
\\  &\quad
+\frac{\epsilon^2}{2}\log\frac{\sqrt{\xc x}-1}{\sqrt{\xc-1}\sqrt{x}}
+\zeta'(-1) + \log \sqrt{2\pi}.
\end{align}
If $x\in[0,\xc^{-1}]$, then  
\begin{equation}
\log P_{N,M,L}\left(x^{-1}\right) 
= N^2 f_2^\mathrm{III}(x)+ Nf_1^\mathrm{III}(x) 
+ \frac{1-\epsilon^2}{2} \log N + f_0^\mathrm{III}(x) + O\left(N^{-1}\right),
\end{equation}
where
\begin{align}
f_2^\mathrm{III}(x)&=r^2\log\frac{1}{1-x}-r\log x,
\\
f_1^\mathrm{III}(x)&=|\epsilon|(2r+1)
\log\frac{\sqrt{\xc}\sqrt{1-x}+\sqrt{1-\xc x}}{\sqrt{\xc-1}}
-|\epsilon|\log\frac{\sqrt{1-x}+\sqrt{1-\xc x}}{\sqrt{\xc-1}\sqrt{x}}
\\ &\quad
+\log\frac{2\sqrt{x}}{\sqrt{\xc}+1}+r\log\frac{\sqrt{\xc}-1}{\sqrt{\xc}+1},
\\
f_0^\mathrm{III}(x)&=\frac{1}{4}\log(1-x)-\frac{\epsilon^2}{4}\log(1-\xc x)
+(1-|\epsilon|)\log\sqrt{2\pi}+\log G(1+|\epsilon|).
\end{align}
In these expressions, $\zeta(z)$ and $G(z)$ stand for the Riemann zeta-function and    
the Barnes G-function, respectively; $\zeta'(-1)=-0.165142...$ and $G(1)=G(2)=G(3)=1$, 
$G(n+2)=1!2!\cdots n!$. 
\end{theorem}

In what follows we refer to the three intervals of values of $x$, namely, $[\xc,\infty)$,
$[\xc^{-1},\xc]$, and $[0,\xc^{-1}]$ as Regimes I, II, and III, respectively. We have 
three remarks concerning the result formulated in Thm.~\ref{th:square}.   

\begin{remark}
From \eqref{def-r-e} it follows that $M=(r+1)N+\frac{\epsilon-1}{2}$ and hence 
\begin{multline}
\log \binom{M}{N} =
N \big((r+1) \log(r+1)- r\log r\big) 
- \frac{1}{2} \log N 
\\
- \log \sqrt{2\pi} 
-\frac{\epsilon}{2}\log\frac{r}{r+1}+ O\left(N^{-1}\right),
\end{multline}
that implies that the partition function $\wt Z$ defined in \eqref{wtZ}  
in each Regime $i$, $i=\mathrm{I,II,III}$, has the form 
\begin{equation}
\log \wt Z = N^2 F_2^{i}+N F_1^i+\kappa^i \log N 
+ F_0^i +O(N^{-1}),
\end{equation}
where  
\begin{align}
F_2^{i}&=f_2^i(x),
\\
F_1^i&=f_1^i(x)-\log\frac{2}{\sqrt{\xc}+1}-r\log\frac{\sqrt{\xc}-1}{\sqrt{\xc}+1},
\\
F_0^i&=f_0^i(x)-\log\sqrt{2\pi}-\frac{\epsilon}{2}\log\frac{\sqrt{\xc}-1}{\sqrt{\xc}+1},
\end{align}
and 
\begin{equation}
\kappa^\mathrm{I}=-\frac{1}{2},\qquad 
\kappa^\mathrm{II}=-\frac{1}{12},\qquad
\kappa^\mathrm{III}=-\frac{\epsilon^2}{2}.
\end{equation}
\end{remark}

\begin{remark}
In the Regime III at $\epsilon=0$, it can be shown that
\begin{equation}\label{e=0RIII}
\log \wt Z 
= N^2 f_2^\mathrm{III}(x)
+ N\log \sqrt{x} + \frac{1}{4}\log(1-x)	+ O\left(N^{-\infty}\right),
\end{equation}
where the symbol $O(N^{-\infty})$ denote terms decaying faster 
than any integer power of $1/N$. 
\end{remark}

We will explain the origin of \eqref{e=0RIII} after 
the proof of Thm.~\ref{th:square} in Sect.~\ref{sec:TD}. The terms which 
we have denoted by $O(N^{-\infty})$ are in fact exponentially small and they 
can also be treated by the method of \cite{KP-16}.  

\begin{remark}\label{rem:egrow}
In the expressions in Thm.~\ref{th:square} the parameter $\epsilon$ 
can be replaced by some quantity of magnitude of $o(N)$ without
altering the leading term behavior governed by the function $f_2$ in all 
the three regimes. This implies that the picture with two phase transitions  
obtained for the square-shaped domain is also valid in the large $N$ limit
such that $M/L\to 1$ with $M-L$ allowed to be of $o(N)$. 
\end{remark}

The second theorem concerns the case of a rectangular, but not square, domain.

\begin{theorem}\label{th:rectangle}
In the case $p\ne q$ there exist two asymptotic regimes which are separated by 
the critical value
\begin{equation}
\xc = \left(\sqrt{(p+1)(q+1)}+\sqrt{pq}\right)^2.
\end{equation}
If $x\in [\xc,\infty)$, then 
\begin{equation}
\log P_{N,M,L}\left(x^{-1}\right) =
N^2 f^{\mathrm{I}}_2(x) 
+ N f^{\mathrm{I}}_1(x)
+ f^{\mathrm{I}}_0(x)
+ O\left( N^{-1} \right),
\end{equation}
where
\begin{align}
f^{\mathrm{I}}_2(x) & = 
p q\log \frac{x}{x-1}, 
\\
f^{\mathrm{I}}_1(x) &= 
\frac{2p+1}{2} \log  \frac{(2p+1) x -p-q-1+\sqrt{s(x)}}{2(p+1)\sqrt{x(x-1)}}
\\ & \quad	
+ \frac{2q+1}{2} \log  \frac{(2q+1) x -p-q-1+\sqrt{s(x)}}{2(q+1)\sqrt{x(x-1)}}
\\ & \quad
-\frac{1}{2}\log\frac{x-2pq-p-q-1+\sqrt{s(x)}}{2 x},
\\ 
f^{\mathrm{I}}_0(x) 
& =  \frac{1}{4} \log \frac{x(x-1)}{s(x)},
\end{align}
and 
\begin{equation}
s(x)=x^2-2(2pq+p+q+1)x+(p+q+1)^2,\qquad s(\xc)=0.
\end{equation}
If $x\in [0,\xc]$, then 
\begin{equation}
\log P_{N,M,L}\left(x^{-1}\right) =
N^2 f^{\mathrm{II}}_2(x) 
+ N f^{\mathrm{II}}_1(x)
+ \frac{5}{12} \log N + f^{\mathrm{II}}_0(x)+ O\left(N^{-1}\right),
\end{equation}
where
\begin{align}
f^{\mathrm{II}}_2(x) & = 
- \frac{(p+q)^2}{2} \log y
- \frac{(p-q)^2+2p+2q+1}{2} \log (y+1)
- p \log(y+p-q)
\\ &\quad
- q \log(y+q-p)
+p(p+1) \log \big((2p+1)y+p-q\big)
\\ &\quad
+q(q+1) \log \big((2p+1)y+q-p\big)
+(p+q+1) \log (y+p+q+2)
\\ &\quad
-\frac{1}{2}\Big\{(p+1)^2\log2(p+1)
+(q+1)^2\log2(q+1)+p^2\log 2p+q^2\log 2q\Big\}
\\
f^{\mathrm{II}}_1(x) &=\log \sqrt{x}
-\frac{1}{2}\Big\{(p+1)\log(p+1)
+(q+1)\log(q+1)-p\log p-q\log q\Big\},
\\
f^{\mathrm{II}}_0(x) &=
\frac{1}{8}\bigg\{\log y+\log(y+1)-2\log\big((2p+1)y+p-q\big)
\\ &\quad
-2\log\big((2q+1)y+q-p\big)
+3\log\big(y^2-2(2pq+p+q)y+(p-q)^2\big)
\\ &\quad
+\frac{1}{3}\log\Big((2p+1)(2q+1)y^3-(p-q)^2\left[3y^2+3y-(p+q+1)^2+1\right]\Big)
\bigg\}
\\ &\quad
-\frac{1}{24}\log\big(16p(p+1)q(q+1)\big)+\zeta'(-1)
+\log\sqrt{2\pi}.
\end{align}
The function $y=y(x)$ is the root of the quartic equation   
\begin{equation}\label{thxy}
x = \frac{(y+1)^2(y-p+q)(y+p-q)}{\big((2p+1)y+p-q\big)\big((2q+1)y+q-p\big)},
\end{equation}
which takes the values $y\in [|p-q|,\yc]$ 
for $x\in \left[0,x_c\right]$, where, moreover, 
$y(0)=|p-q|$ and  
\begin{equation}
\yc \equiv y(\xc)=\left(\sqrt{p(q+1)}+\sqrt{q(p+1)}\right)^2=\xc-1.
\end{equation}
\end{theorem} 

In what follows in the non-symmetric case we will refer 
the intervals $[\xc,\infty)$ and $[0,\xc]$ of values of the variable $x$ 
as Regime I and Regime II, respectively. Note that there is no Regime III here.  

\begin{remark}
In the case $q=p=:r$ the parametrization $x=x(y)$ defined by \eqref{thxy} becomes
\begin{equation}
x= \frac{(y+1)^2}{(2r+1)^2}, 
\quad 	x: \left[0, \yc\right] 
\mapsto \left[ (2r+1)^{-2}, (2r+1)^2 \right],
\end{equation}
and the functions $f_2^{\mathrm{II}}(x)$ and $f_1^{\mathrm{II}}(x)$ 
defined in Thm.~\ref{th:square} are recovered. The function $f_0^{\mathrm{II}}(x)$ 
of Thm.~\ref{th:square} is recovered at $\epsilon=0$; the $\epsilon^2$-term of this 
function is recovered by setting $q=r+\epsilon/2N$ and $p=r-\epsilon/2N$ in the function 
$f_2^\mathrm{II}(x)$ and re-expanding it in $1/N$.
\end{remark}

The purpose of the remaining part of the paper is to give proofs 
of Thms.~\ref{th:square} and \ref{th:rectangle}. In brief, we apply the method of paper  
\cite{KP-16} to the results of paper \cite{BP-21}. We use 
the parameterization related to a rational elliptic curve proposed in 
\cite{CP-15} to obtain the assertion of Thm.~\ref{th:rectangle} for Regime II.   
Along the proofs, we have found that the sub-leading corrections can be treated in 
a simplified way, in a comparison to the original approach of \cite{KP-16}, 
by splitting the $\sigma$-form of the sixth Painlev\'e equation on two factors. 

We start with exposing the main ingredients of our analysis, namely, determinant 
representations for the polynomial $P_{N,M,L}(x^{-1})$ and its connection with 
the sixth Painlev\'e equation in Sect.~\ref{sec:PVI}. In Sect.~\ref{sec:AE} 
we obtain expansions of $P_{N,M,L}(x^{-1})$ 
at the singular points of the sixth Painlev\'e
equation for finite values of $N,M,L$. 
In Sect.~\ref{sec:TD} we show how to construct the leading order term of 
the asymptotic expansion in the large $N$ limit 
and how to treat the sub-leading corrections 
in the symmetric case. In Sect.~\ref{sec:TD2} we consider the 
solution of the same problem in the non-symmetric case. 
In Conclusion (Sect.~\ref{sec:Conc}) we briefly discuss our results 
and illustrate a connection with the 
so-called ``merger transition'' in the square-shaped domain case.   

\section{Exact results for the partition function}
\label{sec:PVI}

Here, we collect known results about the partition function of the five-vertex model 
with scalar-product boundary conditions, which we use below in our proofs of 
Thms.~\ref{th:square} and \ref{th:rectangle}.

\subsection{Basic properties of the model}
\label{sec:FV}

We begin with commenting each factor in formulas \eqref{Z=EP} and \eqref{wtZ} 
describing  the partition function $Z$.
These formulas follow from the relations satisfied by
the numbers $l_i(\mathcal{C})$, discussed after \eqref{Z-def}.
In turn, the indicated relations can also be easily understood using the outlined connection of
model with the 3D Young diagrams (see Fig.~\ref{fig-3DYoung}). 

\begin{figure}

\begin{tikzpicture}[scale=.45]
\draw [semithick] (0.2,1)--(8.8,1);
\draw [semithick] (0.2,2)--(8.8,2);
\draw [semithick] (0.2,3)--(8.8,3);
\draw [semithick] (0.2,4)--(8.8,4);
\draw [semithick] (0.2,5)--(8.8,5);
\draw [semithick] (0.2,6)--(8.8,6);
\draw [semithick] (0.2,7)--(8.8,7);
\draw [semithick] (0.2,8)--(8.8,8);
\draw [semithick] (0.2,9)--(8.8,9);
\draw [semithick] (1,0.2)--(1,9.8);
\draw [semithick] (2,0.2)--(2,9.8);
\draw [semithick] (3,0.2)--(3,9.8);
\draw [semithick] (4,0.2)--(4,9.8);
\draw [semithick] (5,0.2)--(5,9.8);
\draw [semithick] (6,0.2)--(6,9.8);
\draw [semithick] (7,0.2)--(7,9.8);
\draw [semithick] (8,0.2)--(8,9.8);
\draw [line width=3] (1,0.2)--(1,9)--(6,9)--(6,9.8);
\draw [line width=3] (2,0.2)--(2,2)--(7,2)--(7,9.8);
\draw [line width=3] (3,0.2)--(3,1)--(8,1)--(8,9.8);
\end{tikzpicture}
\caption{One of the $\binom{M}{N}$ ferroelectric ground states, $M=9$, $N=3$.}
\label{fig-FEGroundState}
\end{figure}
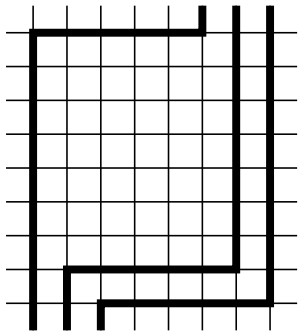

The first relation $l_1(\mathcal{C})=(M-N)\times (L-N)$ gives the number of
elementary squares that obviously is conserved for the plane partitions.  The
factor $E=E(x;\Delta,\alpha)$ in \eqref{Z=EP} 
is the Boltzmann weight of the configuration corresponding to
the ``empty'' partition. The binomial coefficient in \eqref{wtZ} corresponds
to the degeneracy of this weight as there are exactly $\binom{M}{N}$ 
configurations with $N$ horizontal lines at $M$ rows, see 
Fig.~\ref{fig-FEGroundState}. These configurations describe
ferroelectrically ordered states. Since the number of the weights $w_1$ is
fixed, from \eqref{weights} it follows that this is a typical form of configurations 
for sufficiently large values of $x$. 

Moderate values of $x$ correspond to a situation where 
vertices with turn paths (vertices of types $5$ and $6$) are mixed 
with those having straight paths (vertices of types $3$ and $4$). These are disordered 
states. This is an interesting regime because it can be characterized
be appearance of nontrivial limit shapes. 

An important example of such a situation is the case $x=1$ 
which corresponds to the free fermion point of the model. Recall that 
in the general six-vertex model the free-fermion condition is $w_1w_2+w_3w_4=w_5w_6$
which is for the five-vertex model implies $\Delta=0$, where $\Delta$ is defined 
in \eqref{Delta}. In the five-vertex it corresponds to the weights 
\begin{equation}
w_1=\lambda,\qquad
w_3=\alpha,\qquad
w_4=\alpha^{-1},\qquad
w_5=w_6=1,
\end{equation}
where $\lambda$ is some parameter, $\lambda>0$. This model can be obtained 
(recall that $x \lessgtr 1$ for $\Delta \lessgtr 0$) 
upon setting $x=\exp(\lambda \Delta)$ and 
taking the limit $\Delta\to 0$ in \eqref{weights}. 
The correspondence with the boxed plane partitions
means that the partition function $Z=Z(x;\Delta,\alpha)$ at all 
weights equal to $1$ has the following value: 
\begin{equation}\label{ZDelta=0}
\lim_{\Delta\to 0}Z(\rme^{\Delta};\Delta,1)=\PL(L-N,N,M-N).
\end{equation}
Here, $\PL(a,b,c)$ is the number of the boxed plane partitions in 
a box of the size $a\times b \times c$.  
It is well-known to be given by the famous MacMahon triple-product formula
\begin{equation}
\PL(a,b,c) = \prod_{i=1}^{a}\prod_{j=1}^{b}\prod_{k=1}^{c} \frac{i+j+k-1}{i+j+k-2}.
\end{equation}
We use below a less symmetric but more practical single-product expression  
\begin{equation}\label{BPP-prod}
\PL(a,b,c)=\prod_{j=0}^{a-1}\frac{(b+c+j)!j!}{(b+j)!(c+j)!}.
\end{equation}
Analogous expressions can be obtained by permuting cyclically $a$, $b$, $c$. 
Formulas \eqref{wtZ}  and \eqref{ZDelta=0} imply that
\begin{equation}
P_{N,M,L}(1)=\binom{M}{N}^{-1}\,\PL(L-N,N,M-N).
\end{equation}
Using \eqref{BPP-prod} one can check relation \eqref{Psym} at $x=1$. 

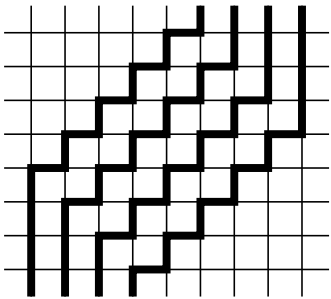
\begin{figure}


\begin{tikzpicture}[scale=.45]
\draw [semithick] (0.2,1)--(9.8,1);
\draw [semithick] (0.2,2)--(9.8,2);
\draw [semithick] (0.2,3)--(9.8,3);
\draw [semithick] (0.2,4)--(9.8,4);
\draw [semithick] (0.2,5)--(9.8,5);
\draw [semithick] (0.2,6)--(9.8,6);
\draw [semithick] (0.2,7)--(9.8,7);
\draw [semithick] (0.2,8)--(9.8,8);
\draw [semithick] (1,0.2)--(1,8.8);
\draw [semithick] (2,0.2)--(2,8.8);
\draw [semithick] (3,0.2)--(3,8.8);
\draw [semithick] (4,0.2)--(4,8.8);
\draw [semithick] (5,0.2)--(5,8.8);
\draw [semithick] (6,0.2)--(6,8.8);
\draw [semithick] (7,0.2)--(7,8.8);
\draw [semithick] (8,0.2)--(8,8.8);
\draw [semithick] (9,0.2)--(9,8.8);
\draw [line width=3] 
(1,0.2)--(1,4)--(2,4)--(2,5)--(3,5)--(3,6)--(4,6)--(4,7)--(5,7)--(5,8)--(6,8)--(6,8.8);
\draw [line width=3] 
(2,0.2)--(2,3)--(3,3)--(3,4)--(4,4)--(4,5)--(5,5)--(5,6)--(6,6)--(6,7)--(7,7)--(7,8.8);
\draw [line width=3] 
(3,0.2)--(3,2)--(4,2)--(4,3)--(5,3)--(5,4)--(6,4)--(6,5)--(7,5)--(7,6)--(8,6)--(8,8.8);
\draw [line width=3] 
(4,0.2)--(4,1)--(5,1)--(5,2)--(6,2)--(6,3)--(7,3)--(7,4)--(8,4)--(8,5)--(9,5)--(9,8.8);
\end{tikzpicture}
\caption{The anti-ferroelectric ground state, $M=L-1=8$.}
\label{fig-AFGroundState}
\end{figure}

As $x$ is small, configurations 
with maximally possible number of 
the vertices of types $5$ and $6$ should dominate. However, such a dominance 
may be affected by relations between the geometric parameters of the 
domain. Indeed, for $M$ and $L$ such that $M=L-1$ and arbitrary $N$ there exists
an anti-ferroelectric ground state, which can be characterized by the presence of a
rectangle region inside of the domain containing only vertices 
of types $5$ and $6$, see Fig.~\ref{fig-AFGroundState}. 
For arbitrary values of $M$ and $L$
there exists no particular state dominating over the other states; 
there are many states which can contribute equally well into the partition function. 
This simple observation hints at the fact that in the thermodynamic 
limit the partition function may demonstrate a different 
behavior in the case of a generic rectangular, but not square, domain, 
in comparison with the case of the domain having a perfect square shape. 

In this relation it is useful to outline some detail on the meaning of the 
limit $x\to 0$. It is closely related to another special case of the five-vertex model,
which is known as the four-vertex model \cite{LPW-90,B-08a}. 
In this model $w_4=0$ that can be 
achieved by setting $x=\alpha^2u^2$ in \eqref{weights}, where $u$ is a new parameter, 
and taking the limit $\alpha\to 0$ \cite{BP-21}. 
Recalling that $\Delta<0$, the nonzero weights are then  
\begin{equation}
w_1=\frac{1}{(-\Delta) u},\qquad
w_3=u,\qquad
w_5=w_6=1.
\end{equation}
As it follows from \eqref{Z=EP} and \eqref{degP},
the partition function $Z=Z(x;\Delta,\alpha)$ 
is nonzero only if $L\leq M+1$. It is known (see, e.g., \cite{BCMP-23}) that 
for all the four weights equal to $1$, it can be 
expressed in terms of the number of boxed plane partitions: 
\begin{equation}
\lim_{\alpha\to 0}Z(\alpha^2;-1,\alpha)=\PL(N,M-L+1,L-N).
\end{equation}
This gives the leading term of the polynomial $P_{N,M,L}(x^{-1})$ in the 
case $L\leq M+1$, 
\begin{equation}
\lim_{x\to0} x^{N(L-N-1)} P_{N,M,L}(x^{-1})=
\binom{M}{N}^{-1}\PL(N,M-L+1,L-N).
\end{equation}  
As we show in the next section, it is also possible to obtain 
a similar formula for the leading term when $L>M+1$ (see Prop.~\ref{pr:xto0}).

\subsection{Hankel determinant representations} 

In what follows we need to use two statements about 
representations for the partition function in terms of determinants. 

As it has been shown in \cite{BP-21}, the partition function 
$Z$ can be written in terms of Hankel determinants of $(L-N)\times (L-N)$ or 
$N\times N$ matrices. This result, rephrased for the 
polynomial $P_{N,M,L}(x^{-1})$, reads as follows. 

\begin{theorem}\label{Th2}
The polynomial $P_{N,M,L}(x^{-1})$ can be given 
in terms of the $(L-N)\times (L-N)$ Hankel determinant,
\begin{multline}\label{Zhom1}
P_{N,M,L}(x^{-1}) = (-1)^{\frac{(L-N)(L-N-1)}{2}} 
\prod_{j=0}^{L-N-1} 
\frac{M!(M+j)!}{(M-N)!(M+L-N-1)!(N+j)!}
\\  \times
\frac{1}{(x-1)^{(L-N)(M-N)}x^{\frac{(L+N)(L-N-1)}{2}}}
\\ \times
\det_{1\leq i,j\leq L-N}
\bigg[(x\pd_x)^{i+j-2}
(x-1)^{M+L-2N-1}
\\ \times
\Ftwoone{-N}{L-N-1}{-M}{x}
\bigg],
\end{multline}
or in terms of the $N\times N$ determinant,
\begin{multline}\label{Zhom2}
P_{N,M,L}(x^{-1}) = 
\prod_{j=0}^{N-1}\frac{(L+M-2N)!}{(L-N+j)!(M-N+j)!}
\\ \times
\frac{(x-1)^{N(M+L-N)}}{
x^{N(L-1)-\frac{N(N+1)}{2}}}
\det_{1\leq i,j\leq N}
\bigg[(x\pd_x)^{i+j-2}
\frac{1}{(x-1)^{M+L-2N+1}}
\\ \times
\Ftwoone{-L+N+1}{-L+N}{-L-M+2N}{1-x}
\bigg].
\end{multline}
\end{theorem}

Note that because of the relation 
$(x\partial_x) x^a =x^a (x\partial_x+a)$, the 
Hankel determinants which appear here, possess the property  
\begin{equation}\label{det=det}
\det_{1\leq i,j \leq N} \left[(x\partial_x)^{i+j-2} x^a f(x)\right]
= x^{aN} \det_{1\leq i,j \leq N} \left[(x\partial_x)^{i+j-2} f(x)\right].
\end{equation}
Below we often use this freedom in writing the Hankel determinants. 

In addition to the two representations given above, we present 
here one more  Hankel determinant formula for $P_{N,M,L}(x^{-1})$. 
\begin{proposition}\label{PropPnew}
The polynomial $P_{N,M,L}(x^{-1})$ admits the following representation
\begin{multline}\label{Pnew}
P_{N,M,L}(x^{-1})=N! \prod_{j=0}^{N-1}\frac{(L-N-1+j)!(M-N+j)!}{(L-2)!(M-1)!}
x^{\frac{N(N-1)}{2}}
\\ \times
\det_{1\leq i,j\leq N}
\bigg[(x\pd_x)^{i+j-2}
\Ftwoone{-L+2}{-M+1}{2}{\frac{1}{x}}
\bigg].
\end{multline}
\end{proposition}
We prove this proposition below in two steps. At the first step 
we show that both representations \eqref{Zhom1} and \eqref{Pnew} satisfy 
the same differential equation, which is 
essentially the sixth Painlev\'e equation in its $\sigma$-form.
At the second step we show that they provide the same solution of this equation.
The solution can be identified in a unique way, say, as $x\to\infty$ by the 
first three terms of the expansion. Specifically, as far as 
the equivalence of \eqref{Zhom1} and \eqref{Zhom2} is established 
(see \cite{BP-21}), we obtain such an expansion from \eqref{Zhom2}. 
The first step is given in Sect.~\ref{sec:SPE} and 
the second step is explained in Sect.~\ref{sec:xatinfty}.

As shown in Sect.~\ref{sec:AE}, representation \eqref{Pnew} 
is useful in obtaining expansions of the polynomial 
$P_{N,M,L}(x^{-1})$ also at the points $x=0,1$, besides the point $x=\infty$.   
Both \eqref{Zhom1} and \eqref{Zhom2} can hardly be used for this purpose. 

\subsection{Connection with the sixth Painlev\'e equation}
\label{sec:SPE}

We begin with recalling some useful facts from the 
theory of sixth Painlev\'e equation; for a detailed exposition, see 
\cite{O-87,JM-81,FW-04}. An important role in the theory 
is played by the $\tau$-function, $\tau=\tau(t)=\tau(t;b_1,b_2,b_3,b_4)$ 
where $t$ is the time variable, and $b_1,\ldots,b_4$ are parameters. 
It is connected to the canonical Hamiltonian by
the relation $H=\partial_t \log \tau(t)$,	 
and hence defined up to a multiplicative constant. 
For our purposes we need 
only an explicit form of the $\tau$-function corresponding to the so-called 
classical solutions related to the Gauss hypergeometric function (see, e.g., 
\cite{FW-04}, Sect.~2.3),  
\begin{multline}\label{tau-f}
\tau(t)= (t(t-1))^{-(b_3+b_4)n/2} 
\left(\frac{t-1}{t}\right)^{(A-B)n/2}
\\ \times
\det_{1\leq i,j \leq n}
\left[\left((t-1)t\partial_t\right)^{i+j-2}
t^A(t-1)^B
\Ftwoone{b_1+b_4}{1-b_1+b_4}{1+b_2+b_4}{t}
\right], 
\end{multline}
where the parameters are subject to the constraints
\begin{equation}
b_1+b_3=n,\qquad A+B=1-b_1+b_4. 
\end{equation}
Note that due to a relation similar to \eqref{det=det}, the 
expression in \eqref{tau-f} is independent of $A-B$, so one can always 
set $A=B=\frac{1}{2}(1-b_1+b_4)$. 
In our cases below, we use $A=0$, $B=1-b_1+b_4$. 

An important property of the $\tau$-function is that, 
for generic values of the parameters $b_1,\ldots,b_4$, the function 
\begin{equation}\label{sigma-t}
\sigma(t)=t(t-1) \partial_t\log \tau(t) +(b_1 b_3+b_1b_4+b_3b_4)t
-\frac{1}{2}\sum_{1\leq j<k \leq 4}b_j b_k 
\end{equation}
satisfies the equation
\begin{equation}\label{sigmaPVI}
\sigma'\big(t(t-1)\sigma''\big)^2
+\big(\sigma'[2\sigma+(1-2t)\sigma']+b_1 b_2 b_3 b_4\big)^2 
=\prod_{j=1}^{4}\big(\sigma'+b_j^2\big).
\end{equation}
The function \eqref{sigma-t} is called $\sigma$-function and 
\eqref{sigmaPVI} is usually referred to as the sixth Painleve equation in its 
$\sigma$-form.

To show that the polynomial $P_{N,M,L}(x^{-1})$ is nothing but the 
$\tau$-function, let us consider the representation \eqref{Zhom1}. 
We apply the Euler transformation of the Gauss hypergeometric function,
\begin{equation}
\Ftwoone{a}{b}{c}{x}=(1-x)^{-a}\Ftwoone{a}{c-b}{c}{\frac{x}{x-1}},
\end{equation}
to the function standing in the determinant in \eqref{Zhom1}, that yields
\begin{equation}
\Ftwoone{-N}{L-N-1}{-M}{x}=
(1-x)^N \Ftwoone{-N}{-M-L+N+1}{-M}{\frac{x}{x-1}}. 
\end{equation}
Ignoring an overall constant factor, 
we then get\footnote{Recall that $f(x)\propto g(x)$ means that $f(x)=Cg(x)$ for 
some constant $C$.}
\begin{multline}
P_{N,M,L}(x^{-1})\propto 
\frac{1}{(x-1)^{(L-N)(M-N)}
x^{\frac{(L+N)(L-N-1)}{2}}
}
\\ \times
\det_{1\leq i,j\leq L-N}
\bigg[
(x\pd_x)^{i+j-2}
(x-1)^{M+L-N-1}
\\ \times
\Ftwoone{-N}{-M-L+N+1}{-M}{\frac{x}{x-1}}
\bigg].
\end{multline}
If we make change of the variable 
\begin{equation}\label{xtot1}
x=\frac{t}{t-1},
\end{equation}
then a comparison with \eqref{tau-f} shows that 
\begin{equation}\label{Psimtau1}
P_{N,M,L}(x^{-1})
\propto \frac{1}{t^{N(L-N-1)}(t-1)^N} \tau(t),
\end{equation}
where the parameters of the $\tau$-function are   
\begin{equation}\label{b-params}
b_1=\frac{L+M}{2}-N,\quad
b_2=\frac{L-M}{2}-1,\quad
b_3=\frac{L-M}{2},\quad
b_4=-\frac{L+M}{2}.
\end{equation}
Note that $n=L-N$, $A=0$ and $B=-L-M+N+1$.

Let us now consider the representation \eqref{Pnew}. Recall that at the moment
it is unproven and we need to show that it is identical to \eqref{Zhom1}. 
We make the change of the variable 
\begin{equation}\label{xtotnew}
x=\frac{t-1}{t},
\end{equation}
so that \eqref{Pnew} takes the following form: 
\begin{multline}
P_{N,M,L}(x^{-1})\propto 
\left(\frac{t-1}{t}\right)^{\frac{N(N-1)}{2}}
\\ \times
\det_{1\leq i,j\leq N}
\bigg[\big((t-1)t\partial_t\big)^{i+j-2}
\Ftwoone{-L+2}{-M+1}{2}{\frac{t}{t-1}}
\bigg].
\end{multline}
After the Euler transformation we get 
\begin{multline}
P_{N,M,L}(x^{-1})\propto 
\left(\frac{t-1}{t}\right)^{\frac{N(N-1)}{2}}
\\ \times
\det_{1\leq i,j\leq N}
\bigg[\big((t-1)t\partial_t\big)^{i+j-2}(t-1)^{-L+2}
\Ftwoone{M+1}{-L+2}{2}{t}
\bigg].
\end{multline}
A comparison with \eqref{tau-f} shows that 
\begin{equation}\label{Psimtaunew}
P_{N,M,L}(x^{-1})\propto \frac{1}{(t-1)^{N(L-N-1)}} \,\tau(t), 
\end{equation}
where the parameters of the $\tau$-function are 
\begin{equation}\label{b-params-new}
b_1=\frac{L+M}{2},\quad
b_2=\frac{L-M}{2},\quad
b_3=N-\frac{L+M}{2},\quad
b_4=\frac{M-L}{2}+1.
\end{equation}
In this case $n=N$, $A=0$, and $B=-L+2$.

A crucial observation which can be made by inspecting 
\eqref{b-params} and \eqref{b-params-new} is that 
these two sets of the parameters can be obtained one from another, modulo 
signs of the elements. To get more insight on the relation between these 
$\tau$-functions, it is useful to consider the 
corresponding $\sigma$-functions appearing in both cases. 
In the first case, described by \eqref{xtot1}, 
\eqref{Psimtau1} and \eqref{b-params}, the $\sigma$-function constructed by 
\eqref{sigma-t} reads
\begin{multline}\label{st1}
\sigma(t)=\frac{t-1}{t}\frac{P_{N,M,L}'(x^{-1})}{P_{N,M,L}(x^{-1})}\bigg|_{x=\frac{t}{t-1}}
-\left(N-\frac{L+M}{2}\right)^2t
\\
+N^2-\frac{3N(M+L)}{4}+\frac{N+ML}{2}+\frac{L-M}{4}.
\end{multline}    
In the second case, described by \eqref{xtotnew}, 
\eqref{Psimtaunew} and \eqref{b-params-new}, the $\sigma$-function constructed by 
\eqref{sigma-t} reads
\begin{multline}\label{stnew}
\sigma(t)=\frac{t}{t-1}\frac{P_{N,M,L}'(x^{-1})}{P_{N,M,L}(x^{-1})}\bigg|_{x=\frac{1-t}{t}}
-\left(N-\frac{L+M}{2}\right)^2t
\\
-\frac{N(M+L)}{4}-\frac{N}{2}+\frac{L^2+M^2-L+M}{4}.
\end{multline}    
It is easy to see that the two $\sigma$-functions \eqref{st1} and \eqref{stnew} 
are related by the map 
\begin{equation}\label{autom}
\sigma(t)\mapsto -\sigma (1-t),\qquad b_1b_2b_3b_4\mapsto - b_1b_2b_3b_4.  
\end{equation}  
The map \eqref{autom} leaves the $\sigma$-form \eqref{sigmaPVI} 
intact and it is an example of symmetry transformations of 
the sixth Painleve equation \cite{O-87}. 

Furthermore, using these transformations (for further details, see \cite{O-87}, Sect.~4) 
one can obtain the $\sigma$-function directly 
in terms of our initial variable $x$. This can be done by making the corresponding change 
of the variables $t\mapsto x$ in each of the two cases \eqref{xtot1} and \eqref{xtotnew}. 
In each case the set of parameters $\{b_1,b_2,b_3,b_4\}$ is mapped into another 
set of parameters $\{\nu_1,\nu_2,\nu_3,\nu_4\}$. 
In our construction the map \eqref{autom} guarantees that the resulting 
$\sigma$-form appears to be the same in both cases, that is  
the two expressions for the polynomial $P_{N,M,L}(x^{-1})$ provided by \eqref{Zhom1} 
and \eqref{Pnew} satisfy the same equation. 

This result in terms of the variable $x$ can be formulated as follows. 
\begin{proposition}\label{pr:sigma-function}
The $\sigma$-function 
\begin{equation}\label{sigmaLogP}
\sigma(x) = x(x-1) \partial_x \log P_{N,M,L}\left(x^{-1}\right)
-\wt A\,x+\wt B,
\end{equation} 
with 
\begin{equation}\label{wtAB}
\wt A = \frac{(N+1)^2}{4},\qquad
\wt B = \frac{L(M+1)}{2}-\frac{(L+M)(3N+1)}{4}+\frac{N}{2}+N^2.
\end{equation}
satisfies the  sixth Painlev\'e equation in its $\sigma$-form
\begin{equation}\label{PVI}
\sigma'\big(x(x-1)\sigma''\big)^2
+\big(\sigma'[2\sigma+(1-2x)\sigma']+\nu_1 \nu_2 \nu_3 \nu_4\big)^2 
=\prod_{j=1}^{4}\big(\sigma'+\nu_j^2\big),
\end{equation}
where $\sigma'\equiv \partial_x\sigma(x)$, $\sigma''\equiv \partial_x^2\sigma(x)$,
and the parameters $\nu_1,\ldots,\nu_4$ can be chosen to be 
\begin{equation}\label{nus}
\nu_1=M-\frac{N-1}{2},\quad 
\nu_2=-L+\frac{N+1}{2},\quad
\nu_3=\frac{N+1}{2},\quad
\nu_4=\frac{N-1}{2}.
\end{equation}
\end{proposition}

In \cite{BP-21} (see Prop.~9 therein) this proposition have been formulated 
for the partition function $Z$, related to the  
$\sigma$-function as 
\begin{equation}
\sigma(x)=x(x-1) \partial_x\log Z-A\, x+ B,
\end{equation} 
with 
\begin{equation}
A=\frac{LM}{2}+\frac{(N-1)^2}{4},\qquad
B=\frac{(N+1)(L+M-2N)}{4}+\frac{N^2-M}{2}. 
\end{equation}

For a later use we also note that instead of the parameters $M$ and $L$, 
one can use the parameters $\nu_1$ and $\nu_2$ given in \eqref{nus} together with $N$ 
(which, in turn, is related to $\nu_3$ and $\nu_4$). For example, 
the constant $\wt B$ in \eqref{wtAB} can be written as 
\begin{equation}\label{wtBnu12N}
\wt B = - \frac{\nu_1\nu_2}{2}-\frac{N(\nu_1-\nu_2)}{2}+\frac{3N^2+1}{8}+\frac{N}{2}.
\end{equation} 
Below we will often use this way of writing for various expressions in addressing 
their behavior as $N,M,L\to \infty$.

Given function \eqref{sigmaLogP} one can reconstruct the 
polynomial $P_{N,M,L}(x^{-1})$ by integrating the $\sigma$-function, 
\begin{equation}\label{P=intsigma}
\log P_{N,M,L}(x^{-1}) =  \int \left(\sigma(x)+\wt A x-\wt B\right) \frac{\rmd x}{x(x-1)} + \wt C,
\end{equation}
where $\wt C$ is some integration constant.

\section{Asymptotic expansions at the singular points}
\label{sec:AE} 

To uniquely identify the solution of the sixth Painlev\'e equation \eqref{PVI}
as being governed by the polynomial $P_{N,M,L}(x^{-1})$, we  
use the asymptotic expansions of this polynomial at the singular points of the 
sixth Painleve equation, namely, at the points 
$x=0,1,\infty$. It is known from the general theory (see, e.g., \cite{J-82}) 
that a solution is uniquely determined by at least first three  terms of 
the asymptotic expansion. 
In this section we construct these expansions 
of $P_{N,M,L}(x^{-1})$ and obtain those for the $\sigma$-function.

\subsection{Expansion as $x\to\infty$}
\label{sec:xatinfty}

Here we have the following result.

\begin{proposition}\label{pr:xtoinfty}
As $x\to\infty$,
\begin{equation}\label{Pxinf}
P_{N,M,L}(x^{-1}) = 1 + \frac{\kappa_1}{x} + \frac{\kappa_2}{x^2} + O\left(x^{-3}\right),
\end{equation}
where the coefficients are
\begin{equation}\label{kappasinf}
\kappa_1= \frac{a b c}{a+1},\qquad
\kappa_2 = \frac{bc\left[a(a+1)(bc+1)-(b+1)(c+1)\right]}{2(a+1)(a+2)},
\end{equation}
with 
\begin{equation}\label{abcxinf}
a=N, \qquad 
b=L-N-1, \qquad 
c=M-N.
\end{equation}
\end{proposition}

We first show how this result follows from the representation \eqref{Zhom2}. 
Next, we will show that the same result follows from \eqref{Pnew}, that completes  
the proof of Prop.~\ref{PropPnew}. 

We start with a standard calculation from the random matrix theory. Let 
$\mu(m)$ denote an arbitrary measure on $\mathbb{Z}_{\geq 0}$.  
We represent the Hankel determinant as a multiple sum
\begin{multline}\label{randommatrix}
\det_{1\leq i,j\leq N}\left[\sum_{m=0}^{\infty}m^{i+j-2}\frac{\mu(m)}{x^m}\right]
=\sum_{m_1,\ldots,m_N=0}^{\infty}
\det_{1\leq i,j\leq N}\left[m_j^{i+j-2}\right]
\prod_{j=1}^N \frac{\mu(m_j)}{x^{m_j}}
\\ 
=\sum_{0\leq m_1<\ldots<m_N\leq \infty}
\prod_{1\leq i<j\leq N}(m_j-m_i)^2\prod_{j=1}^N \frac{\mu(m_j)}{x^{m_j}}.
\end{multline}
From this expression it is clear that the leading term 
of the $x\to\infty$ expansion of the determinant 
corresponds to the values $m_i=i-1$, $i=1,\ldots,N$. 
The first order correction to the leading term comes from the values 
\begin{equation}
m_i =i-1,\quad i=1,\ldots,N-1,\qquad m_N =N.
\end{equation}
The second-order correction is the sum of two contributions, 
which corresponds to the values:
\begin{equation}
m_i=i-1,\quad i=1,\ldots,N-1,\qquad m_N=N+1,
\end{equation}
and 
\begin{equation}
m_i=i-1,\quad  i=1,\ldots,N-2, \qquad 
m_{N-1}=N,\qquad 
m_N=N+1.
\end{equation}
Hence, as $x\to\infty$, we have 
\begin{equation}\label{detxinf}
\det_{1\leq i,j\leq N}\left[\sum_{m=0}^{\infty}m^{i+j-2}\frac{\mu(m)}{x^m}\right]
=\frac{C}{x^{N(N-1)/2}} 
\left(1+\frac{\gamma_1}{x}+\frac{\gamma_2}{x^2}+O\left(\frac{1}{x^{3}}\right)\right),
\end{equation}
where 
\begin{equation}
C=\prod_{1\leq i<j\leq N}(j-i)^2\prod_{j=0}^{N-1}\mu(j)
=\prod_{j=0}^{N-1}(j!)^2\mu(j).
\end{equation}
The coefficients $\gamma_1$ and $\gamma_2$ can be readily computed to be
\begin{equation}\label{gamma12}
\begin{split}
\gamma_1&=\frac{\mu(N)}{\mu(N-1)}N^2,
\\
\gamma_2&=\frac{\mu(N+1)}{\mu(N-1)} \left(\frac{N(N+1)}{2}\right)^2+
\frac{\mu(N)}{\mu(N-2)} \left(\frac{N(N-1)}{2}\right)^2.
\end{split}
\end{equation}

Let us consider the determinant in \eqref{Zhom2}, namely, we focus 
on the Gauss hypergeometric function determining the elements of the matrix. 
Using the identity 
\begin{equation}\label{Gauss}
\Ftwoone{a}{b}{c}{z}=(1-z)^{-a} \Ftwoone{a}{c-b}{c}{\frac{z}{z-1}},
\end{equation}
we first rewrite it in the form 
\begin{equation}
\Ftwoone{-L+N+1}{-L+N}{-L-M+2N}{1-x}
=x^{L-N-1}\Ftwoone{-L+N+1}{-M+N}{-L-M+2N}{1-\frac{1}{x}}.
\end{equation}
The ${}_2F_1$-function in the right-hand side in 
the above relation is a polynomial in $x^{-1}$ of the degree 
$\min(L-N-1,M-N)$. It can further rewritten in the form
\begin{multline}
\Ftwoone{-L+N+1}{-M+N}{-L-M+2N}{1-\frac{1}{x}}
\\
=\frac{(L-N)!(M-N+1)!}{(L+M-2N)!} \Ftwoone{-L+N+1}{-M+N}{2}{\frac{1}{x}}.
\end{multline}
Let us now take into account the $(1-x)$-factor standing in the determinant in 
\eqref{Zhom2}.  
Applying  \eqref{Gauss} twice, one has the relation
\begin{equation}
\Ftwoone{a}{b}{c}{z}= (1-z)^{c-a-b} \Ftwoone{c-a}{c-b}{c}{z}.
\end{equation}
This relation implies that
\begin{multline}
\frac{1}{(1-x^{-1})^{M+L-2N+1}}
\Ftwoone{-L+N+1}{-M+N}{2}{\frac{1}{x}}
\\ 
=\Ftwoone{L-N+1}{M-N+2}{2}{\frac{1}{x}}. 
\end{multline} 
In total, we have thus obtained the identity
\begin{multline}
\frac{1}{(1-x)^{M+L-2N+1}}\Ftwoone{-L+N+1}{-L+N}{-L-M+2N}{1-x}
\\
=(-1)^{M+L+1}\frac{(L-N)!(M-N+1)!}{(L+M-2N)!}
\\
\times
\frac{1}{x^{M-N+2}}
\Ftwoone{L-N+1}{M-N+2}{2}{\frac{1}{x}}.
\end{multline}

As a result, using also identity \eqref{det=det} to 
move the $x$-factor from the determinant in \eqref{Zhom2}, 
we find that the polynomial $P_{N,M,L}(x^{-1})$ admits the following 
representation: 
\begin{multline}\label{Pxinflong}
P_{N,M,L}(x^{-1})= 
N!\prod_{i=0}^{N-1}\frac{(L-N)!(M-N+1)!}{(L-N+i)!(M-N+1+i)!}
\left(1-\frac{1}{x}\right)^{N(M+L-N)}
\\ \times
x^{\frac{N(N-1)}{2}} 
\det_{1\leq i,j\leq N}
\bigg[(x\pd_x)^{i+j-2}
\Ftwoone{L-N+1}{M-N+2}{2}{\frac{1}{x}}
\bigg].
\end{multline}
The determinant here is of the form \eqref{randommatrix}, with 
\begin{equation}
\mu(m)=\frac{(L-N+1)_m(M-N+2)_m}{(m+1)!m!},
\end{equation} 
where the standard notation for the Pochhammer symbol have been used,
\begin{equation}
\qquad 
(z)_m:= z(z+1)\cdots(z+m-1).
\end{equation}
Hence, as $x\to\infty$,
\begin{multline}\label{detasxinf}
\det_{1\leq i,j\leq N}
\bigg[(x\pd_x)^{i+j-2}
\Ftwoone{L-N+1}{M-N+2}{2}{\frac{1}{x}}
\bigg]
\\
=
\frac{\prod_{j=0}^{N-1} (L-N+1)_j(M-N+2)_j}{N!\, x^{N(N-1)/2}}
\bigg\{1+\frac{L(M+1)N}{N+1}\frac{1}{x}
\\
+\frac{L(M+1)}{4}\left(\frac{(L+1)(M+2)}{N+2}+
\frac{(L-1)M}{N+1}\right)\frac{1}{x^2}+O(x^{-3})\bigg\}.
\end{multline}
Clearly, the leading term here cancels the prefactor in \eqref{Pxinflong}, so 
that $P_{N,M,L}(0)=1$. Furthermore, expanding 
the factor $(1-x^{-1})^{N(M+L-N)}$ in \eqref{Pxinflong} in the Taylor series, 
from \eqref{detasxinf} one can easily obtain
the coefficients $\kappa_1$ and $\kappa_2$ in \eqref{Pxinf}. 
This finalizes the proof of Prop.~\ref{pr:xtoinfty}.

Let us now comment that exactly the same result follows from the 
new representation \eqref{Pnew}. Indeed, for the determinant in \eqref{Pnew} we 
have the expansion \eqref{detxinf}, with 
\begin{equation}\label{muPnew}
\mu(m)=\frac{(-L+2)_m(-M+1)_m}{(m+1)!m!}. 
\end{equation}
More explicitly, we have 
\begin{multline}
\det_{1\leq i,j\leq N}
\bigg[(x\pd_x)^{i+j-2}
\Ftwoone{-L+2}{-M+1}{2}{\frac{1}{x}}
\bigg]
\\
=\frac{\prod_{j=0}^{N-1} (-L+2)_j(-M+1)_j}{N!\, x^{N(N-1)/2}}
\left\{1 + \frac{\kappa_1}{x} + \frac{\kappa_2}{x^2} + O\left(x^{-3}\right)\right\},
\end{multline}
where $\kappa_1$ and $\kappa_2$ are exactly those given by \eqref{kappasinf}.
Clearly, the overall constant here cancels the 
prefactor in \eqref{Pnew}, and the property 
$P_{N,M,L}(0)=1$ is recovered. In total, Prop.~\ref{pr:xtoinfty} follows from 
\eqref{Pnew}, as it should. This finalizes the proof of Prop.~\ref{PropPnew}. 

Using Prop.~\ref{pr:xtoinfty} together with Prop.~\ref{pr:sigma-function} 
one can compute an expansion of 
the corresponding $\sigma$-function near the point $x=\infty$, namely 
\begin{equation}
\sigma(x)
=-\wt A x+\wt B-\kappa_1 +\frac{\kappa_1+\kappa_1^2-2\kappa_2}{x} +O\big(x^{-2}\big), 
\end{equation}
where $\kappa_1$ and $\kappa_2$ are to be taken from \eqref{kappasinf}
and $\wt A$ and $\wt B$ are given by \eqref{wtAB}. More explicitly, 
we have the following.  
\begin{corollary}\label{cor:xtoinfty}
The $\sigma$-function \eqref{sigmaLogP} has the following $x\to\infty$ behavior
\begin{multline}\label{sigmatoinf}
\sigma(x)
=
-\frac{(N+1)^2}{4}x+\frac{N-1}{2(N+1)}\nu_1\nu_2+\frac{(N+1)^2}{8}
\\ 
+\frac{\Big[\nu_1^2-\big(\frac{N+1}{2}\big)^2\Big]
\Big[\nu_2^2-\left(\frac{N+1}{2}\right)^2\Big]}{(N+1)^2(N+2)}x^{-1}
+O\big(x^{-2}\big),
\end{multline}
where $\nu_1$ and $\nu_2$ are given by \eqref{nus}.  
\end{corollary}

\subsection{Expansion as $x\to0$}

Next we consider asymptotic behavior of the polynomial $P_{N,M,L}(x^{-1})$ 
near the point $x=0$. We have the following result.
\begin{proposition}\label{pr:xto0}
As $x\to0$,
\begin{equation}\label{P(x)=xto0}
P_{N,M,L}(x^{-1}) = 
\frac{C}{x^{ac}} 
\left\{1+\kappa_1 x+\kappa_2x^2+O(x^{-3})\right\},
\end{equation} 
where
\begin{equation}
C=\binom{a+c}{a}^{-1} \PL(a,b,c) =\binom{a+b+c}{a}^{-1} \PL(a,b,c+1)
\end{equation}
and the coefficients are 
\begin{equation}\label{kappazero}
\kappa_1=\frac{ac(c+1)}{a+b}, \qquad
\kappa_2=\frac{ac(c+1)}{a+b} 
\frac{(c^2+c+1)(a^2+ab-1)-b-2bc}{2(a+b-1)(a+b+1)},
\end{equation}
with 
\begin{equation}\label{abc=xto0}
a = N, \qquad b = | M-L+1 |, \qquad c = \min(L-N-1,M-N).
\end{equation}
\end{proposition}

We will prove Prop.~\ref{pr:xto0} using representation \eqref{Pnew}.  
Constructively, the proof will be based again on formula \eqref{detxinf}
in which we make the change $x^{-1}\mapsto x$. The details however
depend on whether $M\geq L-1$ or $M\leq L-1$. 

Let us perform calculations assuming that $M\geq L-1$. We first transform the 
${}_2F_1$-function in the determinant in \eqref{Pnew} such that it will become 
a polynomial in $x$. For this end one can use the following 
relation valid  for $m$ and $n$ positive integers, $n\leq m$, and 
$c$ real (not to be confused with that in Prop.~\ref{pr:xto0}), $c\geq 1$:   
\begin{equation}
\Ftwoone{-n}{-m}{c}{x}=\frac{m!}{(m-n)!\,(c)_n}x^n
\Ftwoone{-n}{-n-c+1}{m-n+1}{\frac{1}{x}}.
\end{equation}
For $M\geq L-1$, one has  
\begin{multline}
\Ftwoone{-L+2}{-M+1}{2}{\frac{1}{x}}
=\frac{(M-1)!}{(M-L+1)!(L-1)!}
\\ \times
x^{-L+2}\Ftwoone{-L+2}{-L+1}{M-L+2}{x}.
\end{multline}
Removing the factor $x^{-L+2}$ from determinant by relation \eqref{det=det}, 
we then have
\begin{multline}
P_{N,M,L}(x^{-1})=N! \prod_{j=0}^{N-1}\frac{(L-N-1+j)!(M-N+j)!}{(L-2)!(M-L+1)!(L-1)!}
\\ \times
x^{\frac{N(N-1)}{2}-(L-2)N}
\det_{1\leq i,j\leq N}
\bigg[(x\pd_x)^{i+j-2}
\Ftwoone{-L+2}{-L+1}{M-L+2}{x}
\bigg].
\end{multline}
The determinant here can be written, up to the change $x\mapsto x^{-1}$, in the form 
\eqref{detxinf}, with 
\begin{equation}\label{muLLML}
\mu(m)= \frac{(-L+2)_m(-L+1)_m}{(M-L+2)_m m!}.
\end{equation}
Hence, for $M\geq L-1$, as $x\to 0$, 
\begin{multline}
\det_{1\leq i,j\leq N}
\bigg[(x\pd_x)^{i+j-2}
\Ftwoone{-L+2}{-L+1}{M-L+2}{x}
\bigg]
=\prod_{j=0}^{N-1}\frac{(-L+2)_j(-L+1)_j j!}{(M-L+2)_j}
\\ \times 
x^{\frac{N(N-1)}{2}}\left\{1+\kappa_1 x+\kappa_2 x^2+O(x^3)\right\},
\end{multline}
where the constants $\kappa_1$ and $\kappa_2$ can be computed by the formulas
\eqref{gamma12} for $\gamma_1$ and $\gamma_2$, respectively, 
with $\mu(m)$ in \eqref{muLLML}, and they are 
are given by the expressions in \eqref{kappazero} with
\begin{equation}
a=N,\qquad b=M-L+1, \qquad c=L-N-1.
\end{equation}
As a result, we obtain that for $M\geq L-1$, as $x\to 0$, 
\begin{equation}
P_{N,M,L}(x^{-1})=
\frac{C}{x^{(L-N-1)N}} \left\{1+\kappa_1 x+\kappa_2 x^2+O(x^3)\right\},
\end{equation}
where 
\begin{equation}
C= N! \prod_{j=0}^{N-1}\frac{(M-N+j)!j!}{(L-N+j)!(M-L+1+j)!}.
\end{equation}
This constant can also be written as follows:
\begin{equation}
C= \frac{N!(L-N-1)!}{(L-1)!} \prod_{j=0}^{N-1}\frac{(M-N+j)!j!}{(L-N-1+j)!(M-L+1+j)!}.
\end{equation}
Here the product can be recognized as the numbers of the boxed 
plane partitions $\PL(N,M-L+1,L-N-1)$, see \eqref{BPP-prod}. Equivalently, 
one can write
\begin{equation}
C=\frac{N!(M-N)!}{M!} \prod_{j=0}^{N-1}\frac{(M-N+1+j)!j!}{(L-N+j)!(M-L+1+j)!},
\end{equation}
where the product equals $\PL(N,M-L+1,L-N)$, which 
is essentially the partition function of the four-vertex model, see 
\eqref{wtZ} and the discussion in Sect.~2.1.

In the case $M\leq L-1$, the calculations are essentially similar.  
In fact all formulas in this case can be obtained from those given above 
by formal substitution $M\leftrightarrow L-1$.
As a result, this lead to the expansion \eqref{P(x)=xto0}, where 
the parameters $a$, $b$, $c$ are given by \eqref{abc=xto0}. This finalize the proof
of Prop.~\ref{pr:xto0}.  

From Props.~\ref{pr:xto0} and \ref{pr:sigma-function} it follows that  
the $\sigma$-function as $x\to 0$ is given by 
\begin{equation}
\sigma(x)
=ac+\wt B- \big(ac+ \wt A +\kappa_1\big)x 
+\big(\kappa_1+\kappa_1^2-2\kappa_2\big)x^2+O\big(x^{3}\big), 
\end{equation}
where $\kappa_1$ and $\kappa_2$ are given in \eqref{kappazero} and
$a$ and $c$ are defined in \eqref{abc=xto0}. More explicitly, 
we have the following.  

\begin{corollary}\label{cor:xto0}
The $\sigma$-function \eqref{sigmaLogP} has the following $x\to 0$ behavior:
\begin{multline}\label{sigmatozero}
\sigma(x)=-\frac{\nu_1\nu_2}{2}-\frac{N |\nu_1+\nu_2|}{2}-\frac{N^2-1}{8}
+\frac{N\nu_1\nu_2+\frac{N^2-1}{4}|\nu_1+\nu_2|}{|\nu_1+\nu_2|+N}x
\\
+\frac{N|\nu_1+\nu_2|\Big[\big(\nu_1\nu_2+\frac{N}{2}|\nu_1+\nu_2|+\frac{N^2+1}{4}\big)^2-
\frac{1}{4}(|\nu_1+\nu_2|+N)^2\Big]}
{(|\nu_1+\nu_2|+N)^2\big[(|\nu_1+\nu_2|+N)^2-1\big]}x^2
\\
+O\big(x^{3}\big).
\end{multline}
\end{corollary}
It is interesting to note, that if $\nu_1+\nu_2>0$, that is $M>L-1$, 
this expression can be conveniently written in the form
\begin{equation}\label{SigmaSym}
\sigma(x)=-\frac{S_2}{2}+\frac{S_3}{S_1}x
+\frac{\prod_{i<j}(\nu_i+\nu_j)}{S_1^2\big[S_1^2-1\big]}x^2+O\big(x^{3}\big),
\end{equation}
where 
\begin{equation}
S_1=\sum_i\nu_i,\qquad
S_2=\sum_{i<j}\nu_i\nu_j,\qquad
S_3=\sum_{i<j<k}\nu_i\nu_j\nu_k.
\end{equation}
If $\nu_1+\nu_2<0$, then one has to make the replacement $\nu_{1,2}\mapsto -\nu_{2,1}$. 
Finally, if $\nu_1+\nu_2=0$, we just have 
\begin{equation}\label{sigmazeroS}
\sigma(x)=\frac{\nu_1^2}{2}-\frac{N^2-1}{8}-\nu_1^2 x+ O\big(x^{3}\big). 
\end{equation}

\subsection{Expansion as $x\to 1$}

We conclude this section by considering the technically most difficult 
case of an expansion at the point $x=1$. 

\begin{proposition}\label{pr:xto1}
As $x\to1$,
\begin{equation}\label{P(x)=xto1}
P_{N,M,L}(x^{-1}) = C
\left\{1 + \kappa_1 (x-1) + \kappa_2(x-1)^2 + O\left( (x-1)^{3} \right)\right\}, 
\end{equation}
where 
\begin{equation}\label{Cxto1}
C=\binom{a+c}{a}^{-1} \PL(a,b+1,c)
\end{equation}
and the coefficients are
\begin{equation}\label{kappasone}
\kappa_1 = -\frac{abc}{b+c+1},\quad
\kappa_2 =  abc
\frac{abc(b+c+1)+b^2+c^2+3bc+3c+3b+a+1}
{2(b+c)(b+c+1)(b+c+2)},
\end{equation}
with
\begin{equation}\label{abcxto1}
a=N, \qquad b = L-N-1, \qquad c = M-N.
\end{equation}
\end{proposition}

We will prove this proposition using the connection 
of the Hankel matrix in the representation \eqref{Pnew} at $x=1$ with 
the ensemble of the Hahn polynomials, for a list of properties of the Hahn polynomials, 
see, e.g., \cite{KLS-10}, Sect.~9.5. We will use the Hahn polynomials 
in the normalization with the highest coefficient being equal to one,  
\begin{equation}
p_i(m)= \frac{(a+1)_i(-n)_i}{(i+\alpha+\beta+1)_i}
\Fthreetwo{-i}{i+\alpha+\beta+1}{-x}{\alpha+1}{-n}.
\end{equation}
For $\alpha, \beta>-1$ or $\alpha, \beta< -n$ (see \cite{KLS-10}, formula (9.5.2))
these polynomials satisfy the orthogonality condition
\begin{equation}\label{Hahn-orthcond}
\sum_{m=0}^{n}\binom{\alpha+m}{m}\binom{\beta+n-m}{n-m}p_i(m)p_j(m)=
\delta_{ij}h_i,
\end{equation}
with 
\begin{equation}\label{hi}
h_i=\frac{i!}{(n-i)!}\frac{(i+\alpha+\beta+1)_{n+1} (\alpha+1)_i (\beta+1)_i}
{(i+\alpha+\beta+1)_i(i+\alpha+\beta+1)_{i+1}}.
\end{equation}

To see the connection of the Hankel matrix in \eqref{Pnew} at $x=1$ with the 
ensemble of the Hahn polynomials, we multiply the entries of the matrix 
by the factor $(-1)^{i+j-2}$ that obviously does not alter the value of the determinant, 
and denote the resulting matrix by $H(x)$. Its entries are 
\begin{equation}
(H(x))_{ij}=\sum_{m\geq 0}^{} \mu(m) \frac{m^{i+j-2}}{x^m}\qquad (i,j=1,\ldots,N),
\end{equation}
where the measure $\mu(m)$ is given by \eqref{muPnew}. 
This measure is essentially that of the Hahn polynomials, due to the identity 
\begin{equation}
\frac{(-L+2)_m(-M+1)_m}{(m+1)!m!}=\frac{(-1)^n}{n+1}  
\binom{\alpha+m}{m}\binom{\beta+n-m}{n-m},
\end{equation}
where one should set
\begin{equation}\label{abr}
\begin{split}
\alpha&=\min(-M,-L+1),
\\
\beta&=\max(-L,-M-1),
\\
n&=\min(L-2,M-1).
\end{split}
\end{equation}
Note that we deal with the case $\alpha,\beta<-n$; 
we recall that $\binom{-a+m}{m}=(-1)^m\binom{a-1}{m}$. The three parameters 
$\alpha$, $\beta$ and $n$ are constrained by the condition
\begin{equation}\label{rbeta}
n=-\beta-2.
\end{equation}

To simplify writing below, we denote $H\equiv H(1)$. 
Since the entries of $H$ are independent of $N$, the size of the matrix, 
the orthogonality condition \eqref{Hahn-orthcond} yields
\begin{equation}
\det H = \prod_{i=0}^{N-1}\tilde h_i, \qquad \tilde h_i\equiv \frac{(-1)^n}{n+1} h_i. 
\end{equation}
Plugging \eqref{abr} into \eqref{hi} gives
\begin{equation}
\tilde h_i =\frac{i!(L-2)!(M-1)!(M+L-1-2i)!(M+L-2-2i)!}
{(L-1-i)!(L-2-i)!(M-i)!(M-1-i)!(M+L-1-i)!}, 
\end{equation}
and rearranging the factors in the product, using
\begin{equation}
\prod_{i=0}^{N-1}(a+2i)!(a+1+2i)!=\prod_{i=0}^{2N-2}(a+i)!
=\prod_{i=0}^{N-1}(a+i)!(a+N+i)!,
\end{equation}
one can readily find 
\begin{equation}
\det H= \prod_{i=0}^{N-1} 
\frac{(L-2)!(M-1)!(M+L-2N+i)!i!}
{(L-1-i)!(L-2-i)!(M-i)!(M-1-i)!}.
\end{equation}
As a result, from \eqref{Pnew} it follows that 
the constant $C$ in \eqref{P(x)=xto1} reads
\begin{equation}
\begin{split}
C&=N! \prod_{i=0}^{N-1} 
\frac{(M+L-2N+i)!i!}
{(L-N+i)!(M-N+1+i)!}
\\ &
=\frac{N!(M-N)!}{M!} \prod_{i=0}^{N-1} 
\frac{(M+L-2N+i)!i!}
{(L-N+i)!(M-N+i)!},
\end{split}
\end{equation}
where the product in the second equality 
can easily recognized as the number of 
the boxed plane partitions $\PL(N,L-N,M-N)$, see \eqref{BPP-prod}. 

To compute the coefficients $\kappa_1$ and $\kappa_2$ in 
\eqref{P(x)=xto1} we first compute the coefficients in the Taylor-series expansion
\begin{equation}
\frac{\det H(x)}{\det H}=
1+ \gamma_1 (x-1) + \gamma_2 (x-1)^2+ O\left((x-1)^3\right)
\end{equation}
from the relation $\det H(x)=\exp\{\tr \log H(x)\}$, that gives
\begin{equation}\label{gammatr}
\gamma_1=\tr (H^{-1}H'),\qquad 
\gamma_2=\frac{1}{2}\left\{\left(\tr H^{-1}H'\right)^2
+\tr H^{-1}H''-\tr (H^{-1}H')^2\right\}, 
\end{equation}
where $H'\equiv H'(x)|_{x=1}$ and $H''\equiv H''(x)|_{x=1}$.
The entries 
of the matrix $H^{-1}$ can be expressed (see, e.g., \cite{BS-09}, Thm.~1.1, and also 
\cite{Fe-84}, Thm.~9) in terms of the coefficients  
of the polynomials $p_i(m)$, 
\begin{equation}
\big(H^{-1}\big)_{jk}=\sum_{i=0}^{N-1}\tilde h_i^{-1}p_{i,j-1} p_{i,k-1},\qquad
p_i(m)=\sum_{k=0}^{i} p_{i,k}m^k.
\end{equation}
The traces in \eqref{gammatr} can be evaluated with the help of the recurrence 
relation (see \cite{KLS-10}, formula (9.5.4)) 
\begin{equation}
m p_i(m)=p_{i+1}+B_i p_i(m)+C_i p_{i-1}(m).
\end{equation}
Here,
\begin{equation}\label{Chh}
C_i=\frac{h_i}{h_{i-1}}
\end{equation}
and the coefficient $B_i$ in the case of condition \eqref{rbeta} 
can be written in the form 
\begin{equation}\label{Bi}
B_i=\frac{(\alpha+\beta)(\alpha-\beta)(\alpha-\beta-2)}
{4(2i+\alpha+\beta)(2i+\alpha+\beta+2)} -\frac{\alpha+\beta}{4}-1.
\end{equation}
The first trace in \eqref{gammatr} can be computed as follows:
\begin{equation}
\tr H^{-1}H'= -\sum_{m=0}^{r} m \mu(m)\sum_{i=0}^{N-1}\tilde h_i^{-1} p_i^2(m)
=- \sum_{i=0}^{N-1}B_i.
\end{equation}
Essentially similarly, but slightly more involved calculation gives 
\begin{align}
\tr H^{-1}H''&= \sum_{i=0}^{N-1}\left(B_i + B_i^2\right) +2 \sum_{i=1}^{N-1} C_i+C_N,
\\
\tr (H^{-1}H')^2&= \sum_{i=0}^{N-1} B_i^2 +2 \sum_{i=1}^{N-1} C_i.
\end{align}
Hence,
\begin{equation}
\gamma_1=-\sum_{i=0}^{N-1}B_i,\qquad 
\gamma_2=\frac{1}{2}\left(\gamma_1^2-\gamma_1+C_N\right).
\end{equation}
To compute the 
sum of $B_i$'s we expand the rational part in \eqref{Bi} in elementary ratios, 
that gives
\begin{equation}
\sum_{i=0}^{N-1}B_i
= \frac{N}{4} \left(\frac{(\alpha-\beta)(\alpha-\beta-2)}{2N+\alpha+\beta}
-\frac{\alpha+\beta}{4}-1\right).
\end{equation}
We also have (see \eqref{hi}, \eqref{rbeta} and \eqref{Chh})
\begin{equation}
C_N= -\frac{N(N+\alpha-1)(N+\alpha)(N+\beta+1)(N+\beta)(N+\alpha+\beta)}
{(2N+\alpha+\beta-1)(2N+\alpha+\beta)^2(2N+\alpha+\beta+1)}.
\end{equation}
Expanding the factor 
$x^{\frac{N(N-1)}{2}}$ in \eqref{Pnew} at $x=1$, for the coefficients $\kappa_1$ 
and $\kappa_2$ in \eqref{P(x)=xto1} we obtain 
\begin{equation}
\kappa_1= \gamma_1+\frac{N(N-1)}{2},\qquad
\kappa_2=\frac{1}{2}\left(\kappa_1^2-\kappa_1+C_N\right).
\end{equation}
Finally, plugging $\alpha$ and $\beta$ from \eqref{abr} into the above 
expressions and simplifying,
we arrive at \eqref{kappasone}. This completes the proof 
of Prop.~\ref{pr:xto1}.

From Props.~\ref{pr:sigma-function} and \ref{pr:xto1} 
it follows that the $\sigma$-function has the following expression near the point $x=1$:
\begin{equation}
\sigma(x)=\wt B-\wt A + \big(\kappa_1-\wt A\big)(x-1)+\big(\kappa_1+2\kappa_2
-\kappa_1^2\big)(x-1)^2+O\big((x-1)^3\big).
\end{equation}
Here $\kappa_1$ and $\kappa_2$ are given by \eqref{kappasone}, and $\wt A$ and $\wt B$
are given in \eqref{wtAB}; note that 
the coefficient of the second-order term is equal to the constant $C_N$. 
In terms of $\nu_1$ and $\nu_2$ the result reads as follows. 

\begin{corollary}\label{cor:xto1}
The $\sigma$-function \eqref{sigmaLogP} has the following $x\to 1$ behavior:
\begin{multline}\label{sigmatoone}
\sigma(x)=-\frac{\nu_1\nu_2}{2}-\frac{N (\nu_1-\nu_2)}{2}+\frac{N^2-1}{8}
+\frac{N\nu_1\nu_2+\frac{N^2-1}{4}(\nu_1-\nu_2)}{\nu_1-\nu_2-N}(x-1)
\\
+\frac{N(\nu_1-\nu_2)\Big[\big(\nu_1-\frac{N}{2}\big)^2-\frac{1}{4}\Big]
\Big[\big(\nu_2+\frac{N}{2}\big)^2-\frac{1}{4}\Big]}
{(\nu_1-\nu_2-N)^2\big[(\nu_1-\nu_2-N)^2-1\big]}(x-1)^2+
O\big((x-1)^{3}\big).
\end{multline}
\end{corollary}
Note that the coefficients in this expansion can also be written in the 
form analogous to \eqref{SigmaSym} up to the change $\nu_1\mapsto -\nu_1$.

\section{Thermodynamic limit in the symmetric case}
\label{sec:TD} 

In this section we focus on construction of the 
asymptotic expansions for the $\sigma$-function and the corresponding 
polynomial $P_{N,M,L}(x^{-1})$ in the limit 
$N,M,L\to \infty$ such the two parameters $p$ and $q$ defined by \eqref{def-p-q}
are finite and equal to each other, $p=q$. In this case it is suitable to use 
the parameters $r\equiv p=q$ and $\epsilon$ defined in \eqref{def-r-e}.   
This will provide a proof of 
Thm.~\ref{th:square}. The non-symmetric case, $p\ne q$, is considered in the next section. 

\subsection{Preliminaries}\label{sec:CE}

To derive expansions of the function $\log P_{N,M,L}(x^{-1})$ in the 
thermodynamic limit, we start with analyzing the $\sigma$-form of the 
sixth Painlev\'e equation in the large $N$ limit. We recall that the 
$\sigma$-function is given in terms of $\log P_{N,M,L}(x^{-1})$ 
by \eqref{sigmaLogP}. Expressions \eqref{nus} for the parameters 
$\nu_1,\ldots,\nu_4$ and the expansions of the $\sigma$-function 
at the singular points given by 
Cor.~\ref{cor:xtoinfty}, \ref{cor:xto0}, and \ref{cor:xto1},
suggest that the $\sigma$-function may be searched in the form of the following 
asymptotic ansatz in the decaying powers of $N$: 
\begin{equation}\label{sigmaNlargeexp}
\sigma(x)= \sum_{k\geq 0}N^{2-k} \sigma_{2-k}(x).
\end{equation}
Following \cite{KP-16}, we note that the justification of 
ansatz \eqref{sigmaNlargeexp} is based on the Wasow theorem, see 
\cite{W-87}, Chap. IX, Thm.~36.1. This theorem implies that 
if one succeeds in the construction of the expansion \eqref{sigmaNlargeexp} with  
the functions $\sigma_i(x)$, which are piece-wise \emph{analytic} functions of $x$, 
then there exists a genuine solution of equation \eqref{PVI} 
with the asymptotic expansion \eqref{sigmaNlargeexp}. To justify that this solution 
indeed coincides with the solution given by \eqref{sigmaLogP}, 
one has to verify that they have the same behaviors at the singular 
points, $x=0,1,\infty$, given by Cors.~\ref{cor:xtoinfty}, 
\ref{cor:xto0}, and \ref{cor:xto1}. We recall that for the Painlev\'e equations
expansions of solutions at the singular points fix the solutions \cite{J-82}. 

The Wasow theorem is applicable to the 
sixth Painlev\'e equation in its Hamiltonian formulation. 
Indeed, the Wasow theorem 
deals with the first-order vector ordinary differential equations resolved 
with respect to the derivatives.  
The $\sigma$-function is intimately related to the Hamiltonian 
and there exists a one-to-one correspondence between 
the canonical variables (the coordinate and momentum) and the $\sigma$-function 
\cite{O-87}. The conditions of the Wasow theorem can be verified 
by writing the Hamiltonian equations of motion of the sixth Painlev\'e system 
in a vector form. Details of this calculation can be found in 
\cite{KP-16}, App.~A. For further comments concerning the method 
of obtaining asymptotic expansions for the $\sigma$-function, see \cite{KP-16}, 
Sect.~4.2, Rem.~4.3 and the discussion thereafter.  

The expansion \eqref{sigmaNlargeexp} can be constructed in a standard way by 
substituting it in the sixth Painlev\'e equation \eqref{PVI}. On this way, we 
first obtain the leading term, $\sigma_2(x)$, by requiring that it reproduces 
the conditions at the points $x=0,1,\infty$. Next we derive the   
further terms, and show that they can be obtained recursively. For these terms 
we also have to obtain that they reproduce the conditions at the points $x=0,1,\infty$. 

Having the outlined strategy in mind, we now turn to equation \eqref{PVI} and
consider the construction of the leading term of the large $N$ expansion of
the $\sigma$-function. Our first aim here is to expose how the function
$\sigma_2(x)$ can be found under assumption that, as $N\to \infty$,
\begin{equation}\label{vi-def}
\nu_i=v_i N+O(1),\qquad i=1,\ldots,4,
\end{equation}
where $v_i$ are some parameters to be specified later. Below we drop
the dependence on $x$ in the notation for functions to simplify writing.

Clearly, with \eqref{sigmaNlargeexp} 
and \eqref{vi-def} the right-hand side of \eqref{PVI}
is of $O(N^{8})$ and the same is valid for the second term in the left-hand side. 
The first term in the left-hand side, with the second-order derivative, is just 
of $O(N^{6})$. Excluding the trivial root of the constant solution, $\sigma'=0$
(and hence assuming that $\sigma_2'\ne 0$), we thus find that 
the equation for the $\sigma$-function splits into two first-order equations
for the $\sigma_2$-functions:
\begin{equation}\label{sigma2eq}
\sigma_2=x \sigma_2'-\frac{\sigma_2'}{2}-\frac{v_1v_2v_3v_4}{2\sigma_2'}
\pm\frac{\sqrt{\prod_i\big(\sigma_2'+v_i^2\big)}}{2\sigma_2'}.
\end{equation} 
For later use we introduce two functions $f_\pm(\sigma_2')$ 
by rewriting these equations in the form
\begin{equation}\label{ClairautEq}
\sigma_2 = x \sigma_2' + f_{\pm}(\sigma_2'), 
\end{equation}
where the plus and minus signs match those in \eqref{sigma2eq}.

Equations \eqref{ClairautEq} are the Clairaut equations (see, e.g., \cite{I-56}), 
i.e., they are of the form 
\begin{equation}
y=x y'+\Phi(y'), \qquad y'=y'(x).
\end{equation}
Differentiation with respect to $x$ gives 
\begin{equation}\label{dxClairaut}
\big(x+\Phi'(y')\big)y''=0.
\end{equation}
If $y''=0$, then $y$ is a linear function,
\begin{equation}\label{CgenSol}
y = C x + \Phi(C),
\end{equation}
where $C$ is a constant. This is the so-called general solution of the 
Clairaut equation. If instead the first factor in \eqref{dxClairaut} vanishes, then 
the corresponding solution is called singular solution and it is of the form 
\begin{equation}
y=\big(xy'+\Phi(y')\big)\big|_{y'=\left(\Phi'\right)^{-1}(-x)}.
\end{equation}
Note that there could be many such solutions or none. 

To study our problem, let us first consider the situation where   
the parameters $v_1,\ldots,v_4$ are related by 
\begin{equation}\label{v4321}
v_4=v_3,\qquad v_2=-v_1.
\end{equation}
We also assume that $v_1\geq v_3$. Then,  
\begin{equation}
f_\pm(\sigma_2')=
\begin{cases}
g_\pm(\sigma_2') & \sigma_2'\in (-\infty,-v_1^2]\bigcup	[-v_3^2,\infty)
\\
g_\mp(\sigma_2') & \sigma_2'\in [-v_1^2,-v_3^2],
\end{cases}
\end{equation}
where 
\begin{equation}
g_{+}(\sigma_2')=\frac{v_1^2v_3^2}{\sigma_2'}+\frac{v_1^2+v_3^2}{2},
\qquad	
g_{-}(\sigma_2')=-\sigma_2'-\frac{v_1^2+v_3^2}{2}.
\end{equation}
In the case of the function $g_{+}(\sigma_2')$ we have $g_{+}'(\sigma_2')=-(v_1v_3/\sigma_2')^2$ and hence 
there are two singular solutions corresponding to $\sigma_2'=\pm v_1 v_3/\sqrt{x}$; 
in the case of the function $g_{-}(\sigma_2')$ we have $g_{-}'(\sigma_2')=-1$ and there are no singular solutions. 
Thus equations \eqref{sigma2eq} have two general solutions 
\begin{equation}\label{GenSol}
(\sigma_2)_{\mathrm{g},+}=Cx+\frac{v_1^2v_3^2}{C}+\frac{v_1^2+v_3^2}{2},\qquad
(\sigma_2)_{\mathrm{g},-}=C(x-1)-\frac{v_1^2+v_3^2}{2},
\end{equation}
and two singular ones, 
\begin{equation}\label{SingSol}
(\sigma_2)_{\mathrm{s},\pm}=\pm 2v_1v_3\sqrt{x}+\frac{v_1^2+v_3^2}{2}.
\end{equation}  
Note that the two general solutions may correspond to the same linear function 
if the integration constant $C$ has the same value in both of them and 
satisfies $(C+v_1^2)(C+v_3^2)=0$. 
We meet exactly this situation in our considerations below. 

To proceed, we fix now values of the parameters. Recalling \eqref{nus} and 
\eqref{v4321}, we set
\begin{equation}
v_1=-v_2=r+\frac{1}{2}\equiv w,\qquad v_3=v_4=\frac{1}{2},	
\end{equation}
where we have introduced a new parameter $w$. 

Consider now the function $\log P_{N,M,L}(x^{-1})$.  
In the leading order, 
\begin{equation}
\log P_{N,M,L}(x^{-1}) = f_2 N^2 +O(N).
\end{equation}
From \eqref{P=intsigma} we have
\begin{equation}\label{f2froms2}
f_2 =  \int  \left(\sigma_2+\wt A_2 x-\wt B_2\right)\frac{\rmd x}{x(x-1)} + \wt C_2,
\end{equation}
where $\wt A_2$, $\wt B_2$, $\wt C_2$ are $O(N^2)$ terms of the constants 
$\wt A$, $\wt B$, $\wt C$, respectively, e.g., $\wt A=  N^2 \wt A_2+O(N)$. 
The constants $\wt A$ and $\wt B$ are defined in \eqref{wtAB}; the constant 
$\wt C$ fixes the normalization, $P_{N,M,L}(0)=1$. 

We obtain $\wt A_2$ and $\wt B_2$ from $\wt A=(N+1)^2/4$ (see \eqref{wtAB}) and 
the expression \eqref{wtBnu12N} for $\wt B$, respectively, 
\begin{equation}\label{wtA2B2}
\wt A_2=\frac{1}{4},\qquad \wt B_2=\frac{w^2}{2}-w+\frac{3}{8}.
\end{equation}
As for the constant $\wt C_2$, it can fixed by requiring that
the function $f_2$ attains its values 
at the points $x=\infty,0,1$, as prescribed by the statements of 
Props.~\ref{pr:xtoinfty}, \ref{pr:xto0}, and \ref{pr:xto1}, respectively. 
As we see below, there exists one and only one such a function.

To obtain these values of the function $f_2$, we have to rely on some auxiliary 
asymptotic result. As far as the number of boxed plane partitions is involved 
in Props.~\ref{pr:xto1} and \ref{pr:xto0}, we represent this number in the form
\begin{equation}\label{PLabcGs}
\PL(a,b,c) = \frac{G(a+1)G(b+1)G(c+1)G(a+b+c+1)}{G(a+b+1)G(a+c+1)G(b+c+1)},
\end{equation}
where $G(z)$ is the Barnes G-function defined by the relations
\begin{equation}\label{BarnesG}
G(z+1)=G(z)\Gamma(z), \qquad G(2)=G(1)=1.
\end{equation}
It is well known that \cite{E-81}, as $z\to \infty$, 
\begin{equation}\label{largeBarnes}
\log G(z+1) = \frac{z^2}{2} \log z - \frac{3}{4} z^2 + \frac{\ln 2\pi}{2}z - 
\frac{1}{12} \log z +\zeta'(-1) + O\left(z^{-2}\right),
\end{equation}
where $\zeta'(-1)=-0.165142...$ is the derivative of the Riemann function $\zeta(z)$
at $z=-1$.

Now we ready to compute values of the function $f_2$. 
First, from Prop.~\ref{pr:xtoinfty} it follows that $P_{N,M,L}(0)=1$ and hence, 
for arbitrary values of the parameters (i.e., not just limited to the symmetric case):  
\begin{equation}\label{f2atinfty}
\lim_{x\to\infty}f_2(x)=0. 
\end{equation}

Next, we find the value of $f_2$ at the point $x=1$ using Prop.~\ref{pr:xto1}. 
From \eqref{Cxto1} and \eqref{BarnesG}, we find
\begin{equation}\label{PNMLat1}
P_{N,M,L}(1)=\frac{G(a+2)G(b+2)G(c+2)G(a+b+c+2)}{G(a+b+2) G(a+c+2) G(b+c+2)},
\end{equation}
where $a=N$,  $b = L-N-1$, and $c = M-N$ (see \eqref{abcxto1}). 
In the symmetric case $a=N$, $b=r N+O(1)$, and $c=rN+O(1)$, so from \eqref{PNMLat1} 
and \eqref{largeBarnes}  we find that 
\begin{equation}\label{f2at1}
f_2(1)=\frac{(2r+1)^2}{2}\log(2r+1)
-(r+1)^2\log(r+1)-r^2\log 4r.
\end{equation}

Finally, let us consider the behavior of $f_2$ near the point $x=0$. 
From Prop.~\ref{pr:xto0} it follows that 
\begin{equation}\label{PNMLat0}
x^{ac}P_{N,M,L}(x^{-1})\big|_{x=0}
=\frac{G(a+2)G(b+1)G(c+2)G(a+b+c+1)}{G(a+b+1) G(a+c+2) G(b+c+1)},
\end{equation}
where $a=N$, $b =|M-L+1|$, $c = \min(L-N-1,M-N)$ (see \eqref{abc=xto0}).  
In the symmetric case $a=N$, $c=rN+O(1)$, but $b=O(1)$ as $N\to\infty$, hence
\eqref{largeBarnes} yields
\begin{equation}\label{f2at0}
\big(r\log x+ f_2(x)\big)\big|_{x=0}=0.
\end{equation}

The values of the function $f_2$ at the points $x=1$ and $x=0$ 
in the non-symmetric case ($p\ne q$) 
which follow from \eqref{PNMLat1} and \eqref{PNMLat0}, respectively, 
are computed in Sect.~\ref{sec:TD2-1}  (see \eqref{f2at1pq} and \eqref{f2at0pq}).

\subsection{Construction of the leading term}

Now we ready to address the problem of construction of the function 
$f_2$ describing the leading term of the 
function $\log P_{N,M,L}(x^{-1})$ in the thermodynamic limit. 
The function $f_2$ should satisfy the following properties. First, 
the corresponding $\sigma_2$-function should be given in terms of 
the solutions \eqref{GenSol} and \eqref{SingSol} of the associated Clairaut equation.  
Second, the function $f_2$ obtained from the function $\sigma_2$ by \eqref{f2froms2} 
should be consistent with the statements of Props.~\ref{pr:xtoinfty}--\ref{pr:xto1}. 
In particular, it should satisfy the conditions \eqref{f2atinfty}, \eqref{f2at1},
and \eqref{f2at0}. Third, it should be a \emph{continuous} function of $x$,
or, more exactly, piece-wise continuous, in case if several solutions 
from the first property are involved.     

Let us consider the first property, namely, we intend to identify the function 
$\sigma_2$ by requiring that its expansions at the points $x=\infty,1,0$ are 
consistent with \eqref{sigmatoinf}, \eqref{sigmatoone}, and
\eqref{sigmatozero}, respectively, specified to the symmetric case. 
Note that in doing so, we also involve partially the second property, because 
these expansions follow from the Props.~\ref{pr:xtoinfty}--\ref{pr:xto1}. 
We recall that we deal with the following values of the parameters: 
\begin{equation}
v_1=-v_2=w,\qquad v_3=v_4=\frac{1}{2}.	
\end{equation}

We start with considering the vicinity of the point $x=\infty$, where, 
as it follows from \eqref{sigmatoinf}, we should have 
\begin{equation}
\sigma_2=-\frac{x}{4}-\frac{w^2}{2}+\frac{1}{8}+O(x^{-2}),
\qquad x\to\infty.
\end{equation}
Clearly, the solution of the Clairaut equation which fulfills the 
required $x\to\infty$ behavior is any of the two general solutions 
$(\sigma_2)_{g,\pm}$ with $C=-v_3^2=-1/4$, see \eqref{GenSol}. 

Next, in the vicinity of the point $x=1$, as it follows from 
\eqref{sigmatoone}, we should have
\begin{equation}
\sigma_2=\frac{w^2}{2}-w+\frac{1}{8}-\frac{w}{2}(x-1)+\frac{w}{8}(x-1)^2
+O\left((x-1)^3\right),\qquad x\to1.
\end{equation}
Apparently, the solution which has such an expansion  
is the singular solution 
$(\sigma_2)_{\mathrm{s},-}$, see \eqref{SingSol}, with $C= w^2/2+1/8$.

Finally, in the vicinity of the point $x=0$ from \eqref{sigmazeroS} it follows that 
\begin{equation}
\sigma_2=\frac{w^2}{2}-\frac{1}{8} -w^2 x+ O\big(x^{3}\big),\qquad x\to0. 
\end{equation}
The solution which fulfills the required $x\to 0$ behavior 
is any of the two general solutions in \eqref{GenSol} with $C=-w^2$. 

Let us denote the obtained expressions for the $\sigma_2$-function
by $\sigma_2^\mathrm{I}$, $\sigma_2^\mathrm{II}$, and 
$\sigma_2^\mathrm{III}$, respectively. Summarizing, we thus have obtained
\begin{equation}\label{sigmas}
\begin{split}
\sigma_2^\mathrm{I}
&=
-\frac{x}{4} -\frac{w^2}{2} + \frac{1}{8},
\\	
\sigma_2^\mathrm{II}
&=
- w\sqrt{x}+\frac{w^2}{2} + \frac{1}{8},
\\
\sigma_2^\mathrm{III}
&=
- w^2 x + \dfrac{w^2}{2} - \dfrac{1}{8}.
\end{split}
\end{equation}
We recall that these expressions are valid near 
the points $x=\infty,1,0$, respectively. 

Let us now consider the function $f_2=f_2(x)$. We denote by $f_2^\mathrm{I}$, 
$f_2^\mathrm{II}$, and $f_2^\mathrm{III}$ the functions which 
are related to $\sigma_2^\mathrm{I}$, $\sigma_2^\mathrm{II}$, and $\sigma_2^\mathrm{III}$, 
respectively, via \eqref{f2froms2}. Taking into account \eqref{wtA2B2}, 
we obtain the following expressions
\begin{align}
f_2^\mathrm{I}&=\bigg(w-\frac{1}{2}\bigg)^2 \log \frac{x}{x-1}+\wt C_2^\mathrm{I}.
\\
f_2^\mathrm{II}&=2w \log\big(1+\sqrt{x}\big)-\left(w-\frac{1}{4}\right)\log x
+\wt C_2^\mathrm{II}.
\\
f_2^\mathrm{III} &= -\left(w-\frac{1}{2}\right)^2\log (1-x)
-\left(w-\frac{1}{2}\right)\log x +\wt C_2^\mathrm{III}.
\end{align}
Note that just like for the functions 
$\sigma_2^\mathrm{I}$, $\sigma_2^\mathrm{II}$, and $\sigma_2^\mathrm{III}$, the obtained 
expressions for $f_2^\mathrm{I}$, $f_2^\mathrm{II}$, and $f_2^\mathrm{III}$ are valid near 
the points $x=\infty,1,0$, respectively. 
This finishes consideration of the first property of the function $f_2$.  

The second property of the function $f_2$ in question 
concerns the values of the integration constants $\wt C_2^\mathrm{I}$, 
$\wt C_2^\mathrm{II}$, and $\wt C_2^\mathrm{III}$. 
The conditions  \eqref{f2atinfty} and \eqref{f2at0} 
are fulfilled with 
\begin{equation}\label{C0C0}
\wt C_2^\mathrm{I}=0,\qquad 
\wt C_2^\mathrm{III}=0.
\end{equation}
The condition \eqref{f2at1} means that 
\begin{equation}\label{wtC2IIsib}
\wt C_2^\mathrm{II}=
2 w^2\log 2w-\left(w+\frac{1}{2}\right)^2\log(2w+1)
-\left(w-\frac{1}{2}\right)^2\log(2w-1),
\end{equation}
or
\begin{equation}\label{wtC2IIbis}
\wt C_2^\mathrm{II}=
\frac{1}{2}\log \frac{1}{2w}
-\left(w+\frac{1}{2}\right)^2\log\left(1+\frac{1}{2w}\right)
-\left(w-\frac{1}{2}\right)^2\log\left(1-\frac{1}{2w}\right).
\end{equation}
As a result, we have fixed all the three functions 
$f_2^\mathrm{I}$, $f_2^\mathrm{II}$, $f_2^\mathrm{III}$ completely.

Now we address the third property, namely, that the function $f_2$ must be 
piece-wise continuous. We consider the simplest possible ansatz that
each of these three expressions is valid in some interval which contains 
the corresponding point. Specifically, we require that 
$f_2$ is a piece-wise continuous function of $x\in[0,\infty)$ and there exist 
two critical points $\xc^\pm\gtrless 1$ such that
\begin{equation}
f_2=
\begin{cases}
f_2^\mathrm{I} &x\in[\xc^+,\infty)
\\
f_2^\mathrm{II} &x\in[\xc^-,\xc^+] 
\\
f_2^\mathrm{III} &x\in[0,\xc^-]. 
\end{cases}
\end{equation}
The points $\xc^\pm$ must obey the equations
\begin{equation}\label{f2xpm} 
f_2^\mathrm{I}(\xc^+)=f_2^\mathrm{II}(\xc^+),
\qquad
f_2^\mathrm{II}(\xc^-)= f_2^\mathrm{III}(\xc^-).
\end{equation}
It turns out that despite the fact that these equations are in general 
transcendent, they can be solved, and, furthermore, uniqueness of their 
solutions can be proven.

Let us consider the first equation in \eqref{f2xpm}. Introduce the function
\begin{equation}
\rho_{+}(x)=f_2^\mathrm{II}(x)-f_2^\mathrm{I}(x).
\end{equation}
Using the first relation in \eqref{C0C0}  and \eqref{wtC2IIsib}, we get 
\begin{equation}
\rho_{+}(x)=
\left(w+\frac{1}{2}\right)^2\log\frac{\sqrt{x}+1}{2w+1}
+\left(w-\frac{1}{2}\right)^2\log\frac{\sqrt{x}-1}{2w-1}-
w^2\log \frac{x}{4w^2}.
\end{equation}
Apparently, the equation $\rho_{+}(x)=0$ has the root $x=4w^2$, and 
so we conclude that
\begin{equation}
\xc^+=4w^2=:\xc.
\end{equation}
To show that there are no other roots on the interval $(1,\infty)$,
we evaluate the derivative of the function $\rho_{+}(x)$ and find
\begin{equation}
\rho_{+}'(x)=\frac{\left(\sqrt{x}-2w\right)^2}{4x(x-1)}.
\end{equation}
Thus, the function $\rho_{+}(x)$ is a monotonously growing function 
for $x\in (1,\infty)$, except the point $x=4w^2$. This point is exactly 
the root we have obtained, and so there are no other roots on the interval 
$(1,\infty)$. Note that the second derivative of $\rho_{+}(x)$ 
also vanishes at this point  
but the third one does not, that implies that 
this is a point of the third-order phase transition.    

Let us consider the second equation in \eqref{f2xpm}. Introduce the function 
\begin{equation}
\rho_{-}(x)=f_2^\mathrm{II}(x)-f_2^\mathrm{III}(x).
\end{equation}
Using now the second relation in \eqref{C0C0} and \eqref{wtC2IIbis}, we get 
\begin{equation}
\rho_{-}(x)=\left(w+\frac{1}{2}\right)^2\log\frac{1+\sqrt{x}}{1+(2w)^{-1}}
+\left(w-\frac{1}{2}\right)^2\log\frac{1-\sqrt{x}}{1-(2w)^{-1}}-
\frac{1}{4}\log 4w^2x.
\end{equation}
Obviously, $\rho_{-}(x)=0$ for $x=1/4w^2$, and so  
\begin{equation}
\xc^-=\frac{1}{4w^2}=\xc^{-1}.
\end{equation}
We also have 
\begin{equation}
\rho_{-}'(x)=-\frac{\left(2w\sqrt{x}-1\right)^2}{4x(1-x)},
\end{equation}
so the function $\rho_{-}(x)$ is a monotonously decreasing function 
for $x\in (0,1)$, except the point $x=(4w^2)^{-1}$ where it vanishes 
together with its first and second derivatives. Thus, $x=1/4w^2$ is the 
only root of $\rho_{-}(x)$ for $x\in (0,1)$. This is another point of 
the third-order phase transition.  

As a comment to this calculation, it is useful to note that equations \eqref{f2xpm} 
appear to be elementary if we would assume that 
the function $f_2$ is continuous together with its first derivative. In other 
words, the assumption  
that the system demonstrates no first-order phase transitions 
can be very handy.  
Indeed, it means that  
the function $\sigma_2$ is required to be continuous and therefore 
\eqref{f2xpm} are replaced by the similar equations for the values of 
the function $\sigma_2$ at these points. From 
\eqref{sigmas} then one immediately obtains that $\xc^\pm=(4w^2)^{\pm1}$. 

The assumption that there are no first-order phase transitions 
is in a complete agreement with general properties of discrete random matrix models, 
which are known to exhibit phase transitions not harder than 
third-order ones \cite{MS-14}. 
We recall that $P_{N,M,L}(x^{-1})$ can be regarded as such a model, see 
\eqref{randommatrix} and \eqref{Pnew}. 

\subsection{Sub-leading corrections}

We now address the problem of computing the sub-leading corrections. 
For the $\sigma$-function it means construction
of other terms in the $1/N$ expansion \eqref{sigmaNlargeexp}; we 
limit ourselves here by obtaining the functions $\sigma_1$ and $\sigma_0$ though the 
procedure admits derivation of all the terms recursively \cite{KP-16}. 
The corresponding expansion for the function $\log P_{N,M,L}(x^{-1})$, as we see below, 
may additionally contain a $\log N$ term with the constant coefficient.

To fix the structure of $1/N$ corrections in a unique way, 
we first consider the $O(1)$ terms in \eqref{vi-def}. Indeed, for 
$\nu_3$ and $\nu_4$, see \eqref{nus}, the $O(1)$ terms are equal to $1/2$ 
and $-1/2$, respectively, and there are no further $1/N$ corrections. 
The bulk system parameters (besides $N$) are
contained in $\nu_1$ and $\nu_2$. These bulk system parameters can be identified 
in such a way that $\nu_1$ and $\nu_2$ have no $O(1)$ terms, that is, 
$v_1$ and $v_2$ are defined such that following relations hold exactly:
\begin{equation}\label{ExactN}
\nu_1=v_1N,\qquad \nu_2=v_2N.
\end{equation}
If we further set $v_1=1/2+p$ and 
$v_2=-1/2-q$, then we arrive at $p$ and $q$ defined in 
\eqref{def-p-q}. In the symmetric case one cannot however require absence of 
$O(1)$ terms, but can require, for example, that $v_1$ and $v_2$ have these terms equal,   
\begin{equation}\label{ExactNsym}
\nu_1=w N+\frac{\epsilon}{2},\qquad  \nu_2=-w N+\frac{\epsilon}{2}. 
\end{equation}
An advantage of the choice \eqref{ExactNsym} is that 
the $\sigma$-form of the sixth Painleve equation, \eqref{PVI}, 
then contains only even powers of $N$ in its coefficients, just like in the non-symmetric case 
\eqref{ExactN}. This property of the coefficients is very useful and can imply, 
under further conditions 
to be met, that all terms $\sigma_{1-2k}$, $k=0,1,\ldots$, in \eqref{sigmaNlargeexp} vanish. 

The expansion \eqref{sigmaNlargeexp} can be constructed by plugging 
it in \eqref{PVI} and matching powers in $N$. Instead of operating with 
\eqref{PVI} directly, one can simplify   
calculations by noting that since $\sigma_2'\ne 0$ for all values of 
$x\in[0,\infty)$, a systematic treatment of $1/N$ corrections can be done 
by considering the factorization of \eqref{PVI} on two equations 
\begin{equation}\label{splitPVI}
\sigma=x \sigma'+F_\pm(\sigma',\sigma''),
\end{equation}
where the functions $F_\pm(\sigma',\sigma'')=F_\pm(\sigma',\sigma'';\nu_1,\nu_2,\nu_3,\nu_4)$
are  
\begin{equation}\label{Fpm}
F_\pm(\sigma',\sigma'')
=-\frac{\sigma'}{2}-\frac{\prod_i\nu_i}{2\sigma'}
\pm\frac{\sqrt{\prod_i\big(\sigma'+\nu_i^2\big)
-\sigma'\big(x(x-1)\sigma''\big)^2}}{2\sigma'}.
\end{equation}
In the large $N$ limit the functions $F_\pm(\sigma',\sigma'')$
in the leading order turn into the functions $f_\pm(\sigma_2')$ appearing in the 
Clairaut equations \eqref{ClairautEq}, 
\begin{equation}\label{Fpmfpm}
F_\pm(N^2\sigma_2',0;v_1N,v_2N,v_3N,v_4N)=N^2f_\pm(\sigma_2').
\end{equation}
Thus, the expansion 
\eqref{sigmaNlargeexp} can be constructed by identifying which one of the two equations 
in \eqref{splitPVI} actually is satisfied in all orders in $N$. 

This appears to be straightforward in Regime II where the leading term, $\sigma_2$, 
is given by a singular solution of $\sigma_2=x\sigma_2'+f_{-}(\sigma_2')$ thus 
identifying the equation which contains the function $F_{-}(\sigma',\sigma'')$.
Furthermore, it can also be easily shown that in this case the first sub-leading 
correction vanishes, $\sigma_1=0$. 
Indeed, recalling that the $\sigma''$-term 
and the $O(1)$ corrections of the parameters $\nu_1,\ldots,\nu_4$ 
can contribute only to the order $1/N^2$ with respect to the leading term, 
the substitution  $\sigma=N^2\sigma_2+N\sigma_1+O(1)$ yields
\begin{align}\label{FmtoO(N)}
F_{-}\big(\sigma',\sigma'')
&=F_{-}(N^2\sigma_2'+N\sigma_1',0; wN,-wN,N/2,N/2)\left(1+O(N^{-2})\right)	
\notag\\ &
=N^2 f_{-}(\sigma_2'+\sigma_1'/N)+O(1)
\notag\\ &
=N^2 f_{-}(\sigma_2')+Nf_{-}'(\sigma_2')\sigma_1'+O(1),
\end{align}
where at the second step we have used \eqref{Fpmfpm} specified to the symmetric 
case for a concreteness. Taking into account that 
for the singular solution $x+f_{-}'(\sigma_2')=0$, 
for the function $\sigma_1$ we obtain
\begin{equation}\label{sigma1=0}
\sigma_1=\big(x+f_{-}'(\sigma_2')\big)\sigma_1'=0.
\end{equation}
Note that $\sigma_1=0$ for Regime II implies that in the expansion \eqref{sigmaNlargeexp} 
all the terms of odd powers in $1/N$ also vanish, and  
\eqref{sigmaNlargeexp} becomes an expansion in $1/N^2$  
(see also the discussion in \cite{KP-16}, Sect.~4, where 
the similar phenomenon have been argued differently).   

As for Regime I and Regime III, we have obtained that 
$\sigma_2$-function in these cases is given by 
regular solutions of the Clairaut equation. These solutions are such that 
$f_{-}(\sigma_2')=f_{+}(\sigma_2')$, and so both equations in 
\eqref{splitPVI} vanish in the leading order.
To identify which one of the two equations responsible for the $1/N$ expansion,
below we expand functions $F_\pm(\sigma',\sigma'')$ to find equations for 
the $\sigma_1$-function, and choose the solutions which possess 
the required $x\to \infty$ (for Regime I) and $x\to 0$ (for Regime III) expansions
as prescribed by Cor.~\ref{cor:xtoinfty} and Cor~\ref{cor:xto0}, respectively. 
The $\sigma_1$-functions in both cases appear to be given by singular solutions 
of some other Clairaut equations. This allows us to identify the relevant equation 
among the two in \eqref{splitPVI} just like it has been done above for the Regime II. 
It turns out that the equation with the function $F_{-}(\sigma',\sigma'')$ in \eqref{splitPVI} 
is the relevant one for all the three regimes. 

We now turn to giving details of calculations for each of the regimes separately.

\subsubsection{Regime I}

We start with noting that plugging \eqref{ExactNsym} in \eqref{sigmatoinf} gives 
the following expression for the function $\sigma_1$ as $x\to\infty$: 
\begin{equation}\label{s1xtoinf}
\sigma_1=-\frac{x}{2}+w^2+\frac{1}{4}+\left(w^2-\frac{1}{4}\right)^2\frac{1}{x} 
+O\left(x^{-2}\right).
\end{equation}

To obtain the $\sigma_1$-function it is sufficient to consider 
large $N$ expansion of the functions
$F_\pm(\sigma',\sigma'')$  to order $N$; we expand them to order $N^0$, 
so that we can obtain next the function 
$\sigma_0$. We recall that in Regime I $\sigma_2'=-1/4$
and hence the square root term in \eqref{Fpm} is not contributing to the leading, $N^2$, 
order. More 
exactly, the term $\prod_i(\sigma'+\nu_i^2)$ is of $O(N^6)$ (instead of $O(N^8)$); explicitly  
\begin{multline}
\prod_i\big(\sigma'+\nu_i^2\big)
=N^6\left(w^2-\frac{1}{4}\right)^2\left[(\sigma_1')^2-\frac{1}{4}\right]
\\
+N^5 \left\{\left(w^2-\frac{1}{4}\right)\left[(\sigma_1')^2-\frac{1}{4}\right]+
\left(w^2-\frac{1}{4}\right)^2\left(\sigma_0'+\frac{1}{4}\right)\right\}2\sigma_1'
+O(N^4).
\end{multline}
Furthermore, since $\sigma_2''=0$, the term 
$\sigma'(\sigma'')^2\sim N^4 \sigma_2'(\sigma_1'')^2$ contributes to the large 
$N$ expansion of $F_\pm(\sigma',\sigma'')$ starting from order $N^{-1}$. 
As a result, we obtain   
\begin{multline}\label{FpmRI}
F_\pm(\sigma',\sigma'')
=N^2\left(\frac{1}{8}-\frac{w^2}{2}\right)
+N\left(-\frac{\sigma_1'}{2}-2w^2\sigma_1'
\mp 2\left(w^2-\frac{1}{4}\right)\sqrt{(\sigma_1')^2-\frac{1}{4}}\right)
\\
+\left(-\frac{1}{2}-2w^2
\mp 
\frac{2\left(w^2-\frac{1}{4}\right)\sigma_1'}
{\sqrt{(\sigma_1')^2-\frac{1}{4}}}\right)\sigma_0'
-8w^2 (\sigma_1')^2+\frac{w^2}{2}+\frac{\epsilon^2}{8}
\\
\mp \left(8w^2\sqrt{(\sigma_1')^2-\frac{1}{4}}
+\frac{w^2-\frac{1}{4}}
{2\sqrt{(\sigma_1')^2-\frac{1}{4}}}\right)\sigma_1'
+O(N^{-1}).
\end{multline}

Hence, $\sigma_1$ must satisfy one of the two equations
\begin{equation}\label{s1Clairaut}
\sigma_1=  \left(x  
-\frac{1}{2} - 2 w^2\right) \sigma_1'
\mp 2\left(w^2 - \frac{1}{4}\right) \sqrt{(\sigma_1')^{2}-\frac{1}{4}},
\end{equation}
where the signs correspond to $F_\pm(\sigma',\sigma'')$. 
These equations are the Clairaut equations. The presence of the 
$1/x$ term in  \eqref{s1xtoinf} indicates that we deal here with 
a singular solution. All such solutions of 
\eqref{s1Clairaut} satisfy
\begin{equation}
\left(\sigma_1'\right)^2= \frac{(2x-4w^2-1)^2}{16(x-1)(x-4w^2)}.
\end{equation}  
Specifically, the solution which obeys \eqref{s1xtoinf} is
\begin{equation}
\sigma_1 = -\frac{1}{2}\sqrt{(x-1)(x-4w^2)}, 
\end{equation}
and it can be easily checked that it corresponds to the plus sign in \eqref{s1Clairaut}, 
that is, to the function $F_{-}(\sigma',\sigma'')$.  

As far as the function in \eqref{splitPVI} is determined, 
the function $\sigma_0$ can be computed. One can easily see a remarkable property of the 
expansion \eqref{FpmRI}: the coefficient of the $\sigma_0'$ term is exactly the derivative 
with respect to $\sigma_1'$ of the $N$-order term. Since $\sigma_1$ is given by 
the singular solution of \eqref{s1Clairaut}, the $\sigma_0'$ term exactly vanishes at 
order $N^0$ in \eqref{splitPVI}, that yields 
\begin{equation}
\sigma_0=
-8w^2 (\sigma_1')^2+\frac{w^2}{2}+\frac{\epsilon^2}{8}
+8w^2\sigma_1'\sqrt{(\sigma_1')^2-\frac{1}{4}}
+\frac{\left(w^2-\frac{1}{4}\right)\sigma_1'}{2\sqrt{(\sigma_1')^2-\frac{1}{4}}}.
\end{equation}
Explicitly, the result reads
\begin{equation}
\sigma_0 = 
-\frac{w^2(x-1)}{x-4w^2}-\frac{x}{4}+\frac{1+\epsilon^2}{8}.
\end{equation}

Let us now consider the function $\log P_{N,M,L}(x^{-1})$. 
From Props.~\ref{pr:xtoinfty} and \ref{pr:sigma-function}, it follows that 
\begin{equation}\label{logPN2N1N0}
\log P_{N,M,L}(x^{-1})=N^2 f_2+ Nf_1+f_0+\ldots,
\end{equation} 
where, since $P_{N,M,L}(0)=1$, 
all the terms must vanish as $x\to\infty$. In particular,
\begin{equation}
\lim_{x\to\infty}f_1(x)=0,\qquad 
\lim_{x\to\infty} f_0(x)=0. 
\end{equation}
We compute $f_1$ by 
\begin{equation}\label{intf1}
f_1 =\int \left(\sigma_1+\wt A_1 x - \wt B_1\right)\frac{\rmd x}{x(x-1)} +\wt C_1,
\end{equation}
where, $\wt A_1$ and $\wt B_1$ are $O(N)$ terms of the large $N$ expansion of $\wt A$ and 
$\wt B$ in \eqref{wtAB}, respectively. Using \eqref{wtBnu12N} for $\wt B$ 
and taking into account \eqref{ExactNsym}, we find  
\begin{equation}
\wt A_1=\frac{1}{2},\qquad \wt B_1=\frac{1}{2}.
\end{equation}
Choosing $\wt C_1$ to ensure that $\lim_{x\to\infty}f_1(x)=0$, we get
\begin{equation}
f_1
= 2w \log \frac{2w\sqrt{x-1} + \sqrt{x-4w^2}}{(2w+1)\sqrt{x}}
- \log \frac{\sqrt{x-1} + \sqrt{x-4w^2}}{2\sqrt{x}}.
\end{equation}
Essentially similarly, for $f_0$, using 
\begin{equation}\label{intf0}
f_0 =\int \left(\sigma_0+\wt A_0 x - \wt B_0\right)\frac{\rmd x}{x(x-1)} +\wt C_0,
\end{equation}
where  
\begin{equation}
\wt A_0=\frac{1}{4},\qquad \wt B_0=\frac{1-\epsilon^2}{8},	
\end{equation} 
and choosing $\wt C_0$ such that $\lim_{x\to\infty}f_0(x)=0$, we obtain 
\begin{equation}
f_0= -\frac{1}{4} \log \left(1-\frac{4w^2}{x}\right)
+\frac{\epsilon^2}{4}\log\left(1-\frac{1}{x}\right).
\end{equation}
Finally, rewriting these formulas in terms of $\xc=4w^2$ 
we arrive at the expressions for the functions $f_1^\mathrm{I}$ and $f_0^\mathrm{I}$ 
appearing in Thm.~\ref{th:square}.

\subsubsection{Regime II}

Given that $\sigma_1=0$, we are left with the task of obtaining the $\sigma_0$-function. 
This can be done directly by expanding the function $F_{-}(\sigma',\sigma'')$ around the 
leading term by setting $\sigma=N^2\sigma_2+\sigma_0$ and taking into account that 
$\sigma_2'\in (-w^2,-1/4)$. As it can be anticipated
from the considerations above for the $\sigma_1$-function, see \eqref{FmtoO(N)}, 
all terms at $N^0$ order in \eqref{splitPVI} depending on $\sigma_0'$ vanish, 
just like it takes place in \eqref{sigma1=0} due to the overall 
factor $x+f_{-}'(\sigma_2')=0$.
As a result, we get the following expression the function $\sigma_0$:
\begin{multline}
\sigma_0=
-\frac{[x(x-1)\sigma_2'']^2}{\big(4\sigma_2'+1\big)\big(\sigma_2'+w^2\big)}
-\frac{w^2}{8\sigma_2'}
+\frac{\big(\sigma_2'+w^2\big)\big(4\sigma_2'-1\big)}
{8\sigma_2'\big(4\sigma_2'+1\big)}
\\
-\frac{\epsilon^2}{32\sigma_2'}
+\frac{\epsilon^2\big(4\sigma_2'+1\big)\big(\sigma_2'-w^2\big)}
{32\sigma_2'\big(\sigma_2'+w^2\big)}.
\end{multline}
Using $\sigma_2'=-w/2\sqrt{x}$ and $\sigma_2''=w/4x^{3/2}$, 
we get
\begin{equation}
\sigma_0=-\frac{w(x-1)}{8}\left\{\frac{3}{2w-\sqrt{x}}
+\frac{\sqrt{x}}{2w\sqrt{x}-1}\right\}+
\frac{\epsilon^2(x-1)}{4(2w\sqrt{x}-1)}-\frac{1+\epsilon^2}{8}.
\end{equation}

Let us now consider the function  $\log P_{N,M,L}(x^{-1})$. 
We first note that a more detailed calculation with the 
help of \eqref{largeBarnes} applied to \eqref{PNMLat1} 
with $a=N$, $b=rN-\frac{\epsilon+1}{2}$, and $c=rN+\frac{\epsilon-1}{2}$
yields 
\begin{equation}\label{PNMLat1largeN}
\log P_{N,M,L}(1)=N^2 f_2(1)+ N f_1(1)+ \frac{5}{12}\log N +f_0(1)+ O(N^{-1}),
\end{equation}
where $f_2(1)$ is given by \eqref{f2at1}, and the values $f_1(1)$ and
$f_0(1)$ are 
\begin{equation}\label{f1at1}
f_1(1)=-(r+1)\log(r+1)+r\log r
\end{equation}
and
\begin{equation}\label{f0at1}
f_0(1)=
\frac{1}{12}\log\frac{2r^2}{(r+1)(2r+1)}
+\frac{\epsilon^2}{4}\log\frac{r}{r+1}
+\zeta'(-1)+\log \sqrt{2\pi},
\end{equation}
respectively. From Prop.~\ref{pr:xto1} and expansion \eqref{PNMLat1largeN} 
we conclude that for the values of $x$ corresponding to Regime II the following 
expansion is valid:
\begin{equation}\label{logPRegII}
\log P_{N,M,L}(x^{-1})=N^2 f_2(x)+ N f_1(x)+ \frac{5}{12}\log N +f_0(x)+ O(N^{-1}).
\end{equation}
Clearly, the functions $f_1$ and $f_0$ can be found from the functions 
$\sigma_1$ and $\sigma_0$ by \eqref{intf1} and \eqref{intf0}, respectively.

Computing $f_1$ by \eqref{intf1}, where $\wt A_1=\wt B_1=1/2$, we get 
\begin{equation}
f_1=\frac{1}{2}\log x+\wt C_1. 
\end{equation}
Computing $f_0$ by \eqref{intf0}, where $\wt A_0=1/4$ and $\wt B_0=(1-\epsilon^2)/8$, 
we get 
\begin{equation}
f_0=\frac{1}{8}\big\{3\log\big(2w-\sqrt{x}\big)-\log\big(2w\sqrt{x}-1\big)
+\log\sqrt{x}\big\}+\frac{\epsilon^2}{2}\log\frac{2w\sqrt{x}-1}{\sqrt{x}}
+\wt C_0.
\end{equation}
The integration constants can be fixed by using \eqref{f1at1} and \eqref{f0at1}. We obtain
\begin{equation}
\wt C_1=-(1+r)\log(1+r)+r\log r
\end{equation}
and 
\begin{equation}
\wt C_0
=-\frac{1}{12}\log\big(4r(r+1)(2r+1)\big) -\frac{\epsilon^2}{4}\log\big(4r(r+1)\big)
+\zeta'(-1)+ \log\sqrt{2\pi}.
\end{equation}
As a result, using $\sqrt{\xc}=2w=2r+1$ we arrive at 
the expressions for the functions $f_1^\mathrm{II}$
and $f_0^\mathrm{II}$ given in Thm.~\ref{th:square}.

\subsubsection{Regime III}

Considerations in this regime in general are very similar to those in Regime I.
We start with expansion \eqref{sigmatozero}, which with \eqref{ExactNsym} 
gives the following expression for the function $\sigma_1$ as $x\to0$, 
\begin{equation}\label{s1tozero}
\sigma_1=-\frac{|\epsilon|}{2}
+|\epsilon|\left(w^2+\frac{1}{4}\right)x-|\epsilon|\left(w^2-\frac{1}{4}\right)^2 x^2+O(x^3).
\end{equation}

To find the function $\sigma_1$ and which one of the two equations in \eqref{splitPVI} 
is relevant for $1/N$ expansion, and next obtain the function $\sigma_0$, we expand 
the functions $F_\pm(\sigma',\sigma'')$ to order $N^0$: 
\begin{multline}
F_\pm(\sigma',\sigma'')
=N^2\left(\frac{w^2}{2}-\frac{1}{8}\right)
+N\left(-\frac{\sigma_1'}{2}-\frac{\sigma_1'}{8w^2}
\mp \frac{(4w^2-1)\sqrt{(\sigma_1')^2-\epsilon^2w^2}}{8w^2}\right)
\\
+\left(-\frac{1}{2}-\frac{1}{8w^2}
\mp
\frac{(4w^2-1)\sigma_1'}
{8w^2\sqrt{(\sigma_1')^2-\epsilon^2w^2}} 
\right)\sigma_0'
-\frac{(\sigma_1')^2}{8w^4}+\frac{1}{8}+\frac{\epsilon^2}{32w^2}
\\
\pm \left(\frac{\sqrt{(\sigma_1')^2-\epsilon^2w^2}}{8w^4}
-\frac{\epsilon^2(4w^2-1)}
{32w^2\sqrt{(\sigma_1')^2-\epsilon^2w^2}}\right)\sigma_1'
+O(N^{-1}).
\end{multline}
Hence, the function $\sigma_1$ must be a singular solution (as far as \eqref{s1tozero} 
contains $x^2$ term) of one of the following two Clairaut equations: 
\begin{equation}\label{s1Clairaut3}
\sigma_1=x\sigma_1'-\frac{\sigma_1'}{2}-\frac{\sigma_1'}{8w^2}
\mp \frac{(4w^2-1)\sqrt{(\sigma_1')^2-\epsilon^2w^2}}{8w^2}.
\end{equation}
The solution which obeys \eqref{s1tozero} is
\begin{equation}
\sigma_1=-\frac{|\epsilon|}{2}\sqrt{(1-4w^2 x)(1-x)},
\end{equation}
and it corresponds to the plus sign in \eqref{s1Clairaut3}, that is, to the function 
$F_{-}(\sigma',\sigma'')$ in \eqref{splitPVI}.

As a result, for $\sigma_0$ we get
\begin{equation}
\sigma_0=-\frac{(\sigma_1')^2}{8w^4}+\frac{1}{8}+\frac{\epsilon^2}{32w^2}
-\frac{\sqrt{(\sigma_1')^2-\epsilon^2w^2}}{8w^4}
+\frac{\epsilon^2(4w^2-1)\sigma_1'}
{32w^2\sqrt{(\sigma_1')^2-\epsilon^2w^2}}
\end{equation}
and substitution of the function $\sigma_1'$ gives
\begin{equation}
\sigma_0=-\frac{\epsilon^2(1-x)}{4(1-4w^2 x)}-\frac{\epsilon^2 x}{4}+\frac{1+\epsilon^2}{8}.
\end{equation}

Let us now turn to the function $\log P_{N,M,L}(x^{-1})$. We begin with 
addressing its $x\to0 $ behavior, using \eqref{PNMLat0}. 
We have $a=N$, $b=|\epsilon|$, and  
$c=rN-\frac{1+|\epsilon|}{2}$, so that, as $N\to \infty$, \eqref{largeBarnes} 
yields   
\begin{multline}\label{PLMNat0largeN}
x^{ac}P_{N,M,L}(x^{-1})\big|_{x=0}
=
N(1-|\epsilon|)\big(r\log r -(r+1)\log (r+1)\big)+\frac{1-\epsilon^2}{2}\log N
\\
+(1-|\epsilon|)\log \sqrt{2\pi}+\log G(1+|\epsilon|)+O(N^{-1}).
\end{multline}
This formula implies that in the Regime III the following expansion is valid:
\begin{equation}
\log P_{N,M,L}(x^{-1}) = N^2f_2(x)+Nf_1(x)+\frac{1-\epsilon^2}{2}\log N +f_0(x)+O(N^{-1}).
\end{equation} 
The $x\to 0$ behavior of the function $f_2(x)$ is given by \eqref{f2at0}, and
from \eqref{PLMNat0largeN} we also get 
\begin{equation}\label{f1at0}
\left(-(1+|\epsilon|)\log \sqrt{x} +f_1(x)\right)\big|_{x=0} 
=(1-|\epsilon|)\big(r\log r -(r+1)\log (r+1)\big)
\end{equation}
and 
\begin{equation}\label{f0at0}
f_0(0)=(1-|\epsilon|)\log \sqrt{2\pi}+\log G(1+|\epsilon|).
\end{equation}

Computing $f_1$ by \eqref{intf1}, where $\wt A_1=\wt B_1=1/2$, we get 
\begin{multline}
f_1=2|\epsilon|w\log\left(2w\sqrt{1-x}+\sqrt{1-4w^2x}\right)
-|\epsilon|\log\left(\sqrt{1-x}+\sqrt{1+4w^2x}\right)
\\
+ \frac{1+|\epsilon|}{2}\log x
+\wt C_1. 
\end{multline}
Computing $f_0$ by \eqref{intf0}, where $\wt A_0=1/4$ and $\wt B_0=(1-\epsilon^2)/8$, 
we get 
\begin{equation}
f_0=-\frac{\epsilon^2}{4}\log(1-4w^2 x)+\frac{1}{4}\log(1-x)+\wt C_0.
\end{equation}
From \eqref{f1at0} we find
\begin{equation}
\wt C_1=r\log r-(r+1)\log(r+1)
-|\epsilon|r\log\big(4r(r+1)\big)
\end{equation}
and from \eqref{f0at0} we find
\begin{equation}
\wt C_0=(1-|\epsilon|)\log \sqrt{2\pi}+\log G(1+|\epsilon|).
\end{equation}
Finally, rewriting the arguments of the logarithms in terms of $\sqrt{\xc}=2w=2r+1$, 
we arrive at the expressions for the functions $f_1^\mathrm{III}$ and $f_0^\mathrm{III}$
provided in Thm.~\ref{th:square}. 

This finalizes the proof of Thm.~\ref{th:square}. 

Let us now consider the special case of $\epsilon=0$ of Regime III, 
and show how the expansion \eqref{e=0RIII} follows from our 
considerations above. Indeed, in this case  
$\sigma_1=0$ and $\sigma_0=1/8$, so \eqref{sigmaNlargeexp} reads
\begin{equation}\label{sigmae0}
\sigma=N^2\left(- w^{2}x +\frac{w^2}{2} -\frac{1}{8} \right)+ \frac{1}{8} + O(N^{-1}).
\end{equation} 
This expression has to be compared with a trivial solution of \eqref{PVI} valid for 
$\nu_1=-\nu_2$, and $\nu_1,\nu_3,\nu_4$ arbitrary, of the form
\begin{equation}\label{exactsigma}
\sigma_\textrm{triv} =-\nu_1^2 x+\frac{\nu_1^2-\nu_3\nu_4}{2}.
\end{equation}
Setting $\nu_1=wN$, $\nu_3=(N+1)/2$, and $\nu_4=(N-1)/2$ in \eqref{exactsigma}
one can reproduce the terms shown in \eqref{sigmae0}. 
This means that in Regime III at $\epsilon=0$ 
all terms in the expansion \eqref{sigmaNlargeexp} beyond $\sigma_0$ vanish. Note that 
does not mean that $\sigma=\sigma_\textrm{triv}$ but it implies that 
\begin{equation}
\sigma=\sigma_\mathrm{triv}+O(N^{-\infty}),
\end{equation}
where $O(N^{-\infty})$ stands for terms which are less than any given degree in 
$1/N$. In fact, this term corresponds to exponentially small
corrections. These corrections can also be tackled  
though some additional information is necessary. The situation here 
is similar to that considered in \cite{KP-16} for the so-called ordered regime 
(Sect.~5 therein).

It is worth mentioning that the phenomenon of absence of the 
$1/N$ corrections in \eqref{e=0RIII} can already be anticipated from 
the $x\to 0$ expansion for the polynomial $P_{N,M,L}(x^{-1})$ given in Prop.~\ref{pr:xto0}. 
The case of $\epsilon=0$ corresponds to $b=0$ in \eqref{kappazero}, and 
\eqref{P(x)=xto0} says that, as $x\to 0$,    
\begin{equation}\label{Pat0atb0}
P_{N,M,L}(x^{-1}) = 
\frac{1}{\binom{a+c}{a}x^{ac}} 
\left\{1+c(c+1) x+ \frac{c(c+1)(c^2+c+1)}{2} x^2+O(x^{3})\right\},
\end{equation}
or
\begin{equation}\label{logxto0}
\log\left(\binom{a+c}{a}P_{N,M,L}(x^{-1}) \right)
=ac \log x +c(c+1)\left(x+\frac{x^2}{2}\right)+O(x^{3}).
\end{equation}
Recalling that $a=N$ and $c=rN-1/2$, we thus see that the right-hand side of \eqref{logxto0}
in the large $N$ limit contains no $1/N$ corrections; there are only terms of 
orders $N^2$, $N$ and $N^0$. Furthermore, the expression in the braces in \eqref{Pat0atb0} 
is nothing but a truncated expansion of $(1-x)^{-c(c+1)}$ (one can check by expanding 
further in $x$) and so 
the corresponding $c(c+1)$ term in \eqref{logxto0} is $\log(1-x)$.  
One can therefore expect that for $x$ taking positive values at some interval 
attached to the origin, the following must hold: 
\begin{equation}\label{logbinomP}
\log\left(\binom{a+c}{a}P_{N,M,L}(x^{-1}) \right)
=ac \log x +c(c+1)\log(1-x)+O(N^{-\infty}).
\end{equation}
As we have established, Regime III corresponds to 
$x\in [0,\xc^{-1})=[0,(2r+1)^{-2})$ and so \eqref{logbinomP} holds for these 
values of $x$.  

As a final comment here, we mention that \eqref{logbinomP} admits a simple interpretation 
when translated back to the partition function using \eqref{Z=EP} and \eqref{wtZ}. 
Reverting to the weights \eqref{weights}, one can write
\begin{equation}\label{ZIII}
Z=\frac{w_1^{(M-N)(L-N)}w_3^{(M-L+N)N}(w_5w_6)^{M(L-N)}}{(w_5w_6-w_3w_4)^{(M-N)(L-N)}}
\left(1+O(N^{-\infty})\right) 
\end{equation}
where $M-L+1=0$. The weights are subject to the constraint that they must 
obey the condition of the Regime III, which now reads 
\begin{equation}
\frac{w_3w_4}{w_5w_6}\in \big[0, N^2/(M+L-N)^2\big).
\end{equation} 
If one expands the 
denominator in \eqref{ZIII} in Taylor series in $w_3w_4/w_5w_6$, then the leading 
term gives the weight of the anti-ferroelectric ground state configuration 
shown in Fig.~\ref{fig-AFGroundState}. Thus, formula \eqref{ZIII} 
can be interpreted as the result of summation over relevant perturbations 
from this ground state, valid up to 
exponentially small corrections in the large $N$ limit.   

\section{Thermodynamic limit in the non-symmetric case}
\label{sec:TD2}

In this section we focus on construction of the 
asymptotic expansions for the $\sigma$-function and the corresponding 
polynomial $P_{N,M,L}(x^{-1})$ in the limit 
$N,M,L\to \infty$ such that the parameters $p$ and $q$ defined by \eqref{def-p-q}
are finite and \emph{not} equal to each other, $p\ne q$. This will provide 
a proof of Thm.~\ref{th:rectangle}.

\subsection{Preliminaries}
\label{sec:TD2-1}

In Sect.~\ref{sec:CE} it is shown that the leading term of the $\sigma$-function 
in the thermodynamic limit, the function $\sigma_2$, see \eqref{sigmaNlargeexp}, 
can be found as a solution of the Clairaut equations \eqref{ClairautEq}. 
The non-symmetric case mean that the two parameters 
$v_1$ and $v_2$ are unrelated, thought the relation $v_4=v_3$ holds. 
Henceforth we set 
\begin{equation}\label{v-u12}
v_1:=v,\qquad v_2:=-u,\qquad v_3=v_4=\frac{1}{2}.
\end{equation}
Note that $v=p+1/2$ and $u=q+1/2$ where $p$ and $q$ are defined in \eqref{def-p-q}.  
Since $p,q>0$, we have $v,u>1/2$. 
We focus our attention on the case where the function $\sigma_2'$ satisfies
\begin{equation}\label{minu2v2}
\sigma_2'\in(-\min(v^2,u^2),-1/4].
\end{equation}
For the functions $f_\pm(\sigma_2')$ in \eqref{ClairautEq} we then have
\begin{equation}\label{sigma2eq34}
f_\pm(\sigma_2')=-\frac{\sigma_2'}{2}+\frac{v u}{8\sigma_2'}
\mp\frac{\big(\sigma_2'+1/4\big)\sqrt{\big(\sigma_2'+v^2\big)\big(\sigma_2'+u^2\big)}}
{2\sigma_2'}.
\end{equation} 
As we see below, the condition \eqref{minu2v2} is indeed always fulfilled in our problem.    

As usual, we have general solutions given by linear functions \eqref{CgenSol}. 
In the non-symmetric case \eqref{v-u12} some concerns may arise in
dealing with the singular solutions. Recall that these are the solutions
which correspond to vanishing of the first factor in \eqref{dxClairaut}. In
the case of the functions \eqref{sigma2eq34} one has to find roots of 
\emph{quartic} equations. 

Instead of dealing with these roots explicitly, which are given by bulky expressions,
one can search the solutions in a parametric form \cite{CP-15}.  
To solve the equations $x+f_\pm'(\sigma_2')=0$ for 
the function $\sigma_2'$ in terms of $x$, we introduce function $y=y(x)$
by defining it such that
\begin{equation}\label{sqrt-y}
\sqrt{\frac{\sigma_2'+u^2}{\sigma_2'+v^2}}=\frac{\alpha y+ \beta}{\gamma y +\delta},
\end{equation}
where $\alpha,\ldots,\delta$ are some functions of $v$ and $u$ only. 
One can set $\alpha(v,u)=\gamma(u,v)$ and $\beta(v,u)=\delta(u,v)$, so that 
\eqref{sqrt-y} holds identically at $u=v$. In our calculations 
below we make a particular
choice of these functions, though this choice is not essential for 
obtaining a solution in the parametric form. 

To proceed, we introduce the notation
\begin{equation}
X_\pm=\alpha y+ \beta\pm (\gamma y+\delta),\qquad Y_\pm=v (\alpha y+\beta)\pm u (\gamma y+\delta).
\end{equation}
From \eqref{sqrt-y} we get
\begin{equation}\label{YYoverXX}
\sigma_2'=-\frac{Y_{+}Y_{-}}{X_{+}X_{-}}.
\end{equation}
Substituting \eqref{YYoverXX} into the relation $x+f_\pm'(\sigma_2')=0$, 
and using \eqref{sqrt-y}, we obtain
\begin{equation}\label{x-funof-y}
x=\frac{X_\pm^2 \big(X_\mp+2Y_\pm\big)\big(X_\mp-2Y_\pm\big)}
{16 Y_\pm^2 (\alpha y +\beta)(\gamma y + \delta)}.
\end{equation}
Expression \eqref{x-funof-y} together with \eqref{sqrt-y} and \eqref{YYoverXX} 
substituted in \eqref{ClairautEq} yields
\begin{equation}\label{s-funof-y}
\sigma_2=-\frac{v}{4u}+\frac{(v^2-u^2)(\alpha y+\beta)}{4uY_\pm}
\mp \frac{(4u^2-1)(\gamma y+\delta)}{16(\alpha y+\beta)}
\mp \frac{(4v^2-1)(\alpha y+\beta)}{16(\gamma y+\delta)}.
\end{equation}
In these expressions the $\pm$-signs corresponds to the functions $f_\pm(\sigma_2')$.
We also note that $x-1$ has a factorized form as well:
\begin{equation}\label{x-1-funof-y}
x-1=\frac{X_\mp^2\big(X_\pm+2Y_\pm\big)\big(X_\pm-2Y_\pm\big)}
{16 Y_\pm^2 (\alpha y +\beta)(\gamma y + \delta)}.
\end{equation}
This is a remarkable property of the parametrization \eqref{sqrt-y} because 
we need to integrate the function $\sigma_2$ to obtain the corresponding
function $f_2$.

Indeed, according to \eqref{P=intsigma}, we have 
\begin{equation}\label{f2intdy}
f_2 =\int \left(\sigma_2+\wt A_2 x -\wt B_2\right)
\frac{\rmd x}{x(x-1)}  +\wt C_2,
\end{equation}
where (see \eqref{wtAB}, \eqref{wtBnu12N}, and \eqref{v-u12})
\begin{equation}
\wt A_2=\frac{1}{4},\qquad \wt B_2=\frac{vu-u-v}{2}+\frac{3}{8}.
\end{equation}
From expressions \eqref{x-funof-y}, \eqref{s-funof-y}, and \eqref{x-1-funof-y}
it is clear that if we change of the integration variable $x\mapsto y$ and take into 
account that $\rmd x =(\partial_y x) \rmd y$, then
the integrand in \eqref{f2intdy} appears to be an algebraic function 
of $y$. Hence, we can compute $f_2$ explicitly in terms of $y$. 

To construct the function $f_2$ as a piece-wise continuous function 
one has to take into account its values at the points $x=\infty,1,0$, 
for generic values of $p$ and $q$. 
We recall that at the point $x=\infty$ 
the function $f_2$ vanishes, see \eqref{f2atinfty}.
The value $f_2(1)$ can be found from \eqref{PNMLat1} and \eqref{largeBarnes},
\begin{multline}\label{f2at1pq}
f_2(1)=\frac{1}{2}\Big\{p^2\log p+q^2\log q+(p+q+1)^2\log(p+q+1)
\\
-(p+1)^2\log(p+1)-(q+1)^2\log(q+1)-(p+q)^2\log(p+q)
\Big\}.
\end{multline}
Concerning the point $x=0$, from Prop.~\ref{pr:xto0} one can find
\begin{multline}\label{f2at0pq}
\big[\min(p,q) \log x + f_2(x)\big]\big|_{x=0}=
\frac{(p-q)^2}{2}\log|p-q|-\frac{(|p-q|+1)^2}{2}\log(|p-q|+1)
\\
+\frac{\sgn(p-q)}{2}\Big\{(p+1)^2\log(p+1)-p^2\log p
\\
-(q+1)^2\log(q+1)+q^2\log q\Big\},
\end{multline}
where formulas \eqref{PLabcGs} and \eqref{largeBarnes} have been used.

\subsection{Construction of the leading term}

We start with listing 
properties of the function $\sigma_2$ near the points $x=\infty,1,0$
assuming that $p\ne q$. 

Behavior of the $\sigma$-function at the point $x=\infty$ 
is established in Cor.~\ref{cor:xtoinfty}. 
From \eqref{sigmatoinf} in the parameterization \eqref{v-u12} we have 
\begin{equation}\label{gotoinf}
\sigma_2= -\frac{x}{4} - \frac{uv}{2}+\frac{1}{8}+ O\left(x^{-2}\right),
\qquad x\to\infty.
\end{equation}
The case of the point $x=1$ is considered in  
Cor.~\ref{cor:xto1}. From \eqref{sigmatoone} it follows that
\begin{multline}\label{gotoone}
\sigma_2 = 
\frac{vu-v-u}{2}+\frac{1}{8}-\frac{4vu-v-u}{4(v+u-1)} (x-1)
\\
+\frac{\left(v-\frac{1}{2}\right)^2\left(u-\frac{1}{2}\right)^{2} (v+u)}
{(v+u-1)^4} (x-1)^2 +O\left( (x-1)^{3} \right),\qquad x\to 1.
\end{multline}
The case of the point $x=0$ is considered in Cor.~\ref{cor:xto0}.
From \eqref{sigmatozero} it follows that
\begin{multline}\label{gotozero}
\sigma_2=\frac{vu-|v-u|}{2}-\frac{1}{8} -\frac{4vu-|v-u|}{4(|v-u|+1)} x 
\\ 
+\frac{\left(4vu-2|v-u|+1\right)^2 |v-u|}{16(|v-u|+1)^4} x^2
+ O\left(x^3\right),  \qquad x\to 0.
\end{multline}
We now construct the function $\sigma_2$ satisfying all these properties.

The linear growth at infinity of the function $\sigma_2$ and the absence of an 
$1/x$ term in \eqref{gotoinf} 
imply that it is given by the general solution \eqref{CgenSol} where we have to 
choose $C=-1/4$. Denoting this function by $\sigma_2^\mathrm{I}$, we conclude that 
the function $\sigma_2$ for sufficiently large values of $x$ is given by 
$\sigma_2^\mathrm{I}$, which reads
\begin{equation}\label{sigmaIpq}
\sigma_2^\mathrm{I}= -\frac{x}{4} - \frac{uv}{2}+\frac{1}{8}.
\end{equation}
More exactly, we have $\sigma_2=\sigma_2^\mathrm{I}$ for $x\in [\xc,\infty)$, where 
the value of the critical point $\xc$ needs to be determined. The interval 
$[\xc,\infty)$ corresponds to Regime I. 

Let us now consider the case of the vicinity of the point $x=1$.
Expression \eqref{gotoone} imply a non-linear behavior of the function
$\sigma_2$ near $x=1$ and hence we have to search this function among the
singular solutions of the Clairaut equations \eqref{ClairautEq} where the
functions $f_{\pm}(\sigma_2')$ are given by \eqref{sigma2eq34}. We first identify which
equation, with $f_{+}(\sigma_2')$ or $f_{-}(\sigma_2')$, may possess the required
asymptotics \eqref{gotoone}. Substituting the values of $\sigma_2$ and
$\sigma_2'$ in \eqref{ClairautEq} at $x=1$ we find that it is the equation with 
the function $f_{-}(\sigma_2')$.

Next we pass to the solution in the parametric form defined by the
relation \eqref{sqrt-y}, which we choose in the following form:
\begin{equation}
\sqrt{\frac{\sigma_2' + u^2}{\sigma_2' + v^2} }=\frac{2uy+u-v}{2vy+ v-u}.
\end{equation}
Hence, 
\begin{equation}\label{sigma2prime}
\sigma_2' = \frac{(u-v)^2-4uvy}{4y(1+y)}
\end{equation}
and therefore
\begin{equation}\label{x(eta)}
x = \frac{(y+1)^2 (y+u-v)(y+v-u)}{(2vy+v-u)(2uy+u-v)}.
\end{equation}
To indicate that in the vicinity of the point  
$x=1$ the function $\sigma_2$ belongs to Regime II, 
we denote it $\sigma_2^\mathrm{II}$. We have
\begin{equation}\label{sigma2(eta)}
\sigma_2^\mathrm{II} = \frac{vu}{2}  - \frac{v^2+u^2}{16vu}-\frac{1}{4}
-\frac{y}{2}
+\frac{\left(4v^2-1\right)\left(u^2-v^2\right)}{16v(2vy+v-u)}
+\frac{\left(4u^2-1\right)\left(v^2-u^2\right)}{16u(2uy+u-v)}.
\end{equation}
Now the crucial step in the whole procedure is to identify which one of the
four roots of the quartic equation \eqref{x(eta)} corresponds to the
asymptotic expansion \eqref{gotoone}. 

To do this, we use \eqref{x-1-funof-y} to obtain
\begin{equation}\label{x-1=yyyy}
x-1= \frac{y^2 (y-v-u+1)(y+v+u+1)}{(2vy+v-u)(2uy+u-v)}.
\end{equation}
Computing the values of the expression in \eqref{sigma2(eta)} at 
$y=-v-u-1,0,v+u-1$ we find that the required value $\sigma_2(1)=(uv-u-v+4)/2$, see
\eqref{gotoone}, is attained at $y=v+u-1$. Furthermore, 
representing $y$ near this value as  
\begin{equation}\label{yatx=1}
y=v+u-1 + \gamma_1 \eps+ \gamma_2\eps^2+O(\eps^3)
\end{equation}
and choosing the coefficients $\gamma_1$ and $\gamma_2$ 
such that \eqref{x-1=yyyy} becomes 
\begin{equation}
x-1=\eps+O(\eps^3), 
\end{equation} 
we find 
\begin{align}
\gamma_1&=\frac{\left(2v-1\right) \left(2u-1\right)(v+u) }{2(v+u-1)^2},
\\
\gamma_2&= \frac{\left(2v-1\right) \left(2u-1\right)(v+u) 
\left(1+6(v^2+u^2)-(4uv+3)(v+u)\right)}{8(v+u-1)^5}.
\end{align}
Clearly, expression \eqref{yatx=1} provides an expansion of 
the function $y=y(x)$ near the point $x=1$. This function is uniquely defined as the root
of the quartic equation \eqref{x(eta)} by its value $y(1)=u+v-1$. 
Substitution of \eqref{yatx=1} into \eqref{sigma2(eta)} exactly reproduces
the required expansion \eqref{gotoone}. This means that we have constructed the 
function $\sigma_2^\mathrm{II}$ which corresponds to the function 
$\sigma_2$ in Regime II. 

Let us now address the position of the critical value of $x=\xc$ at which 
Regime I changes into Regime II. At the moment, we do this under the assumption that 
the function $\sigma_2$ is continuous at this point,       
\begin{equation}
\sigma_2^\mathrm{I}(\xc)=\sigma_2^\mathrm{II}(\xc).
\end{equation}
Below we lift this assumption and derive the result for $\xc$ from the
similar equation for the function $f_2$, that proves absence of 
first-order phase transitions. 
Denoting the corresponding value of $y$ at the critical point by 
$\yc\equiv y(\xc)$ we find from \eqref{sigmaIpq}, \eqref{sigma2(eta)}, and 
\eqref{x(eta)} that $\yc$ is one of the two roots of the equation    
\begin{equation}\label{yceq}
y^2+(1-4vu)y+(v-u)^2=0.
\end{equation}
Choosing the root which lies on the right from the value 
$y(1)=u+v-1$ (note that $\gamma_1>0$ in \eqref{yatx=1}, so $y(x)$ is expected 
to be an increasing function), we get 
\begin{equation}\label{yc}
\yc=\frac{4vu-1+\sqrt{(4v^2-1)(4u^2-1)}}{2}. 
\end{equation}
The corresponding value of $\xc$ is
\begin{align}\label{xcsqrt}
\xc &= \frac{4vu+1+\sqrt{(4v^2-1)(4u^2-1)}}{2}
\\ &
=\frac{1}{4}\big(\sqrt{(2v-1)(2u-1)} +\sqrt{(2v+1)(2u+1)}\big)^2.
\end{align}
Thus, for the values $x\in[1,\xc]$ the function $y(x)$ monotonously increases from 
the value $u+v-1$ up to the value $\yc$, given in \eqref{yc}. Note also that 
$\xc \to 4 w^2$ as $v,u\to w$, in agreement with the symmetric case. 

Let us now consider the values of $y$ on the left from the point $x=1$. As it follows from 
\eqref{x(eta)}, as $y$ decreases from the value $u+v-1$ down to the value 
$|u-v|$, the variable $x$ runs its values from $1$ to $0$. We thus conclude that our 
function $y$ is the monotonous bijective map on the interval $[0,\xc]$: 
\begin{equation}
y(x): \left[0, \xc \right] \mapsto [|v-u|,\yc],
\end{equation}
where $y(0)=|v-u|$ and $y(\xc)=\yc$. Furthermore, since 
$x\to\infty$ as $y\to\infty$ in \eqref{x(eta)}, this map extends to the 
whole domain $x\in[0,\infty)$ and it corresponds to $y\in [|v-u|,\infty)$.

The obtained property of the function $y$ makes it possible 
to study the function $\sigma_2^\mathrm{II}$ near the point $x=0$. 
Essentially similarly, as we have found the expansion near the point $x=1$ above, 
we find that, as $x\to0$, 
\begin{multline}
y = |v-u|+ 
\frac{|v-u|(4vu-2|v-u|+1)}{2(|v-u|+1)^2} x
\\
+ \frac{|v-u|(4 uv-2|v-u|+1)\big(1+ 6(v^2 + u^2)-(4vu-3)|v-u|\big)}
{8(|p-q|+1)^5}x^2+O(x^3).
\end{multline}
Substituting this expansion into \eqref{sigma2(eta)}, we immediately 
arrive at the expression given in \eqref{gotozero}. This means that 
our solution obtained for Regime II also satisfies the conditions near the point 
$x=0$ and no analogue of Regime III arises in the non-symmetric case, $p\ne q$ 
(or $u\ne v$). 

Now turning to the function $f_2$, we conclude that 
all these considerations imply that  
\begin{equation}\label{f2pq}
f_2=
\begin{cases}
f_2^\mathrm{I} &x\in[\xc,\infty)
\\
f_2^\mathrm{II} &x\in[0,\xc]. 
\end{cases}
\end{equation}
As for the function $f_2^\mathrm{I}$, by substituting \eqref{f2intdy} into  
\eqref{sigmaIpq} and fixing the constant of integration to match the condition 
\eqref{f2atinfty}, we obtain
\begin{equation}\label{f2Ipq}
f_2^\mathrm{I}=\frac{(2v-1)(2u-1)}{4}\log \frac{x}{x-1}.
\end{equation}  
As for the function $f_2^\mathrm{II}$, we make the change of the integration 
variable $x\mapsto y$ in \eqref{f2intdy}, where, due to \eqref{sigma2(eta)},
we have
\begin{equation}\label{rmdx}
\frac{\rmd x}{x} =\left\{\frac{2}{y+1}+\frac{1}{y+u-v}+\frac{1}{y+v-u}
-\frac{2v}{2vy+v-u}-\frac{2u}{2uy+u-v}\right\}\rmd y.
\end{equation}
Taking into account \eqref{x-1=yyyy}, from \eqref{x(eta)} we obtain  
\begin{multline}\label{f2IIpq}
f_2^\mathrm{II}=
-\frac{(u+v-1)^2}{2}\log y
-\frac{(u-v)^2+2v+2u-1}{2}\log (y+1)
\\
+\frac{4 u^2-1}{4} \log (2uy+u-v)
+\frac{4 v^2-1}{4}  \log (2vy+v-u)
\\
-\frac{2u-1}{2}\log(y+u-v)
-\frac{2v-1}{2}\log(y+v-u)
\\
+ (u+v)\log(y+u+v+1)+\wt C_2.
\end{multline}
The constant of integration $\wt C_2$ can be fixed by imposing the 
condition \eqref{f2at1pq}. Using that $y(1)=u+v-1$ from 
\eqref{f2IIpq} we obtain
\begin{multline}
f_2^\mathrm{II}(1)=
-\frac{(u+v-1)^2}{2}\log (u+v-1)
+\frac{(u+v)^2}{2}\log (u+v)
\\
+\frac{(2u-1)^2}{4}\log (2u-1)
+\frac{(2v-1)^2}{4}\log (2v-1)
+(u+v)\log2 +\wt C_2,
\end{multline}
and a comparison (recall that $v=p+1/2$ and $u=q+1/2$) with \eqref{f2at1pq} yields 
\begin{multline}\label{wtC2pq}
\wt C_2=-\frac{(2v+1)^2}{8}\log (2v+1)-\frac{(2v-1)^2}{8}\log (2v-1)
\\
-\frac{(2u+1)^2}{8}\log (2u+1)-\frac{(2u-1)^2}{8}\log (2u-1).
\end{multline}{}
This fixes the function $f_2^\mathrm{II}$. 

Using \eqref{f2Ipq}, \eqref{f2IIpq}, and \eqref{wtC2pq}, one can  
now show directly that \eqref{f2pq} indeed holds. Namely, we prove 
that there exists one and only one solution $\xc\in (1,\infty)$ of the 
equation $f_2^\mathrm{II}(\xc)=f_2^\mathrm{I}(\xc)$.
Calculation goes along the same lines 
as in the symmetric case (see end of the Sect.~4.2). Introduce the function 
\begin{equation}
\rho(x)= f_2^\mathrm{II}(x)-f_2^\mathrm{I}(x).
\end{equation}
Substituting  \eqref{x(eta)} and \eqref{x-1=yyyy} into \eqref{f2Ipq}, 
we get 
\begin{multline}\label{rhopq}
\rho(x)=
\left(\frac{v^2}{2}+\frac{1}{8}\right) 
\log \frac{(2vy+v-u)^2}{\left(4v^2-1\right) y (y+1)}
+\left(\frac{u^2}{2}+\frac{1}{8}\right) 
\log\frac{(2uy+u-v)^2}{\left(4 u^2-1\right)y (y+1)}
\\
+\left(u v+\frac{1}{4}\right) 
\log\frac{y (y+u+v+1) (y-v-u+1)}{(y+1) (y+u-v) (y+v-u)}
\\
+\frac{1}{2} v 
\log\frac{(2v-1)(y+u-v)(y+v+u+1)}{(2v+1)(y+v-u)(y-v-u+1)}
\\
+\frac{1}{2} u \log\frac{(2u-1)(y+v-u)(y+v+u+1)}{(2u+1)(y+u-v)(y-v-u+1)}
\\
+\frac{1}{2} 
\log\frac{(y+u-v) (y+v-u) (y+1)}{(2 u y+u-v) (2vy+v-u)}.
\end{multline}
It is not difficult to see, that all the six logarithms in \eqref{rhopq} 
vanish as soon as \eqref{yceq} holds, that leads us to \eqref{xcsqrt}. Hence,  
our result for $\xc$ obtained above under the assumption of absence of the first order 
transition is recovered. 
Let us now show that there are no other roots of the equation $\rho(x)=0$ 
on the interval $(1,\infty)$. Evaluating the derivative of the function $\rho(x)$, from 
\eqref{rhopq} and \eqref{x(eta)} we get
\begin{align}
\rho'(x)&
=\left(\frac{\partial x(y)}{\partial y}\right)^{-1}\frac{\partial \rho(x(y))}{\partial y}
\notag\\  &
=\frac{(2vy+v-u)(2uy+u-v)\left[y^2+(1-4vu)y+(v-u)^2\right]^2}
{4y^2(y+1)^2(y+v+u+1)(y-v-u+1)(y+v-u)(y+u-v)},
\end{align}
and, again using \eqref{x(eta)}, 
\begin{equation}\label{rhoprime}
\rho'(x)=\frac{\left[y^2+(1-4vu)y+(v-u)^2\right]^2}
{4x y^2(y+v+u+1)(y-v-u+1)}.
\end{equation}
Recalling that $x\to 1$ as $y\to v+u-1$ (see \eqref{x-1=yyyy}) and
$x\to \infty$ as $y\to \infty$ (see \eqref{x(eta)}), we conclude from 
\eqref{rhoprime} that the function $\rho(x)$ is an increasing function 
on the interval $(1,\infty)$ except the point $x=\xc$ where it has 
a simple zero, and where its first and second derivatives vanish, but the third one 
does not. This means that, besides that the point $x=\xc$ is the only possible 
point of the phase transition, our system undergoes a third-order phase transition
at this point.

It can also be directly checked that the resulting 
function $f_2$, given by \eqref{f2pq}, has the expected $x\to 0$ behavior
as prescribed by \eqref{f2at0pq}.

\subsection{Sub-leading corrections}

Now we address calculation of the corrections to the leading term. 
Recall that we fix our parameters such that $\nu_1$ and $\nu_2$ has no 
$O(1)$ terms as $N$ is large, 
\begin{equation}
\nu_1=v N,\qquad 
\nu_2=-u N, 
\end{equation}
and $v$ and $u$ are related to the parameters $p$ and $q$ 
defined in \eqref{def-p-q} by $v=p+1/2$ and $u=q+1/2$. 

Just like in the symmetric case, the $1/N$ expansion \eqref{sigmaNlargeexp} 
in the non-symmetric case can be constructed by using a relevant equation 
among the two in \eqref{splitPVI}. Again, such an equation can be easily 
identified provided the leading term, the function $\sigma_2$, is given by a singular 
solution of the Clairaut equation. We meet such a situation in our problem 
in Regime II, and it is the equation containing the function $F_(\sigma',\sigma'')$. 
Calculation \eqref{FmtoO(N)} can be repeated without modifications for the non-symmetric case, 
again providing the result $\sigma_1=0$, see \eqref{sigma1=0}. In fact, 
all the terms of odd powers in $1/N$ also vanish, and  
\eqref{sigmaNlargeexp} is an expansion in $1/N^2$ in Regime II.  

In Regime I the situation in the non-symmetric case repeats that in the symmetric case, 
since the leading term $\sigma_2$ is given by a regular solution of the Clairaut equation. 
This term solves both equations in \eqref{splitPVI} in the leading order, 
since $f_{+}(\sigma_2)=f_{-}(\sigma_2)$.
To identify which one of the two equations in \eqref{splitPVI} is actually 
responsible for the $1/N$ expansion, we should search for a suitable solution for the function 
$\sigma_1$. In total, the recipe of derivation of sub-leading corrections 
in the non-symmetric case goes along the same lines as in the symmetric one. 

We turn to the details of calculation for Regime I and Regime II separately.

\subsubsection{Regime I} 

We start with $x\to\infty$
asymptotic expansion \eqref{s1xtoinf} yielding
\begin{equation}\label{s1xtoinf-pq}
\sigma_1= -\frac{x}{2}+ u v+\frac{1}{4}+\frac{(4v^2-1)(4u^2-1)}{16x}
+ O\left(x^{-2}\right),\qquad x\to\infty.
\end{equation}

We recall that in this regime $\sigma_2'=-1/4$ and so both 
equations in \eqref{splitPVI} vanish in the leading order.  
To identify which one of the two equations in \eqref{splitPVI} is responsible to 
$1/N$ expansion and find the functions $\sigma_1$ and $\sigma_0$, 
we expand the functions $F_\pm(\sigma',\sigma'')$ to order $N^0$:
\begin{multline}
F_\pm(\sigma',\sigma'')
=N^2\left(\frac{1}{8}-\frac{vu}{2}\right)
\\
+N\left(-\frac{\sigma_1'}{2}-2vu\sigma_1'
\mp 2\sqrt{\left(v^2-\frac{1}{4}\right)\left(u^2-\frac{1}{4}\right)
\left((\sigma_1')^2-\frac{1}{4}\right)}\right)
\\
+\left(-\frac{1}{2}-2uv
\mp 2\sigma_1'
\frac{\sqrt{\left(v^2-\frac{1}{4}\right)\left(u^2-\frac{1}{4}\right)}}{
\sqrt{(\sigma_1')^2-\frac{1}{4}}}
\right)\sigma_0'
-8vu (\sigma_1')^2+\frac{vu}{2}
\\ 
\mp
\left[
\frac{\left(8v^2u^2-v^2-u^2\right)\sqrt{(\sigma_1')^2-\frac{1}{4}}}{
\sqrt{\left(v^2-\frac{1}{4}\right)\left(u^2-\frac{1}{4}\right)}}
+ 
\frac{\sqrt{\left(v^2-\frac{1}{4}\right)\left(u^2-\frac{1}{4}\right)}}{
2\sqrt{(\sigma_1')^2-\frac{1}{4}}}
\right]
\sigma_1'
+O(N^{-1}).
\end{multline}
Hence,  $\sigma_1$ must be a solution of one the following two equations:
\begin{equation}\label{s1Clairaut-pq}
\sigma_1= x\sigma_1'
-\frac{\sigma_1'}{2} - 2 v u\sigma_1' 
\mp 2\sqrt{\left(v^2-\frac{1}{4}\right)\left(u^2-\frac{1}{4}\right)
\left((\sigma_1')^2-\frac{1}{4}\right)}.
\end{equation}
These are the Clairaut equations and we have to search for a  
singular solution that matches the asymptotic expansion \eqref{s1xtoinf-pq}. 
The proper solution reads
\begin{equation}\label{sigma1-pq}
\sigma_1 =
-\frac{1}{2}\sqrt{s(x)},\qquad s(x) = x^2-(1+4vu)x+(v+u)^2.
\end{equation}
Note that the critical value $x=\xc$ given by \eqref{xcsqrt} is one of the roots 
of the polynomial $s(x)$, the second root is always smaller than $\xc$.  
The solution \eqref{sigma1-pq} corresponds to the plus sign in \eqref{s1Clairaut-pq} and 
hence the equation in \eqref{splitPVI} relevant for the $1/N$ expansion   
is the one involving the function $F_{-}(\sigma',\sigma'')$. 

The function $\sigma_0$ can be directly computed. Since $\sigma_1$ is given by 
a singular solution, the $\sigma_0'$ term vanishes in the equation in the order $N^0$, that 
yields 
\begin{equation}
\sigma_0 =-8vu (\sigma_1')^2
+\frac{vu}{2}+
\frac{\left(8v^2u^2-v^2-u^2\right)\sigma_1'\sqrt{(\sigma_1')^2-\frac{1}{4}}}{
\sqrt{\left(v^2-\frac{1}{4}\right)\left(u^2-\frac{1}{4}\right)}}
+\frac{\sqrt{\left(v^2-\frac{1}{4}\right)\left(u^2-\frac{1}{4}\right)}}{
2\sqrt{(\sigma_1')^2-\frac{1}{4}}}\sigma_1'.
\end{equation}
Substituting \eqref{sigma1-pq}, we get 
\begin{equation}
\sigma_0 =-\frac{x}{4}+\frac{1}{8}-
\frac{4vu x^2 -2(v+u)^2x+(v+u)^2}{4s(x)}.
\end{equation}

Let us now consider the function $\log P_{N,M,L}(x^{-1})$. 
For this quantity we have the expansion \eqref{logPN2N1N0}. 
Note that all the terms must vanish at infinity, $\lim_{x\to\infty}f_1=0$, 
$\lim_{x\to\infty} f_0=0$, etc. Computing $f_1$ by \eqref{intf1} 
with $\wt A_1=\wt B_1=1/2$ and choosing $\wt C_1$ such that $f_1(\infty)=0$, we get
\begin{multline}
f_1= 
v\log\left(\frac{2vx-v-u+\sqrt{s(x)}}{(2v+1)\sqrt{x(x-1)}}\right)
+u\log\left(\frac{2ux-v-u+\sqrt{s(x)}}{(2u+1)\sqrt{x(x-1)}}\right)
\\ 
-\frac{1}{2}\log\frac{2x-4uv-1+2\sqrt{s(x)}}{4 x}.
\end{multline}
Essentially similarly, for $f_0$, using \eqref{intf0}  
with $\wt A_0=1/4$ and $\wt B_0=1/8$ and 
choosing $\wt C_0$ such that $f_0(\infty)=0$, we obtain 
\begin{equation}
f_0= \frac{1}{4} \log\frac{x(x-1)}{s(x)}.
\end{equation}
Finally, rewriting these formulas in terms of $p$ and $q$
we arrive at the expressions for the functions $f_1^\mathrm{I}$ and $f_0^\mathrm{I}$ 
appearing in Thm.~\ref{th:rectangle}.

\subsubsection{Regime II}

We recall that here we have $\sigma_1=0$. The function 
$\sigma_0$ can be found by expanding the function $F_{-}(\sigma',\sigma'')$
to order $N^0$, that yields 
\begin{equation}
\sigma_0=
-\frac{vu}{8\sigma_2'}
+\frac{\big(\sigma_2'-\frac{1}{4}\big)\sqrt{\big(\sigma_2'+v^2\big)\big(\sigma_2'+u^2\big)}}
{8\sigma_2'\big(\sigma_2'+\frac{1}{4}\big)}
-\frac{[x(x-1)\sigma_2'']^2}{4\big(\sigma_2'+\frac{1}{4}\big)
\sqrt{\big(\sigma_2'+v^2\big)\big(\sigma_2'+u^2\big)}}.
\end{equation}
To obtain an explicit expression for $\sigma_0$ in terms of 
$y=y(x)$, one can use the expression \eqref{sigma2prime} for $\sigma_2'$. 
Taking into account (see \eqref{rmdx}) that 
\begin{equation}
\partial_yx=\frac{2y(y+1)\Big(4vuy^3-(v-u)^2\left[3y^2+3y-(v+u)^2+1\right]\Big)}
{(2uy+u-v)^2(2vy+v-u)^2}
\end{equation}
one can also express $\sigma_2''$ in terms of $y$ and hence find $\sigma_0$ 
as a function of $y$.  

Let us now consider the function $\log P_{N,M,L}(x^{-1})$. As in the symmetric case, for 
$\log P_{N,M,L}(1)$ we have the expansion \eqref{PNMLat1largeN},
where $f_2(1)$ is given by \eqref{f2at1pq}, and the values $f_1(1)$ and
$f_0(1)$ are given by 
\begin{equation}\label{f1at1pq}
f_1(1)=-\frac{1}{2}\Big\{(p+1)\log(p+1)+(q+1)\log(q+1)-p\log p-q\log q\Big\},
\end{equation}
and
\begin{equation}\label{f0at1pq}
f_0(1)=
-\frac{1}{24}\log\frac{(p+1)(q+1)}{pq}
-\frac{1}{12}\log\frac{p+q+1}{p+q}
+\zeta'(-1)+\log \sqrt{2\pi},
\end{equation}
respectively. These values can be used to construct the expansion \eqref{logPRegII} 
where the functions $f_1$ and $f_0$ can be found from 
$\sigma_1$ and $\sigma_0$ by the usual formulas \eqref{intf1} and \eqref{intf0}.
As for the function $f_1$, since $\sigma_1=0$, we just repeat the calculation from the 
symmetric case, now using \eqref{f1at1pq} to fix the integration constant, that yields
\begin{equation}
f_1=\frac{1}{2}\left\{\log x+p\log p-(p+1)\log(p+1)+q\log q-(q+1)\log(q+1)\right\}. 
\end{equation}
As for function $f_0$, one can perform 
the integration \eqref{intf0} by changing the integration 
variable $x\mapsto y$, similarly to the case of the function $f_2$. 
The calculation of $f_0$ appears to be notably involved, 
the final result reads
\begin{multline}
f_0=\frac{1}{8}\bigg\{\log y+\log(y+1)-2\log(2vy+v-u)
\\
-2\log(2uy+u-v)
+3\log\left(y^2+(1-4vu)y+(v-u)^2\right)
\\
+\frac{1}{3}\log\Big(4vuy^3-(v-u)^2\left[3y^2+3y-(v+u)^2+1\right]\Big)
\bigg\}+\wt C_0.  
\end{multline}
The constant of integration can be fixed by computing the value
at $x=1$, or $y=u+v-1$, and we get
\begin{equation}
f_0\big|_{y=u+v-1}=\frac{1}{12}\log\frac{(2v-1)(2u-1)(v+u-1)}{v+u}+
\wt C_0.
\end{equation}
Comparison with \eqref{f0at1pq} gives
\begin{equation}
\wt C_0=-\frac{1}{24}\log \big(16p(1+p)q(1+q)\big)+\zeta'(-1)
+\log\sqrt{2\pi}.
\end{equation}
In total, we arrive at the functions $f_1^\mathrm{II}$ and $f_0^\mathrm{II}$ appearing 
in Thm.~\ref{th:rectangle}, which is now finally proven.

\section{Conclusion}\label{sec:Conc}

In this paper, we have studied the five-vertex model on a rectangular domain
with scalar-product boundary conditions. Relying on the connection between the
partition function of the model and the sixth Painlev\'e equation, we have
derived the expansion of the free energy in the limit where the size of the
domain tends to infinity. The key advantage of this approach lies in its
capability to provide not only the leading term of such an expansion, but
also sub-leading corrections. All terms of the expansion can be computed recursively. 
Here, we limit ourselves by explicit expressions 
to the order of a constant (see Thms.~\ref{th:square} and \ref{th:rectangle}).

\begin{figure}
\begin{subfigure}{.49\textwidth}
\centering
\includegraphics[width =\linewidth]{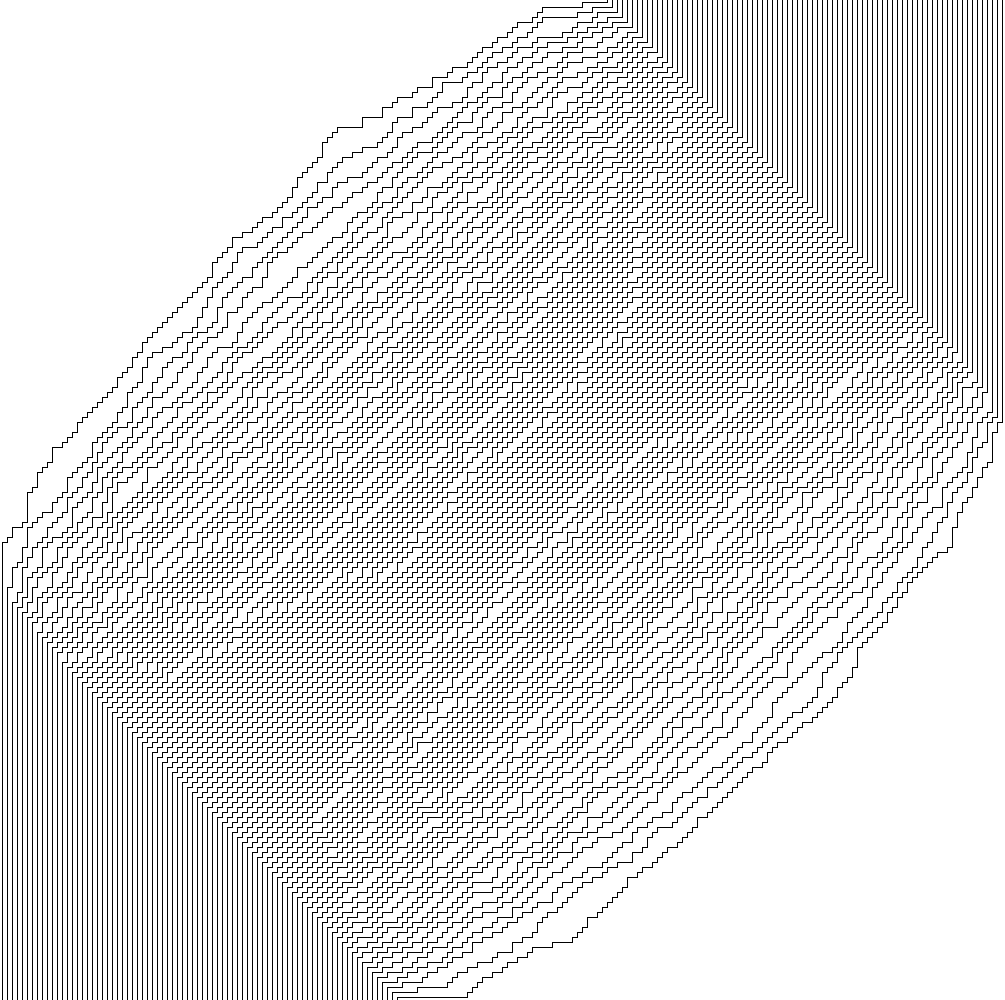}
\end{subfigure}
\begin{subfigure}{.49\textwidth}
\centering
\includegraphics[width =\linewidth]{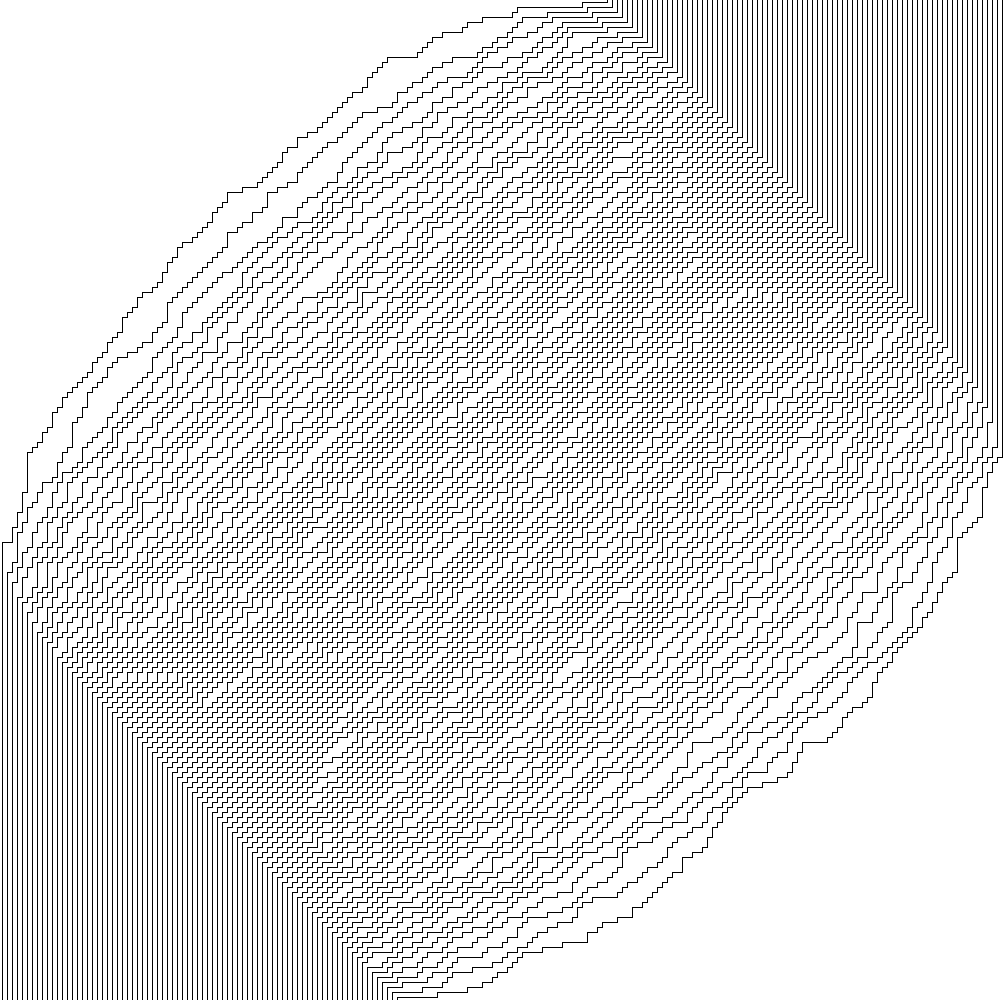}
\end{subfigure}
\caption{Configurations of the five-vertex model at $\sqrt{x}=0.24$ (left) 
and $\sqrt{x}=0.3$ (right) on a `square' domain with $N=80$, $M=200$, and $L=201$.
On the left picture the two disordered regions are separated by a region 
of the anti-ferroelectric order. On the right picture 
these two disordered regions merge with each other.}
\label{fig:square_numerics}
\end{figure}

\begin{figure}
\includegraphics[width=.6\linewidth]{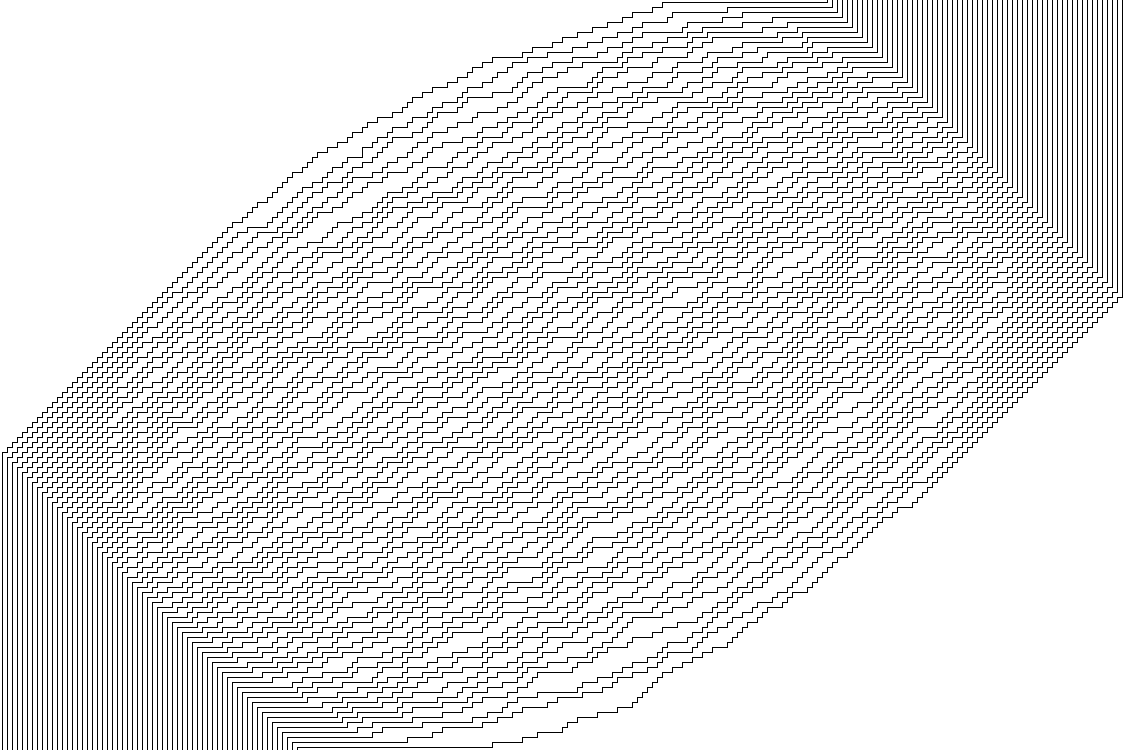}
\caption{A configuration of the five-vertex model at $\sqrt{x}=0.01$ on a 
rectangular domain with $N=60$, $M=150$, and $L=225$. The disordered region 
does not split, and there is no region of the anti-ferroelectric order.}
\label{fig:rec_numerics}
\end{figure}

Our results reveal an interesting feature: in the case of a rectangular
domain, there is no Regime III, in other words, one phase transition
disappears. To gain a better understanding of this phenomenon, we have
generated several configurations of the model numerically. In order to ensure
sampling from the correct probability distribution, we resorted to the
Coupling From the Past Algorithm \cite{PW-96,PW-98}. For the simulation we
use $w_1=w_3=w_4=1$ and $w_5=w_6=1/\sqrt{x}$ (see the comment at the end 
of Sect.~\ref{sec:MD} and the discussion in Sect.~\ref{sec:FV}). 
Examples of configurations are presented in Fig.~\ref{fig:square_numerics}
(square domain) and Fig.~\ref{fig:rec_numerics} (rectangular domain).
Numerical simulation clearly shows the distinction in the behavior of the
model for the square and rectangular domains. 

Specifically, for a square
domain (where $\epsilon=M-L+1$ is of $O(1)$ as $N,M,L\to\infty$) 
if $x>\xc^{-1}$, then there exists a single disordered region. 
For $x=1$, which is known as the free-fermion point of the 
five-vertex model, the configurations are described by random boxed 
plane partitions and the disordered region has a form of the ellipse \cite{CLP-98}. 
In numerical simulations,  
we have been interested in the values of $x$ slightly above and 
below the critical value $x=\xc^{-1}$ separating Regime II and Regime III, 
according to Thm.~\ref{th:square}. Pictures of Fig.~\ref{fig:square_numerics} 
show typical configurations for a `square' domain with $N=80$, $M=200$, and $L=201$. 
In this geometry the phase transition
between Regime III and Regime II occurs at $1/\sqrt{\xc}\approx 0.25$. 
On the left picture where $\sqrt{x}=0.24$, 
which corresponds to the Regime III, one can see two disordered
regions separated by a region with the anti-ferroelectric order. 
On the right picture $\sqrt{x}={0.3}$, which corresponds to the Regime II, 
these two disordered regions merge with each other and 
we observe just a single disordered region. 

Based on this simple illustration one can conclude, that the phase transition
occurring at $x=\xc^{-1}$ corresponds to the split of the disordered region
into two distinct parts and hence it resembles the ``merger transition''
studied in \cite{PGA-22}. This interpretation also aligns perfectly with the
results of \cite{GKW-21}. It is useful also to mention that as $x\to 0$ both 
disordered regions shrink down and  
the model falls into the anti-ferroelectric ground state 
(see Fig.~\ref{fig-AFGroundState}). 

On the other hand, in the case of a rectangular domain, the disordered region
remains connected and the split does not occur. 
The picture of Fig.~\ref{fig:rec_numerics} shows an example of 
configuration for a very small value 
(as small as the algorithm has allowed us to produce the picture 
for a reasonable amount of computing time) of the parameter $\sqrt{x}=0.01$ for the 
rectangle domain with $N=60$, $M=150$, and $L=225$. The disordered region 
changes its shape from the ellipse (which occurs at $x=1$) 
but one can observe no signal of splitting it on two (or whatever) regions.  
Thus, one can conclude that the disordered region remains connected as $x$ decreases. 
On the contrary to the case of the square domain, there is no analogue 
of the anti-ferroelectric ground state, and as a consequence, the disordered 
region does not disappear as $x\to 0$ \cite{BCMP-23}. 

We end up by a brief discussion of the phase transition between 
Regime I and Regime II at $x=\xc$, 
which take place for both rectangular and square shaped domains. 
This transition can be characterized by disappearance of the disordered region 
in the center of the domain as the parameter $x$ increases from $x<\xc$ 
(Regime II) to $x>\xc$ (Regime I). It is clear, 
that the dominance of the $b$-weight vertices (see also discussion 
in Sect.~\ref{sec:FV}) that occurs for large $x$ cannot be spoiled by the 
geometry of the domain. One could expect to see this in 
numerical simulations, but unfortunately for large $x$, especially for
$x > \xc$, the methods such as used above for small $x$ are not able to produce 
a configuration for sufficiently large domains in a reasonable time. It
definitely deserves further study how various algorithms can be adapted to
produce meaningful pictures at large sizes of the domain.

\section*{Acknowledgements}

The authors thanks the anonymous referee for interesting and useful comments. 
This work has been supported by the Theoretical Physics and Mathematics
Advancement Foundation <<BASIS>>.

\bibliography{fvfren_bib.bib}

\begin{bibdiv}
\begin{biblist}

\bib{G-90}{article}{
      author={Garrod, C.},
       title={Stochastic models of crystal growth in two dimensions},
        date={1990},
     journal={Phys. Rev. A},
      volume={41},
       pages={4184\ndash 4194},
         doi={10.1103/PhysRevA.41.4184},
}

\bib{GLT-90}{article}{
      author={Garrod, C.},
      author={Levi, A.C.},
      author={Touzani, M.},
       title={Mapping of crystal growth onto the 6-vertex model},
        date={1990},
     journal={Solid State Comm.},
      volume={75},
       pages={375\ndash 382},
         doi={10.1016/0038-1098(90)90915-X},
}

\bib{GS-92}{article}{
      author={Gwa, L.-H.},
      author={Spohn, H.},
       title={Six-vertex model, roughened surfaces, and an asymmetric spin
  {H}amiltonian},
        date={1992},
     journal={Phys. Rev. Lett.},
      volume={68},
       pages={725\ndash 728},
         doi={10.1103/PhysRevLett.68.725},
}

\bib{GvBL-93}{article}{
      author={Gul\'acsi, M.},
      author={Beijeren, H.~Van},
      author={Levi, A.~C.},
       title={Phase diagram of the five-vertex model},
        date={1993},
     journal={Phys. Rev. E},
      volume={47},
       pages={2473\ndash 2483},
         doi={10.1103/PhysRevE.47.2473},
}

\bib{HWKK-96}{article}{
      author={Huang, H.~Y.},
      author={Wu, F.~Y.},
      author={Kunz, H.},
      author={Kim, D.},
       title={Interacting dimers on the honeycomb lattice: An exact solution of
  the five-vertex model},
        date={1996},
     journal={Physica A},
      volume={228},
       pages={1\ndash 32},
      eprint={cond-mat/9510161},
         doi={10.1016/S0378-4371(96)00057-X},
}

\bib{KZj-00}{article}{
      author={Korepin, V.~E.},
      author={Zinn-Justin, P.},
       title={{Thermodynamic limit of the six-vertex model with domain wall
  boundary conditions}},
        date={2000},
     journal={J. Phys. A: Math. Gen.},
      volume={33},
       pages={7053\ndash 7066},
      eprint={cond-mat/0004250},
         doi={10.1088/0305-4470/33/40/304},
}

\bib{Zj-00}{article}{
      author={Zinn-Justin, P.},
       title={Six-vertex model with domain wall boundary conditions and
  one-matrix model},
        date={2000},
     journal={Phys. Rev. E},
      volume={62},
       pages={3411\ndash 3418},
      eprint={math-ph/0005008},
         doi={10.1103/PhysRevE.62.3411},
}

\bib{BL-13}{book}{
      author={Bleher, P.},
      author={Liechty, K.},
       title={{Random Matrices and the Six-Vertex Model}},
      series={CRM Monograph Series},
   publisher={American Mathematical Society},
     address={Providence, RI},
        date={2013},
      volume={32},
}

\bib{KBI-93}{book}{
      author={Korepin, V.~E.},
      author={Bogoliubov, N.~M.},
      author={Izergin, A.~G.},
       title={{Quantum Inverse Scattering Method and Correlation Functions}},
   publisher={Cambridge University Press},
     address={Cambridge},
        date={1993},
}

\bib{B-10}{article}{
      author={Bogolyubov, N.~M.},
       title={Five-vertex model with fixed boundary conditions},
        date={2010},
     journal={St. Petersburg Math. J.},
      volume={21},
       pages={407\ndash 421},
         doi={10.1090/S1061-0022-10-01100-3},
}

\bib{MSa-13}{article}{
      author={Motegi, K.},
      author={Sakai, K.},
       title={{Vertex models, TASEP and Grothendieck polynomials}},
        date={2013},
     journal={J. Phys. A},
      volume={46},
       pages={355201},
      eprint={1305.3030},
         doi={10.1088/1751-8113/46/35/355201},
}

\bib{GKW-21}{article}{
      author={de~Gier, J.},
      author={Kenyon, R.},
      author={Watson, S.~S.},
       title={Limit shapes for the asymmetric five vertex model},
        date={2021},
     journal={Commun. Math. Phys.},
      volume={385},
       pages={793\ndash 836},
      eprint={1812.11934},
         doi={10.1007/s00220-021-04126-7},
}

\bib{KP-21}{article}{
      author={Kenyon, R.},
      author={Prause, I.},
       title={The genus-zero five-vertex model},
        date={2022},
     journal={Prob. Math. Phys.},
      volume={3},
       pages={707\ndash 729},
      eprint={2101.04195},
         doi={10.2140/pmp.2022.3.707},
}

\bib{KP-22}{article}{
      author={Kenyon, R.},
      author={Prause, I.},
       title={Gradient variational problem in {$\mathbb{R}^2$}},
        date={2022},
     journal={Duke Math. J.},
      volume={171},
       pages={3003\ndash 3022},
      eprint={2006.01219},
         doi={10.1215/00127094-2022-0036},
}

\bib{KP-24}{article}{
      author={Kenyon, R.},
      author={Prause, I.},
       title={Limit shapes from harmonicity: dominos and the five vertex
  model},
        date={2024},
     journal={J. Phys. A: Math. Theor.},
      volume={57},
       pages={035001},
      eprint={2310.06429},
         doi={10.1088/1751-8121/ad17d7},
}

\bib{NP-20}{article}{
      author={Nazarov, A.~A.},
      author={Paston, S.~A.},
       title={Finite-size correction to the scaling of free energy in the dimer
  model on a hexagonal domain},
        date={2020},
     journal={Theor. Math. Phys.},
      volume={205},
       pages={1473\ndash 1491},
      eprint={1812.10274},
         doi={10.1134/S0040577920110069},
}

\bib{BP-21}{article}{
      author={Burenev, I.~N.},
      author={Pronko, A.~G.},
       title={Determinant formulas for the five-vertex model},
        date={2021},
     journal={J. Phys. A: Math. Theor.},
      volume={54},
       pages={055008},
      eprint={2011.01972},
         doi={10.1088/1751-8121/abd785},
}

\bib{KP-16}{article}{
      author={Kitaev, A.~V.},
      author={Pronko, A.~G.},
       title={{Emptiness formation probability of the six-vertex model and the
  sixth Painlev\'e equation}},
        date={2016},
     journal={Comm. Math. Phys.},
      volume={345},
       pages={305\ndash 354},
      eprint={1505.00032},
         doi={10.1007/s00220-016-2636-5},
}

\bib{JM-81}{article}{
      author={Jimbo, M.},
      author={Miwa, T.},
       title={Monodromy preserving deformation of linear ordinary differential
  equations with rational coefficients. {II}},
        date={1981},
     journal={Physica D},
      volume={2},
       pages={407\ndash 448},
         doi={10.1016/0167-2789(81)90021-X},
}

\bib{O-87}{article}{
      author={Okamoto, K.},
       title={{Studies on the Painlev\'e equations. I. Sixth Painlev\'e
  Equation $P_\mathrm{VI}$}},
        date={1987},
     journal={Ann. Mat. Pura Appl.},
      volume={146},
       pages={337\ndash 381},
         doi={10.1007/BF01762370},
}

\bib{J-82}{article}{
      author={Jimbo, M.},
       title={Monodromy problem and the boundary condition for some
  {P}ainlev\'e equations},
        date={1982},
     journal={Publ. Res. Inst. Math. Sci.},
      volume={18},
       pages={1137\ndash 1161},
         doi={10.2977/PRIMS/1195183300},
}

\bib{IN-86}{book}{
      author={Its, A.~R.},
      author={Novokshenov, V.~Yu.},
       title={The isomonodromic deformation method in the theory of
  {P}ainlev\'e equations},
      series={Lecture Notes in Mathematics},
   publisher={Springer-Verlag, Berlin},
        date={1986},
      volume={1191},
}

\bib{DZh-93}{article}{
      author={Deift, P.},
      author={Zhou, X.},
       title={A steepest descent method for oscillatory {R}iemann-{H}ilbert
  problems. {A}symptotics for the {MK}d{V} equation},
        date={1993},
     journal={Ann. of Math.},
      volume={137},
       pages={295\ndash 368},
         doi={10.2307/2946540},
}

\bib{FW-04}{article}{
      author={Forrester, P.~J.},
      author={Witte, N.~S.},
       title={{Application of the $\tau$-function theory of Painlev\'e
  equations to random matrices: $P_\mathrm{VI}$, the JUE, CyUE, cJUE and scaled
  limits}},
        date={2004},
     journal={Nagoya Math. J.},
      volume={174},
       pages={29\ndash 114},
      eprint={math-ph/0204008},
         doi={10.1017/S0027763000008801},
}

\bib{FIK-92}{article}{
      author={Fokas, A.~S.},
      author={Its, A.~R.},
      author={Kitaev, A.~V.},
       title={{The isomonodromy approach to matrix models in 2D quantum
  gravity}},
        date={1992},
     journal={Comm. Math. Phys.},
      volume={183},
       pages={395\ndash 430},
         doi={10.1007/BF02096594},
}

\bib{BKMM-07}{book}{
      author={Baik, J.},
      author={Kriecherbauer, T.},
      author={McLaughlin, K. T.-R.},
      author={Miller, P.~D.},
       title={{Discrete orthogonal polinomials: Asymptotics and applications}},
      series={Annals of Mathematics Studies},
   publisher={Princeton University Press},
     address={Princeton, NJ},
        date={2007},
      volume={164},
}

\bib{W-87}{book}{
      author={Wasow, W.},
       title={Asymptotic expansions for ordinary differential equations},
   publisher={Dover Publications},
     address={New York},
        date={1987},
}

\bib{PGA-22}{article}{
      author={Pallister, J.~S.},
      author={Gangardt, D.~M.},
      author={Abanov, A.~G.},
       title={Limit shape phase transitions: a merger of arctic circles},
        date={2022},
     journal={J. Phys. A: Math. Theor.},
      volume={55},
       pages={304001},
      eprint={2203.05269},
         doi={10.1088/1751-8121/ac79ad},
}

\bib{LW-72}{incollection}{
      author={Lieb, E.~H.},
      author={Wu, F.~Y.},
       title={Two dimensional ferroelectric models},
        date={1972},
   booktitle={{Phase Transitions and Critical Phenomena}},
      editor={Domb, C.},
      editor={Green, M.~S.},
      volume={1},
   publisher={Academic Press},
     address={London},
       pages={331\ndash 490},
}

\bib{B-82}{book}{
      author={Baxter, R.~J.},
       title={{Exactly Solved Models in Statistical Mechanics}},
   publisher={Academic Press},
     address={San Diego, CA},
        date={1982},
}

\bib{CP-15}{article}{
      author={Colomo, F.},
      author={Pronko, A.~G.},
       title={Thermodynamics of the six-vertex model in an {L}-shaped domain},
        date={2015},
     journal={Comm. Math. Phys.},
      volume={339},
       pages={699\ndash 728},
      eprint={1501.03135},
         doi={10.1007/s00220-015-2406-9},
}

\bib{LPW-90}{article}{
      author={Li, W.},
      author={Park, H.},
      author={Widom, M.},
       title={Finite-size scaling amplitudes in a random tiling model},
        date={1990},
     journal={J. Phys. A: Math. Gen.},
      volume={23},
       pages={L573–L580},
         doi={10.1088/0305-4470/23/11/011},
}

\bib{B-08a}{article}{
      author={Bogoliubov, N.~M.},
       title={Four-vertex model and random tilings},
        date={2008},
     journal={Theor. Math. Phys.},
      volume={155},
       pages={523\ndash 535},
      eprint={0711.0030},
         doi={10.1007/s11232-008-0043-6},
}

\bib{BCMP-23}{article}{
      author={Burenev, I.~N.},
      author={Colomo, F.},
      author={Maroncelli, A.},
      author={Pronko, A.~G.},
       title={Arctic curves of the four-vertex model},
        date={2023},
     journal={J. Phys. A: Math. Theor.},
      volume={56},
       pages={465202},
      eprint={2307.03076},
         doi={10.1088/1751-8121/ad02ce},
}

\bib{KLS-10}{book}{
      author={Koekoek, R.},
      author={Swarttouw, R.~F.},
      author={Lesky, P.~A.},
       title={Hypergeometric orthogonal polynomials and their $q$-analogues},
      series={Springer Monographs in Mathematics},
   publisher={Springer-Verlag},
     address={Berlin},
        date={2010},
}

\bib{BS-09}{article}{
      author={Berg, C.},
      author={Szwarc, R.},
       title={The smallest eigenvalue of {H}ankel matrices},
        date={2011},
     journal={Constr. Approx.},
      volume={34},
       pages={107\ndash 133},
      eprint={0906.4506},
         doi={10.1007/s00365-010-9109-4},
}

\bib{Fe-84}{article}{
      author={Fiedler, M.},
       title={{H}ankel and {L}oewner matrices},
        date={1984},
     journal={Linear Algebra Appl.},
      volume={58},
       pages={75\ndash 95},
         doi={10.1016/0024-3795(84)90205-2},
}

\bib{I-56}{book}{
      author={Ince, E.~L.},
       title={Ordinary differential equations},
   publisher={Dover},
     address={New York},
        date={1956},
}

\bib{E-81}{book}{
      author={Erd\'elyi, A.},
       title={{Higher Transcendental Functions, Vol. I}},
   publisher={Robert E. Krieger Publishing Company},
     address={Malabar, FL},
        date={1981},
}

\bib{MS-14}{article}{
      author={Majumdar, S.~N.},
      author={Schehr, G.},
       title={Top eigenvalue of a random matrix: large deviations and third
  order phase transition},
        date={2014},
     journal={J. Stat. Mech.},
      volume={2014},
       pages={P01012},
      eprint={1311.0580},
         doi={10.1088/1742-5468/2014/01/P01012},
}

\bib{PW-96}{article}{
      author={Propp, J.~G.},
      author={Wilson, D.~B.},
       title={Exact sampling with coupled {M}arkov chains and applications to
  statistical mechanics},
        date={1996},
     journal={Random Struct. Algor.},
      volume={9},
       pages={223\ndash 252},
  doi={10.1002/(SICI)1098-2418(199608/09)9:1/2<223::AID-RSA14>3.0.CO;2-O},
}

\bib{PW-98}{incollection}{
      author={Propp, J.},
      author={Wilson, D.},
       title={Coupling from the past: A user's guide},
        date={1998},
   booktitle={Microsurveys in discrete probability},
      editor={Aldous, D.},
      editor={Propp, J.},
      series={DIMACS: Series in Discrete Mathematics and Theoretical Computer
  Science},
      volume={41},
   publisher={AMS},
       pages={181\ndash 192},
         doi={10.1090/dimacs/041},
}

\bib{CLP-98}{article}{
      author={Cohn, H.},
      author={Larsen, M.},
      author={Propp, J.},
       title={{The shape of a typical boxed plane partition}},
        date={1998},
     journal={New York J. Math.},
      volume={4},
       pages={137\ndash 165},
      eprint={math.CO/9801059},
}

\end{biblist}
\end{bibdiv}
\end{document}